\documentclass{ws-ijmpd}

\usepackage{aas_macros}
\usepackage{lscape}
\usepackage{url}

\newcommand{\pasa}{PASA}

\newcommand{\phz}{photo-$z$}
\newcommand{\phzs}{photo-$z$s}
\newcommand{\knn}{$k$NN}
\newcommand{\etal}{{\it et al.}}

\begin{document}

\markboth{N.~M. Ball \& R.~J. Brunner}{Data Mining in Astronomy}

%
\catchline{}{}{}{}{}
%

\title{DATA MINING AND MACHINE LEARNING IN ASTRONOMY}

\author{NICHOLAS M. BALL}

\address{Herzberg Institute of Astrophysics, National Research Council, 5017 West Saanich Road, Victoria, BC V9E 2E7, Canada\\
nick.ball@nrc-cnrc.gc.ca}

\author{ROBERT J. BRUNNER}

\address{Department of Astronomy, University of Illinois at Urbana-Champaign, 1002 West Green Street,\\
Urbana, IL 61801, USA\\
bigdog@illinois.edu}

\maketitle

\begin{history}
\received{Day Month Year}
\revised{Day Month Year}
\comby{Managing Editor}
\end{history}

\begin{abstract} 
We review the current state of data mining and machine learning in astronomy. {\it Data Mining} can have a somewhat mixed connotation from the point of view of a researcher in this field. If used correctly, it can be a powerful approach, holding the potential to fully exploit the exponentially increasing amount of available data, promising great scientific advance. However, if misused, it can be little more than the black-box application of complex computing algorithms that may give little physical insight, and provide questionable results. Here, we give an overview of the entire data mining process, from data collection through to the interpretation of results. We cover common machine learning algorithms, such as artificial neural networks and support vector machines, applications from a broad range of astronomy, emphasizing those where data mining techniques directly resulted in improved science, and important current and future directions, including probability density functions, parallel algorithms, petascale computing, and the time domain. We conclude that, so long as one carefully selects an appropriate algorithm, and is guided by the astronomical problem at hand, data mining can be very much the powerful tool, and not the questionable black box.
\end{abstract}



\section{Introduction} \label{Sec: Intro}

In its broadest sense, data mining is simply the act of turning raw data from an observation into useful information. This information can be interpreted by hypothesis or theory, and used to make further predictions. This scientific method, where useful statements are made about the world, has been widely employed to great effect in the West since the Renaissance, and even earlier in other parts of the world. What has changed in the past few decades is the exponential rise in available computing power, and, as a related consequence, the enormous quantities of observed data, primarily in digital form. The exponential rise in the amount of available data is now creating, in addition to the natural world, a digital world, in which extracting new and useful information from the data already taken and archived is becoming a major endeavor in itself. This action of {\it knowledge discovery in databases} (KDD), is what is most commonly inferred by the phrase data mining, and it forms the basis for our review.

Astronomy has been among the first scientific disciplines to experience this flood of data. The emergence of data mining within this and other subjects has been described\cite{bell:petascale,bell:deluge,hey:4thparadigm} as the {\it fourth paradigm}. The first two paradigms are the well-known pair of theory and observation, while the third is another relatively recent addition, computer simulation. The sheer volume of data not only necessitates this new paradigmatic approach, but the approach must be, to a large extent, automated. In more formal terms, we wish to leverage a computational machine to find patterns in digital data, and translate these patterns into useful information, hence {\it machine learning}. This learning must be returned in a useful manner to a human investigator, which hopefully results in human learning.

It is perhaps not entirely unfair to say, however, that scientists in general do not yet appreciate the full potential of this fourth paradigm. There are good reasons for this of course: scientists are generally not experts in databases, or cutting-edge branches of statistics, or computer hardware, and so forth. What we hope to do in this review, primarily for the data mining skeptic, is to shed light on why this is a useful approach. To accomplish this goal, we emphasize either algorithms that have or could currently be usefully employed, and the actual scientific results they have enabled. We also hope to give an interesting and fairly comprehensive overview to those who do already appreciate this approach, and perhaps provide inspiration for exciting new ideas and applications. However, despite referring to data mining as a whole new paradigm, we try to emphasize that it is, like theory, observation, and simulation, only a part of the broader scientific process, and should be viewed and utilized as such. The algorithms described are {\it tools} that, when applied correctly, have vast potential for the creation of useful scientific results. But, given that it is only part of the process, it is, of course, not the answer to everything, and we therefore enumerate some of the limitations of this new paradigm.

We start in \S \ref{Subsec: Why} with a summary of some of the advantages of this approach. In \S \ref{Sec: Overview}, we summarize the process from the input of raw data to the visualization of results. This is followed in \S \ref{Sec: Uses} by the actual application of data mining tools in astronomy. \S \ref{Sec: Overview} is arranged algorithmically, and  \S \ref{Sec: Uses} is arranged astrophysically. It is likely that the expert in astronomy or data mining, respectively, could infer much of \S \ref{Sec: Uses} from \S \ref{Sec: Overview}, and vice-versa. But it is unlikely (we hope) that the combination of the two sections does not have new ideas or insights to offer to either audience. Following these two sections, in \S \ref{Sec: Future}, we combine the lessons learned to discuss the future of data mining in astronomy, pointing out likely near-term future directions in both the data mining process and its physical application. We conclude with a summary of the main points in \S \ref{Sec: Conclusions}.

\subsection{Why Data Mining?} \label{Subsec: Why}

Of course, what astronomers care about is not a fashionable new computational method for ever more complex data analysis, but the {\it science}. A fancy new data mining system is not worth much if all it tells you is what you could have gained by the judicious application of existing tools and a little physical insight\cite{hand:illusion}. We therefore summarize some of the advantages of this approach:

\begin{itemlist}

\item {\it Getting anything at all}: upcoming datasets will be almost overwhelmingly large. When one is faced with Petabytes of data, a rigorous, automated approach that intelligently extracts pertinent scientific information will be the only one that is tractable.

\item {\it Simplicity}: despite the apparent plethora of methods, straightforward applications of very well-known and well-tested data mining algorithms can quickly produce a useful result. These methods can generate a model appropriate to the complexity of an input dataset, including nonlinearities, implicit prior information, systematic biases, or unexpected patterns. With this approach, {\it a priori} data sampling of the type exemplified by elaborate color cuts, is not necessary. For many algorithms, new data can be trivially incorporated as they become available.

\item {\it Prior information}: this can be either fully incorporated, or the data can be allowed to completely `speak for themselves'. For example, an unsupervised clustering algorithm can highlight new classes of objects within a dataset that might be missed if a prior set of classifications were imposed.

\item {\it Pattern recognition}: an appropriate algorithm can highlight patterns in a dataset that might not otherwise  be noticed by a human investigator, perhaps due to the high dimensionality. Similarly, rare or unusual objects can be highlighted.

\item {\it Complimentary approach}: although there are numerous examples where the data mining approach demonstrably exceeds more traditional methods in terms of scientific return. Even when the approach does not produce a substantial improvement, it still acts as an important complementary method of analyzing data, because different approaches to an overall problem help to mitigate systematic errors in any one approach.

\end{itemlist}

\section{Overview of Data Mining and Machine Learning Methods} \label{Sec: Overview}

In this section, we review the data mining process. Specifically, as described in \S \ref{Sec: Intro}, this data mining review focuses on knowledge discovery in databases (KDD), although our definition of a `database' is somewhat broad, essentially being any machine-readable astronomical data. As a result, this section is arranged algorithmically. To avoid overlap with \S \ref{Sec: Uses} on the astronomical uses, we defer most of the application examples to that section. Nevertheless, all algorithms we describe have been, or are of sufficient maturity that they could immediately be applied to astronomical data. The reader who is expert in astronomy but not in data mining is advised to read this section to gain the full benefit from \S \ref{Sec: Uses}. As in any specialized subject, a certain level of jargon is necessary for clarity of expression. Terms likely to be unfamiliar to astronomers not versed in data mining are generally explained as they are introduced, but for additional background we note that there are other useful reviews of the data mining field\cite{witten:datamining2nd,bishop:pattern,hastie:learning2nd}. Another recent overview of data mining in astronomy by Borne has also been published\cite{borne:datamining}.

\subsection{Data Collection} \label{Subsec: Data}

The process of data collection encompasses all of the steps required to obtain the desired data in a digital format. Methods of data collection include acquiring and archiving new observations, querying existing databases according to the science problem at hand, and performing as necessary any cross-matching or data combining, a process generically described as {\it data fusion}.

A common motivation for cross-matching is the use of multiwavelength data, i.e., data spanning more than one of the regions of the electromagnetic spectrum (gamma ray, X-ray, ultraviolet, optical, infrared, millimeter, and radio). A common method in the absence of a definitive identification for each object spanning the datasets is to use the object's position on the sky with some astrometric tolerance, typically a few arcseconds. Cross-matching can introduce many issues including ambiguous matches, variations of the point spread function (resolution of objects) within or between datasets, differing survey footprints, survey masks, and large amounts of processing time and data transfer requirements when cross-matching large datasets.

A major objective of the {\it Virtual Observatory} (VO, \S \ref{Subsec: VO}) is to make the data collection process more simple and tractable. Future VO webservices are planned that will perform several functions in this area, including cross-matches on large, widely distributed, heterogeneous data.

Common astronomical data formats include FITS\cite{wells:fits}, a binary format, and plain ASCII, while an emerging format is VOTable\cite{ochsenbein:votable}. Commonly used formats from other areas of data mining, such as attribute relation file format (ARFF)\footnote{\url{http://weka.wiki.sourceforge.net/ARFF}}, are generally not widely used in astronomy.

\subsection{Preprocessing of Data} \label{Subsec: Preprocessing}

Some data preprocessing may necessarily be part of the data collection process, for example, sample cuts in database queries. Preprocessing can be divided into steps that make the data to be read meaningful, and those that transform the data in some way as appropriate to a given algorithm. Data preprocessing is often problem-dependent, and should be carefully applied because the results of many data mining algorithms can be significantly affected by the input data. A useful overview of data preprocessing is given by Pyle\cite{pyle:dataprep}.

Algorithms may require the object {\it attributes}, i.e., the values in the data fields describing the properties of each object, to be numerical or categorical, the latter being, e.g. `star', or `galaxy'. It is possible to transform numerical data to categorical and vice versa.

A common categorical-to-numerical method is scalarization, in which different possible categorical attributes are given different numerical labels, for example, `star', `galaxy', `quasar' labeled as the vectors [1,0,0], [0,1,0], and [0,0,1], respectively. Note that for some algorithms, one should {\it not} label categories as, say, 1, 2 and 3, if the output of the algorithm is such that if it has confused an object between 1 and 3 it labels the object as intermediate, in this case, 2. Here, 2 (galaxy) is certainly not an intermediate case between 1 (star) and 3 (quasar). One common algorithm in which such categorical but not ordered outputs could occur is a decision tree with multiple outputs.

Numerical data can be made categorical by transformations such as binning. The bins may be user-specified, or can be generated optimally from the data\cite{hogg:histogram}. Binning can create numerical issues, including comparing two floating point numbers that should be identical, objects on a bin edge, empty bins, values that cannot be binned such as {\it NaN}, or values not within the bin range.

Object attributes may need to be {\it transformed}. A common operation is the differencing of magnitudes to create colors. These transformations can introduce their own numerical issues, such as division by zero, or loss of accuracy.

In general, data will contain one or more types of {\it bad values}, where the value is not correct. Examples include instances where the value has been set to something such as -9999 or NaN, the value appears correct but has been flagged as bad, or the value is not bad in a formatting sense but is clearly unphysical, perhaps a magnitude of a high value that could not have been detected by the instrument. They may need to be removed either by simply removing the object containing them, ignoring the bad value but using the remaining data, or interpolating a value using other information. Outliers may or may not be excluded, or may be excluded depending on their extremity.

Data may also contain {\it missing values}. These values may be genuinely missing, for example in a cross-matched dataset where an object is not detected in a given waveband, or is not in an overlapping region of sky. It is also possible that the data should be present, but are missing for either a known reason, such as a bad camera pixel, a cosmic ray hit, or a reason that is simply not known. Some algorithms cannot be given missing values, which will require either the removal of the object or interpolation of the value from the existing data. The advisability of interpolation is problem-dependent.

As well as bad values, the data may contain values that are correct but are outside the desired range of analysis. The data may therefore need to be {\it sampled}. There may simply be a desired range, such as magnitude or position on the sky, or the data may contain values that are correct but are outliers. Outliers may be included, included depending on their extremity (e.g., $n$ standard deviations), downweighted, or excluded. Alternatively, it may be more appropriate to generate a random subsample to produce a smaller dataset.

Outside any normalization of the data prior to its use in the data mining algorithm, for example, calibration using standard sources, input or target attributes of the data will often be further normalized to improve the {\it numerical conditioning} of the algorithm. For example, if one axis of the $n$-dimensional space created by $n$ input attributes encompasses a range that, numerically, is much larger than the other axes, it may dominate the results, or create conditions where very large and small numbers interact, causing loss of accuracy. Normalization can reduce this, and examples include linear transformations, like scaling by a given amount, scaling using the minimum and maximum values so that each attribute is in a given range such as 0--1, or scaling each attribute to have a mean of 0 and a standard deviation of 1. The latter example is known as {\it standardization}. A more sophisticated transformation with similar advantages is {\it whitening}, in which the values are not only scaled to a similar range, but correlations among the attributes are removed via transformation of their covariance matrix to the identity matrix.

\subsection{Attribute Selection} \label{Subsec: Attributes}

In general, a large number of attributes will be available for each object in a dataset, and not all will be required for the problem. Indeed, use of all attributes may in many cases worsen performance. This is a well-known problem, often called the {\it curse of dimensionality}. The large number of attributes results in a high-dimensional space with many low density environments or even empty voids. This makes it difficult to generalize from the data and produce useful new results. One therefore requires some form of {\it dimension reduction}, in which one wishes to retain as much of the information as possible, but in fewer attributes. As well as the curse of dimensionality, some algorithms work less well with noisy, irrelevant, or redundant attributes. An example of an irrelevant attribute might be position on the sky for a survey with a uniform mask, because the position would then contain no information, and highly redundant attributes might be a color in the same waveband measured in two apertures.

The most trivial form of dimension reduction is simply to use one's judgement and select a subset of attributes. Depending on the problem this can work well. Nevertheless, one can usually take a more sophisticated and less subjective approach, such as principal component analysis (PCA)\cite{karhunen:kl,loeve:kl,jolliffe:pca}. This is straightforward to implement, but is limited to linear relations. It gives, as the principal components, the eigenvectors of the input data, i.e., it picks out the directions which contain the greatest amount of information. Another straightforward approach is {\it forward selection}, in which one starts with one attribute and selectively adds new attributes to gain the most information. Or, one can perform the equivalent process but starting with all of the attributes and removing them, known as {\it backward elimination}.

In many ways, dimension reduction is similar to classification, in the sense that a larger number of input attributes is reduced to a smaller number of outputs. Many classification schemes in fact directly use PCA. Other dimension reduction methods utilize the same or similar algorithms to those used for the actual data mining: an ANN can perform PCA when set up as an autoencoder, and kernel methods can act as generalizations of PCA. A binary genetic algorithm (\S \ref{Subsubsec: Other}) can be used in which each individual represents a subset of the training attributes to be used, and the algorithm selects the best subset. The epsilon-approximate nearest neighbor search\cite{arya:epsilon} reduces the dimensionality of nearest neighbor methods. Other methods include information bottleneck\cite{tishby:ib}, which directly uses information theory to optimize the tradeoff between the number of classes and the information contained, Fisher Matrix\cite{fisher:matrix}, Independent Component Analysis\cite{hyvarinen:ica}, and wavelet transforms. The curse of dimensionality is likely to worsen in the future for a similar reason to that of missing values, as more multiwavelength datasets become available to be cross-matched. Classification and dimension reduction are not identical of course: a classification algorithm may build a model to represent the data, which is then applied to further examples to predict their classes.

\subsection{Selection and Use of Machine Learning Algorithms} \label{Subsec: Algorithm}

Machine learning algorithms broadly divide into {\it supervised} and {\it unsupervised} methods, also known as predictive and descriptive, respectively. These can be generalized to form {\it semi-supervised} methods. Supervised methods rely on a {\it training set}\footnote{For many astronomical applications, one might more properly call it a training {\it sample}, but the term training set is in widespread use, so we use that here to avoid confusion.} of objects for which the target property, for example a classification, is known with confidence. The method is trained on this set of objects, and the resulting mapping is applied to further objects for which the target property is not available. These additional objects constitute the {\it testing set}. Typically in astronomy, the target property is spectroscopic, and the input attributes are photometric, thus one can predict properties that would normally require a spectrum for the generally much larger sample of photometric objects. The training set must be representative, i.e., the parameter space covered by the input attributes must span that for which the algorithm is to be used. This might initially seem rather restrictive, but in many cases can be handled by combining datasets. For example, the zCOSMOS redshift survey\cite{lilly:zcosmos}, at one square degree, provides spectra to the depth of the photometric portion of the Sloan Digital Sky Survey (SDSS)\cite{york:sdss}, $r \sim 22$ mag, which covers over 8000 square degrees. Since SDSS photometry is available for zCOSMOS objects, one can in principle use the 40,000 zCOSMOS galaxies as a training set to assign photometric redshifts to over 200 million SDSS galaxies.

In contrast to supervised methods, unsupervised methods do not require a training set. This is an advantage in the sense that the data can speak for themselves without preconceptions such as expected classes being imposed. On the other hand, if there is prior information, it is not necessarily incorporated. Unsupervised algorithms usually require some kind of initial input to one or more of the adjustable parameters, and the solution obtained can depend on this input.

Semi-supervised methods attempt to allow the best-of-both-worlds, and both incorporate known priors while allowing objective data interpretation and extrapolation. But given their generality, they can be more complex and difficult to implement. They are of potentially great interest astronomically because they could be used to analyze a full photometric survey beyond the spectroscopic limit, without requiring priors, while at the same time incorporating the prior spectroscopic information where it is available.

\subsubsection{Supervised Methods} \label{Subsubsec: Supervised}

\begin{figure}[!ht]
\centerline{\psfig{file=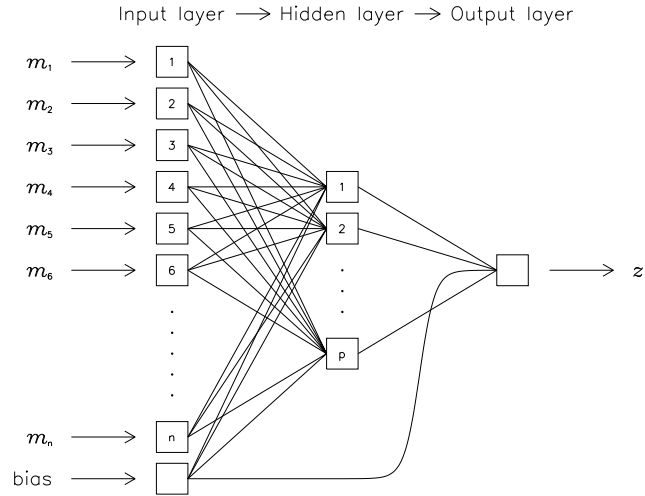,width=4in}}
\vspace*{8pt}
\caption{Schematic of an artificial neural network for an object with $n$ attributes, a hidden layer of size $p$, and a single continuously-valued output, in this case, the redshift, $z$. From Firth, Lahav \& Somerville\protect \cite{firth:annphotoz}. \label{Fig: ANN}}
\end{figure}

The most widely used and well-known machine learning algorithm in astronomy to-date, referred to as far back as the mid 1980s,\cite{jeffrey:annealing} is the {\it artificial neural network} (ANN, Fig. \ref{Fig: ANN})\cite{bishop:ann,ripley:ann,duda:pattern}. This consists of a series of interconnected nodes with weighted connections. Each node has an activation function, perhaps a simple threshold, or a sigmoid. Although the original motivation was that the nodes would simulate neurons in the brain,\cite{mcculloch:ann,hopfield:ann} the ANNs in data mining are of such a size that they are best described as nonlinear extensions of conventional statistical methods.

The supervised ANN takes parameters as input and maps them on to one or more outputs. A set of parameter vectors, each vector representing an object and corresponding to a desired output, or target, is presented. Once the network is trained, it can be used to assign an output to an unseen parameter vector. The training uses an algorithm to minimize a cost function. The cost function, $c$, is commonly of the form of the mean-squared deviation between the actual and desired output: $$c = \frac{1}{N} \sum_{k=1}^N~(o_{k} - t_{k})^{2},$$ where $o_{k}$ and $t_{k}$ are the output and target respectively for the $k$th of $N$ objects.

In general, the neurons could be connected in any topology, but a commonly used form is to have an $a: b_1 : b_2 : \ldots : b_n :c$ arrangement, where $a$ is the number of input parameters, $b_{1,\ldots,n}$ are the number of neurons in each of $n$ one dimensional `hidden' layers, and $c$ is the number of neurons in the final layer, which is equal to the number of outputs. Each neuron is connected to every neuron in adjacent layers, but not to any others. Multiple outputs can each give the Bayesian {\it a posteriori} probability that the output is of that specific class, given the values of the input parameters.

The weights are adjusted by the training algorithm. In astronomy this has typically been either the well-known backpropagation algorithm\cite{werbos:backprop,parker:backprop,rumelhart:backprop} or the quasi-Newton algorithm\cite{bishop:ann}, although other algorithms, such as Levenberg-Marquardt\cite{levenberg:levmarq,marquardt:levmarq} have also been used.

\begin{figure}[!ht]
\centerline{\psfig{file=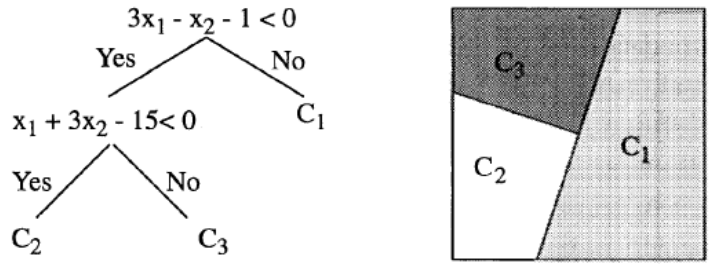,width=4in}}
\vspace*{8pt}
\caption{As Fig. \ref{Fig: ANN}, but showing a decision tree. The oblique planes specified by the division criteria on the input attributes $x_1$ and $x_2$ at the nodes in this case divide the input parameter space into three regions. From Salzberg \etal \protect \cite{salzberg:dt}. \label{Fig: DT}}
\end{figure}

Another common method used in data mining is the {\it decision tree} (DT, Fig. \ref{Fig: DT})\cite{morgan:dt,breiman:dt,quinlan:dt,quinlan:dtbook,rokach:dt}. Decision trees consist first of a root node which contains all of the parameters describing the objects in the training set population along with their classifications. A node is split into child nodes using the criterion that minimizes the classification error. This splitting subdivides the parent population group into children population groups, which are assigned to the respective child nodes. The classification error quantifies the accuracy of the classification on the test set. The process is repeated iteratively, resulting in layered nodes that form a tree. The iteration stops when specific user-determined criteria are reached. Possibilities include a minimum allowed population of objects in a node (the minimum decomposition population), the maximum number of nodes between the termination node and the root node (the maximum tree depth), or a required minimum decrease resulting from a population split (the minimum error reduction). The terminal nodes are known as the leaf nodes. The split is tested for each input attribute, and can be axis-parallel, or oblique, which allows for hyperplanes at arbitrary angles in the parameter space. The split statistic can be the midpoint, mean, or median of the attribute values, while the cost function used is typically the variance, as with ANN.

\begin{figure}[!ht]
\centerline{\psfig{file=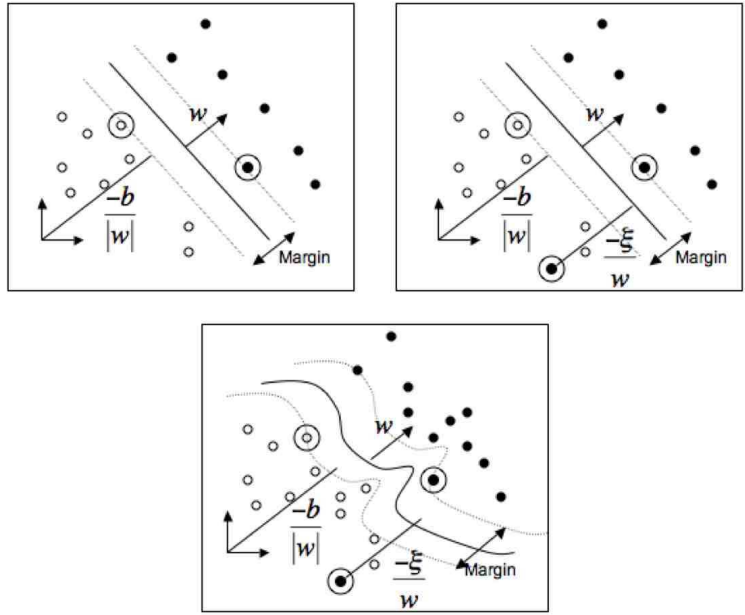,width=4in}}
\vspace*{8pt}
\caption{As Fig. \ref{Fig: ANN}, but showing a support vector machine. The circled points are the support vectors between the two classes of objects, represented by open and filled circles. The cases shown are separable and non-separable data with linear and nonlinear boundaries. $w$ is the normal to the hyperplane, and $b$ is the perpendicular distance. From Huertas-Company \etal\protect \cite{huertascompany:svmmorph}, to which the reader is referred for details of $\xi$. \label{Fig: SVM}}
\end{figure}

In recent years, another algorithm, the {\it support vector machine} (SVM, Fig. \ref{Fig: SVM})\cite{cortes:svm,burges:svm,vapnik:svm,cristianini:svm,kecman:svm,schlkopf:svm,abe:svm,wang:svm,steinwart:svm}, has gained popularity in astronomical data mining. SVM aims to find the hyperplane that best separates two classes of data. The input data are viewed as sets of vectors, and the data points closest to the classification boundary are the support vectors. The algorithm does not create a model of the data, but instead creates the decision boundaries, which are defined in terms of the support vectors. The input attributes are mapped into a higher dimensional space using a kernel so that nonlinear relationships within the data become linear (the `kernel trick')\cite{aizerman:kerneltrick}, and the decision boundaries, which are linear, are determined in this space. Like ANN and DT, the training algorithm minimizes a cost function, which in this case is the number of incorrect classifications. The algorithm has two adjustable hyperparameters: the width of the kernel, and the regularization, or cost, of classification error, which helps to prevent {\it overfitting} (\S \ref{Subsec: Improving}) of the training set. The shape of the kernel is also an adjustable parameter, a common choice being the Gaussian radial basis function. As a result, an SVM has fewer adjustable parameters than an ANN or DT, but because these parameters must be optimized, the training process can still be computationally expensive. SVM is designed to classify objects into two classes. Various refinements exist to support additional classes, and to perform regression, i.e., to supply a continuous output value instead of a classification. Classification probabilities can be output, for example, by using the distance of a data point from the decision boundary.

Another powerful but computationally intensive method is {\it k nearest neighbor} (\knn)\cite{fix:knn,cover:nn,aha:ib,dasarathy:knn,shakhnarovich:knn}. This method is powerful because it can utilize the full information available for each object, with no approximations or interpolations. The training of \knn~is in fact trivial: the positions of each of the objects in the input attribute space are simply stored in memory. For each test object, the same attributes are compared to the training set and the output is determined using the properties of the nearest neighbors. The simplest implementation is to output the properties of the single nearest neighbor, but more commonly the weighted sum of $k$ nearest neighbors is used. The weighting is typically the inverse Euclidean distance in the attribute space, but one can also use adaptive distance metrics. The main drawback of this method is that is it computationally intensive, because for each testing object the entire training set must be examined to determine the nearest neighbors. This requires a large number of distance calculations, since the test datasets are often much larger than the training datasets. The workload can be mitigated by storing the training set in an optimized data structure, such as a kd-tree.

However, in the past few years, novel supercomputing hardware (which is discussed in more detail in \S \ref{Subsec: Hardware}) has become available that is specifically designed to carry out exactly this kind of computationally intensive work, including applications involving a large number of distance calculations. The curve of growth of this technology exceeds that of conventional CPUs, and thus the direct implementation of \knn~using this technology has the potential to exceed the performance of conventional CPUs.

\subsubsection{Unsupervised Methods} \label{Subsubsec: Unsupervised}

{\it Kernel density estimation} (KDE)\cite{parzen:kde,duda:kde,silverman:kde,scott:kde,taylor:kde,wasserman:kde,klemela:kde} is a method of estimating the probability density function of a variable. It is a generalization of a histogram where the kernel function is any shape instead of the top-hat function of a histogram bin. This has the advantages that it avoids the discrete nature of the histogram and does not depend on the position of the bin edges, but the width of the kernel must still be chosen so as not to over- or under-smooth the data. A Gaussian kernel is commonly utilized. In higher numbers of dimensions, common in astronomical datasets, the width of the kernel must be specified in each dimension.

{\it K-means clustering}\cite{steinhaus:kmeans,macqueen:kmeans} is an unsupervised method that divides data into clusters. The number of clusters must be initially specified, but since the algorithm converges rapidly, many starting points can be tested. The algorithm uses a distance criterion for cluster membership, such as the Euclidean distance, and a stopping criterion for iteration, for example, when the cluster membership ceases to change.

{\it Mixture models}\cite{titterington:mixture,mclachlan:mixture} decompose a distribution into a sum of components, each of which is a probability density function. Often, the distributions are Gaussians, resulting in Gaussian mixture models. They are often used for clustering, but also for density estimation, and they can be optimized using either expectation maximization or Monte Carlo methods. Many astronomical datasets consist of contributions from different populations of objects, which allows mixture modeling to disentangle these population groups. Mixture models based on the EM algorithm have been used in astronomy for this purpose\cite{connolly:fast,dolence:lci}.

{\it Expectation maximization} (EM)\cite{dempster:em,watanabe:em,mclachlan:em} treats the data as a sum of probability distributions, which each represent one cluster. This method alternates between an expectation stage and a maximization stage. In the expectation stage, the algorithm evaluates the membership probability of each data point given the current distribution parameters. In the maximization stage, these probabilities are used to update the parameters. This method works well with missing data, and can be used as the unsupervised component in semi-supervised learning (\S \ref{Subsubsec: Semisupervised}) to provide class labels for the supervised learning.

The {\it Kohonen self-organizing map} (SOM)\cite{kohonen:somprev,kohonen:som} is an unsupervised neural network that forms a general framework for visualizing datasets of more than two dimensions. Unlike many methods which seek to map objects onto a new output space, the SOM is fundamentally topological. This is neatly illustrated by the fact that one astronomical SOM application\cite{naim:som} is titled `Galaxy Morphology Without Classification'. A related earlier method is learning vector quantization\cite{kohonen:lvq}.

{\it Independent component analysis} (ICA)\cite{comon:ica,lee:ica,hyvarinen:ica,roberts:ica,stone:ica}, an example of {\it blind source separation}, can separate nonlinear components of a dataset, under the assumption that those components are statistically independent. The components are found by maximizing this independence. Related statistical methods include principal component analysis (\S \ref{Subsec: Attributes}), singular value decomposition, and non-negative matrix factorization.

\subsubsection{Semi-Supervised} \label{Subsubsec: Semisupervised}

\begin{figure}[!ht]
\centerline{\psfig{file=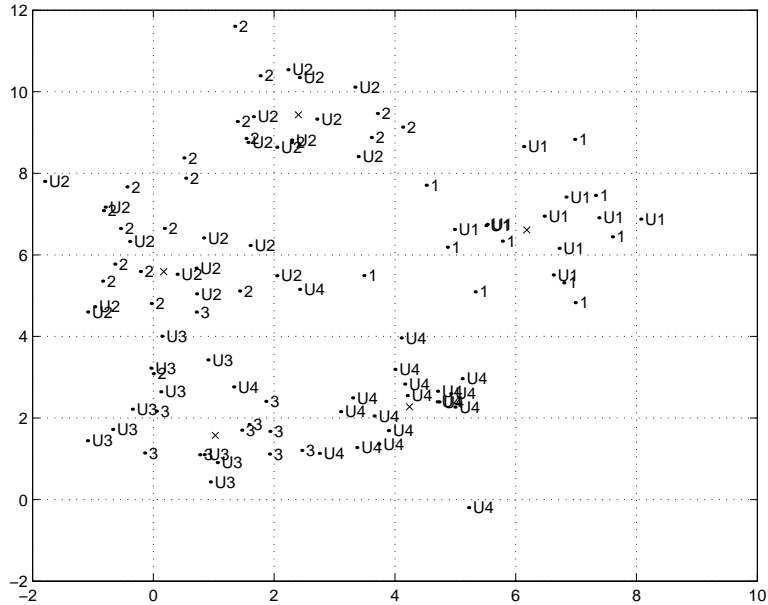,width=4in}}
\vspace*{8pt}
\caption{Dataset amenable to semi-supervised learning, showing labeled and unlabeled classes, denoted by 1--4 and U1--U4, respectively. The axes are arbitrary units. The crosses result from a mixture model applied to the data. From Bazell \& Miller\protect \cite{bazell:classdisc}. \label{Fig: Semi-supervised}}
\end{figure}

The semi-supervised approach\cite{chapelle:semisupervised,zhu:semisupervised} has been somewhat underused to-date, but holds great potential for the upcoming, large, purely photometric surveys. Supervised methods require a labeled training set, but will not assign new classes. On the other hand, unsupervised methods do not require training, but do not use existing known information. Semi-supervised methods aim to capture the best from both of these methods by retaining the ability to discover new classes within the data, and also incorporating information from a training set when available. An example of a dataset amenable to the approach is shown in Fig. \ref{Fig: Semi-supervised}.

This is particularly relevant in astronomical applications using large amounts of photometric and a more limited subsample of spectroscopic data, which may not be fully representative of the photometric sample. The semi-supervised approach allows one to use the spectral information to extrapolate into the purely photometric regime, thereby allowing a scientist to utilize all of the vast amount of information present there.

Semi-supervised learning represents an entire subfield of data mining research. Given the nontrivial implementation requirements, this field is a good area for potential fruitful collaborations between astronomers, computer scientists, and statisticians. As one example of a possible issue, a lot of photometric data are likely to be a direct continuation in parameter space of spectroscopic data, with, therefore, a highly overlapping distribution. This means that certain semi-supervised approaches will work better than others, because they contain various assumptions about the nature of the labeled and unlabeled data.

\subsubsection{Other Algorithms} \label{Subsubsec: Other}

In \S\S \ref{Subsubsec: Supervised}--\ref{Subsubsec: Unsupervised} above, we described the main data mining algorithms used to date in astronomy, however, there are numerous additional algorithms available, which have often been utilized to some extent. These algorithms may be employed at more than one stage in the process, such as attribute selection, as well as the classification/regression stage.

While neural networks in some very broad sense mimic the learning mechanism of the brain, {\it genetic algorithms}\cite{holland:genetic,goldberg:genetic,coley:ga,mitchell:ga,haupt:genetic2nd,sivanandam:ga} mimic natural selection, as the most successful individuals created are those that are best adapted for the task at hand. 

The simplest implementation is the binary genetic algorithm, in which each `individual' is a vector of ones and zeros, which represent whether or not a particular attribute, e.g., a training set attribute, is used. From an initial random population, the individuals are combined to create new individuals. The fitness of each individual is the resulting error in the training algorithm run according to the formula encoded by the individual. This process is repeated until convergence if found, producing the best individual.

A typical method of combining two individuals is one-point crossover, in which segments of two individuals are swapped. To more fully explore the parameter space, and to prevent the algorithm from converging too rapidly on a local minimum, a probability of mutation is introduced into the newly created individuals before they are processed. This is simply the probability that a zero becomes a one, or vice-versa. An approximate number of individuals to use is given by $n_{{\mathrm{in}}} \sim 2n_f~{\mathrm{log}}(n_f),$ where $n_f$ is the number of attributes. The algorithm converges in $n_{{\mathrm{it}}} \sim \alpha n_f~{\mathrm{log}}(n_f)$ iterations, where $\alpha$ is a problem-dependent constant; generally $\alpha > 3$.

Numerous refinements to this basic approach exist, including using continuous values instead of binary ones, and more complex methods for combining individuals. Further possibilities for the design of genetic algorithms exist\cite{goldberg:design}, and it is possible in principle to combine the optimization of the learning algorithm and the attribute set.

The {\it Information bottleneck} method\cite{tishby:ib} is based directly on information theory and is designed to achieve the best tradeoff between accuracy and compression for the desired number of classes. The inputs and outputs are probability density functions. {\it Association rule} mining\cite{adamo:associationrules,zhang:associationrules} is a method of finding qualitative rules within a database such that a rule derived from the occurrence of certain variables together implies something about the occurrence of a variable not used in creating that rule. The {\it false discovery rate}\cite{benjamini:fdr} is a method of establishing a significant discovery from a smaller set of data than the usual $n$ sigma hypothesis test.

This list could continue, broadening into traditional statistical methods such as least squares, and regression, as well as Bayesian methods, which are widely used in astronomy. For brevity we do not consider these additional methods, but we do note that {\it graphical models}\cite{bishop:pattern} are a general way of describing the interrelationships between variables and probabilities, and many of the data mining algorithms described earlier, such as ANNs, are special cases of these models.

\subsubsection{Choice of Algorithm} \label{Subsubsec: Choice}

Unfortunately, there is no simple method to select the optimal algorithm to use, because the most appropriate algorithm can depend not only on the dataset, but also the application for which it will be employed. There is, therefore, no single best algorithm. Likewise, the choice of software is similarly non-trivial. Many general frameworks exist, for example WEKA\cite{witten:datamining2nd} or Data to Knowledge\cite{welge:d2k}, but it is unlikely that one framework will be able to perform all steps necessary from raw catalog to desired science result, particularly for large datasets. In Table \ref{Table: Algorithms}, we summarize some of the advantages and disadvantages of some of the more popular and well-known algorithms used in astronomy. We do not attempt to summarize available software. Various other general comparisons of machine learning algorithms exist\cite{hastie:learning2nd}, as well as numerous studies comparing various algorithms for particular datasets, a field which itself is rather complex\cite{salzberg:classifiercritique}.

\begin{landscape}
\begin{table}[ph]
\tbl{Advantages and disadvantages of well-known machine learning algorithms in astronomy. These algorithms, and others, are described in more detail in \S\S \ref{Subsubsec: Supervised}--\ref{Subsubsec: Other}.}
{\begin{tabular}{@{}lll@{}} \toprule
Algorithm   &Advantages   &Disadvantages\\
\colrule
Artificial Neural Network &Good approximation of nonlinear functions    &Black-box model\\
                          &Easily parallelized                          &Local minima\\
                          &Good predictive power                        &Many adjustable parameters\\
                          &Extensively used in astronomy                &Affected by noise\\
                          &Robust to irrelevant or redundant attributes &Can overfit\\
                          &                                             &Long training time\\
                          &                                             &No missing values\\
\colrule
Decision Tree &Popular real-world data mining algorithm                 &Can generate large trees that require pruning\\
              &Can input and output numerical or categorical variables  &Generally poorer predictive power than ANN, SVM or kNN\\
              &Interpretable model                                      &Can overfit\\
              &Robust to outliers, noisy or redundant attributes        &Many adjustable parameters\\
              &Good computational scalability                           &\\
\colrule
Support Vector Machine &Copes with noise                           &Harder to classify $>2$ classes\\
                       &Gives expected error rate                  &No model is created\\
                       &Good predictive power                      &Long training time\\
                       &Popular algorithm in astronomy             &Poor interpretability\\
                       &Can approximate nonlinear functions        &Poor at handling irrelevant attributes\\
                       &Good scalability with number of attributes &Can overfit\\
                       &Unique solution (no local minima)          &Some adjustable parameters\\
\colrule
Nearest Neighbor &Uses all available information  &Computationally intensive\\
                 &Does not require training       &No model is created\\
                 &Easily parallelized             &Can be affected by noise and irrelevant attributes\\
                 &Few or no adjustable parameters &\\
                 &Good predictive power           &\\
\colrule
Expectation Maximization &Gives number of clusters in the data               &Can be biased toward Gaussians\\
                         &Fast convergence                                   &Local minima\\
                         &Copes with missing data                            &\\
                         &Can give class labels for semi-supervised learning &\\
\botrule
\end{tabular} \label{Table: Algorithms}}
\end{table}
\end{landscape}

\subsection{Improving Results} \label{Subsec: Improving}

Many of the algorithms previously described involve `greedy' optimization. In these cases, the cost function, which is the measure of how well the algorithm is performing in its classification or prediction task, is minimized in a way that does not allow the value of the function to increase much if at all. As a result, it is possible for the optimization to become trapped in a local minimum, whereby nearby configurations are worse, but better configurations exist in a different region of parameter space. Various approaches exist to overcome local minima. One approach is to simply run the algorithm several times from different starting points. Another approach is {\it simulated annealing}\cite{kirkpatrick:annealing,cerny:annealing,vanlaarhoven:annealing,aarts:annealing}, where, in following the metallurgical metaphor, the point in parameter space `heats up', thus perturbing it and allowing it to escape from the local minimum. The point is allowed to `cool', thus having the ability to find a solution closer to the global minimum.

Models produced by data mining algorithms are subject to a fundamental limitation common to many systems in which a predictive model is constructed, the {\it bias-variance tradeoff}. The bias is the accuracy of the model in describing the data, for example, a linear model might have a higher bias than a higher order polynomial. The variance is the accuracy of this model in describing new data. The higher order polynomial might have a lower bias than a linear model, but it might be more strongly affected by variations in the data and thus have a higher variance. The polynomial has {\it overfit} the data. There is usually an optimal point between minimizing bias and minimizing variance. A typical way to minimize overfitting is to measure the performance of the algorithm on a test set, which is not part of the training set, and adjusting the stopping criterion for training to stop at an appropriate location.

To help prevent overfitting, training can also be {\it regularized}, in which an extra term is introduced into the cost function to penalize configurations that add complexity, such as large weights in an ANN. This complexity can cause a function to be less smooth, which increases the likelihood of overfitting. As is the case with supervised learning, unsupervised algorithms can also overfit the data, for example, if some kind of smoothing is employed but its scalelength is too small. In this case, the algorithm will `fit the noise' and not the true underlying distribution.

Another common approach to control overfitting and improve confidence in the accuracy of the results is {\it cross-validation}, where subsets of the data are left out of the training and used for testing. The simplest form is the holdout method, where a single subset of the training data is kept out of the training, and the algorithm error is evaluated by running on this subset. However, this can have a high bias (see bias-variance tradeoff, above) if the training set is small, due to a significant portion of the training information being left out. $K$-fold cross-validation improves on this by subdividing the data into $K$ samples and training on $K-1$ samples, or alternatively using $K$ random subsets. Typically, $K=5$ or $K=10$, as small $K$ could still have high bias, as in the holdout method, but large $K$, while being less biased, can have high variance due to the testing set being small. If $K$ is increased to the size of the dataset, so that each subsample is a single point, the method becomes leave-one-out cross-validation. In all instances, the estimated error is the mean error from those produced by each run in the cross-validation.

Another important refinement to running one algorithm is the ability to use a {\it committee} of instances of the algorithm, each with different parameters. One can allow these different instantiations to vote on the final prediction, so that the majority or averaged result becomes the final answer. Such an arrangement can often provide a substantial improvement, because it is more likely that the majority will be closer to the correct answer, and that the answer will be less affected by outliers. One such committee arrangement is {\it bootstrap aggregating}, or {\it bagging}\cite{breiman:bagging,witten:datamining2nd}, where random subsamples with replacement (bootstrap samples) are taken, and the algorithm trained on each. The created algorithms vote on the testing set. Bagging is often applied to decision trees with considerable success, but it can be applied to other algorithms. The combination of bagging and the random selection of a small subset of features for splitting at each node is known as a Random Forest\texttrademark \cite{breiman:randomforest}.

{\it Boosting}\cite{hastie:learning2nd} uses the fact that several `weak' instances of an algorithm can be combined to produce a stronger instance. The weak learners are iteratively added and misclassified objects in the data gain higher weight. Thus boosting is not the same as bagging because the data themselves are weighted. Boosted decision trees are a popular approach, and many different boosting algorithms are available.

As well as committees of the same algorithm, it is also possible to combine the results of more than one different algorithm on the same dataset. Such a {\it mixture of experts} approach often provides an optimal result on real data. The outcome may be decided by voting, or the output of one algorithm can form the input to another, in a chaining approach.

For many astronomical applications, the results are, or would be, significantly improved by utilizing the full probability density function (PDF) for a predicted property, rather than simply its single scalar value. This is because much more information is retained when using the PDF. Potential uses of PDFs are described further in \S \ref{Subsec: PDF}.

\subsection{Application of Algorithms and Some Limitations} \label{Subsec: Application}

The purpose of this review is not to uncritically champion certain data mining algorithms, but to instead encourage scientific progress by exploiting the full potential of these algorithms in a considered scientific approach. We therefore end this section by outlining some of the limitations of and issues raised by KDD and the data mining approach to current and future astronomical datasets. Several of these problems might be ameliorated by increased collaboration between astronomers and data mining experts.

\begin{itemlist}

\item {\it Extrapolation}: In many astronomical applications, it is common for data with less information content to be available for a greater number of objects over a larger parameter space. The classic example is in surveys where photometric objects are typically observed several magnitudes fainter than spectroscopic objects. For a supervised learning algorithm, it is usually inappropriate to extrapolate beyond the parameter space for which the training set (e.g., the spectroscopic objects) is representative.

\item {\it Non-intuitive results}: It is very easy to run an implementation of a well-known algorithm and output a result that appears reasonable, but is in fact either statistically invalid or completely wrong. For example, randomly subsampled training and testing sets from a dataset will overlap and produce a model that overfits the data.

\item {\it Measurement error}: Most astronomical data measurements have an associated error, but most data mining algorithms do not take this explicitly into account. For many algorithms, the intrinsic spread in the data corresponding to the target property is the measurement of the error.

\item {\it Adjustable parameters}: Several algorithms have a significant number of adjustable parameters, and the optimal configuration of these parameters is not obvious. This can result in large parameter sweeps that further increase the computational requirement.

\item {\it Scalability}: Many data mining algorithms scale, for $n$ objects, as $n^2$, or even worse, making their simple application to large datasets on normal computing hardware intractable. One can often speed up a na\"ive implementation of an algorithm that must access large numbers of data points and their attributes by storing the data in a hierarchical manner so that not all the data need to be searched. A popular hierarchical structure for accomplishing this task is the kd-tree\cite{bentley:kdtree}. However, the implementation of such trees for large datasets and on parallel machines remains a difficult problem\cite{gardner:paralleltree}.

\item{\it Learning Curve}: Data mining is an entire field of study in its own right, with strong connections to statistics and computing. The avoidance of some of the issues we present, such as the selection of appropriate algorithms, collaboration where needed, and the full exploitation of their potential for science return, require overcoming a substantial learning curve.

\item{\it Large datasets}: Many astronomical datasets are larger than can be held in machine memory. The exploitation of these datasets thus requires more sophisticated database technology than is currently employed by most astronomical projects.

\item{\it ``It's not science''}: The success of an astronomical project is judged by the science results produced. The time invested by an astronomer in becoming an expert in data mining techniques must be balanced against the expected science gain. It is difficult to justify and obtain funding based purely on a methodological approach such as data mining, even if such an approach will demonstrably improve the expected science return.

\item{\it It does not do the science for you}: The algorithms will output patterns, but will not necessarily establish which patterns or relationships are important scientifically, or even which are causal. The truism `correlation is not causation' is apt here. The successful interpretation of data mining results is up to the scientist.

\item{\it The result can only be as good as the data}: Related to this, the single largest factor in the success of any data mining algorithm is the quality of the input data. If the data are not sufficient for the task, or are poorly collected or incorrectly treated, the result will not be useful.

\end{itemlist}

\section{Uses in Astronomy} \label{Sec: Uses}

We now turn to the use of data mining algorithms in astronomical applications, and their track record in addressing some common problems. Whereas in \S \ref{Sec: Overview}, we introduced terms for the astronomer unfamiliar with data mining, here for the non-expert in astronomy we briefly put in context the astronomical problems. However, a full description is beyond the scope of this review. Whereas \S \ref{Sec: Overview} was subdivided according to data mining algorithms and issues, here the subdivision is in terms of the astrophysics. Throughout this section, we abbreviate data mining algorithms that are either frequently mentioned or have longer names according to the abbreviations introduced in \S \ref{Sec: Overview}: PCA, ANN, DT, SVM, \knn, KDE, EM, SOM, and ICA.

Given that there is no exact definition of what constitutes a data mining tool, it would not be possible to provide a complete overview of their application. This section therefore illustrates the wide variety of actual uses to date, with actual or implied further possibilities. Uses which exist now but will likely gain greater significance in the future, such as the time domain, are largely deferred to \S \ref{Sec: Future}. Several other overviews of applications of machine learning algorithms in astronomy exist, and contain further examples, including ones for ANN\cite{miller:annapps,lahav:annmethods,bailerjones:ann,li:annapps,tagliaferri:nnast}, DT\cite{white:dts}, genetic algorithms\cite{charbonneau:ga}, and stellar classification\cite{bailerjones:stellar}.

Most of the applications in this section are made by astronomers utilizing data mining algorithms. However, several projects and studies have also been made by data mining experts utilizing astronomical data, because, along with other fields such as high energy physics and medicine, astronomy has produced many large datasets that are amenable to the approach. Examples of such projects include the Sky Image Cataloging and Analysis System (SKICAT)\cite{weir:skicat} for catalog production and analysis of catalogs from digitized sky surveys, in particular the scans of the second Palomar Observatory Sky Survey; the Jet Propulsion Laboratory Adaptive Recognition Tool (JARTool)\cite{burl:jartool}, used for recognition of volcanoes in the over 30,000 images of Venus returned by the Magellan mission; the subsequent and more general Diamond Eye\cite{burl:diamondeye}; and the Lawrence Livermore National Laboratory Sapphire project\cite{kamath:sapphire}. A recent review of data mining from this perspective is given by Kamath in the book {\it Scientific Data Mining}\cite{kamath:scientific}. In general, the data miner is likely to employ more appropriate, modern, and sophisticated algorithms than the domain scientist, but will require collaboration with the domain scientist to acquire knowledge as to which aspects of the problem are the most important.

\subsection{Object classification} \label{Subsec: Classification}

Classification is often an important initial step in the scientific process, as it provides a method for organizing information in a way that can be used to make hypotheses and to compare with models. Two useful concepts in object classification are the {\it completeness} and the {\it efficiency}, also known as recall and precision. They are defined in terms of true and false positives (TP and FP) and true and false negatives (TN and FN). The completeness is the fraction of objects that are truly of a given type that are classified as that type: $${\mathrm{completeness} = \frac{TP}{TP+FN}},$$ and the efficiency is the fraction of objects classified as a given type that are truly of that type $${\mathrm{efficiency} = \frac{TP}{TP+FP}}.$$ These two quantities are astrophysically interesting because, while one obviously wants both higher completeness and efficiency, there is generally a tradeoff involved. The importance of each often depends on the application, for example, an investigation of rare objects generally requires high completeness while allowing some contamination (lower efficiency), but statistical clustering of cosmological objects requires high efficiency, even at the expense of completeness.

\subsubsection{Star-Galaxy Separation} \label{Subsubsec: Star/gal}

Due to their small physical size compared to their distance from us, almost all stars are unresolved in photometric datasets, and thus appear as point sources. Galaxies, however, despite being further away, generally subtend a larger angle, and thus appear as extended sources. However, other astrophysical objects such as quasars and supernovae, also appear as point sources. Thus, the separation of photometric catalogs into stars and galaxies, or more generally, stars, galaxies, and other objects, is an important problem. The sheer number of galaxies and stars in typical surveys (of order $10^8$ or above) requires that such separation be automated.

This problem is a well studied one and automated approaches were employed even before current data mining algorithms became popular, for example, during digitization by the scanning of photographic plates by machines such as the APM\cite{maddox:apmstargal} and DPOSS\cite{djorgovski:dposs}. Several data mining algorithms have been employed, including ANN\cite{odewahn:autostargal,odewahn:annstargal,bazell:preprocessing,andreon:stargal,philip:dbnn,odewahn:dpossstargal,collister:megazlrg}, DT\cite{weir:stargal,ball:dtclassification}, mixture modeling\cite{qin:stargal}, and SOM\cite{miller:som}, with most algorithms achieving over 95\% efficiency. Typically, this is done using a set of measured morphological parameters that are derived from the survey photometry, with perhaps colors or other information, such as the seeing, as a prior. The advantage of this data mining approach is that all such information about each object is easily incorporated. As well as the simple outputs `star' or `galaxy', many of the refinements described in \S \ref{Sec: Overview} have improved results, including probabilistic outputs and bagging\cite{ball:dtclassification}.

\subsubsection{Galaxy Morphology} \label{Subsubsec: Morphology}

\begin{figure}[!ht]
\centerline{\psfig{file=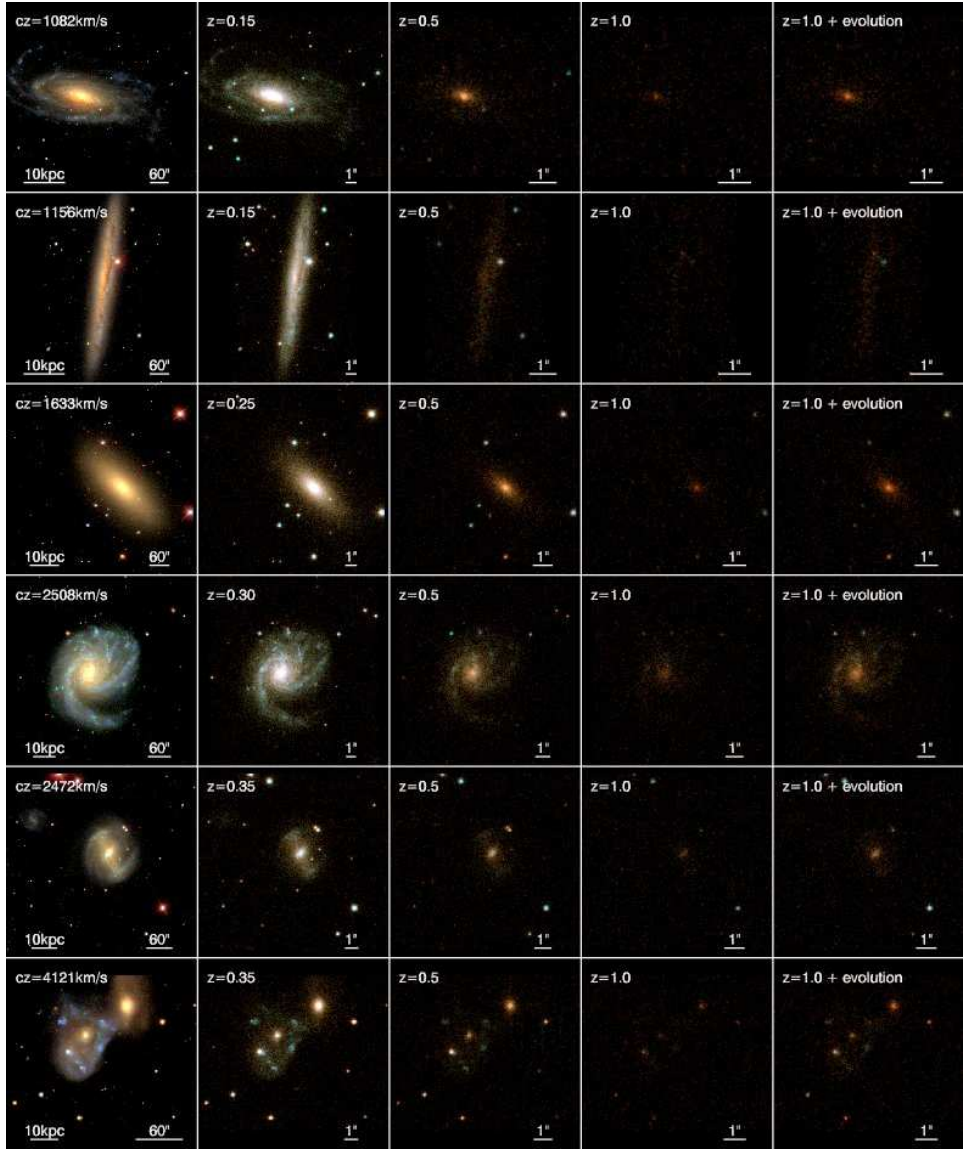,width=5in}}
\vspace*{8pt}
\caption{Examples of galaxy morphology showing many aspects of the information available to, and issues to be aware of for, a data mining process. These include galaxy shape, structure, texture, inclination, arm pitch, color, resolution, exposure, and, from left to right, redshift, in this case artificially constructed. From Barden, Jahnke \& H{\"a}u{\ss}ler\protect \cite{barden:ferengi}.
 \label{Fig: Morph}}
\end{figure}

As shown in Fig. \ref{Fig: Morph}, galaxies come in a range of different sizes and shapes, or more collectively, morphology. The most well-known system for the morphological classification of galaxies is the Hubble Sequence of elliptical, spiral, barred spiral, and irregular, along with various subclasses\cite{hubble:extragalnebulae,hubble:realm,sandage:hubbatlas,sandage:carnegieatlas,vandenbergh:morph,sandage:classfnhistory}. This system correlates to many physical properties known to be important in the formation and evolution of galaxies\cite{roberts:hubble,firmani:hubble}. Other well-known classification systems are the Yerkes system based on concentration index\cite{morgan:ci,morgan:ci2,devaucouleurs:ci}, the de Vaucouleurs\cite{devaucouleurs:devprofile}, exponential\cite{patterson:expprofile,freeman:expprofile}, and S\'ersic index\cite{sersic:australes,graham:sersic} measures of the galaxy light profile, the David Dunlap Observatory (DDO) system\cite{vandenbergh:lumclass,vandenbergh:lumclassb,vandenbergh:ddo}, and the concentration-asymmetry-clumpiness (CAS) system\cite{conselice:cas}.

Because galaxy morphology is a complex phenomenon that correlates to the underlying physics, but is not unique to any one given process, the Hubble sequence has endured, despite it being rather subjective and based on visible-light morphology originally derived from blue-biased photographic plates. The Hubble sequence has been extended in various ways, and for data mining purposes the T system\cite{devaucouleurs:tsystem,devaucouleurs:ttype} has been extensively used. This system maps the categorical Hubble types E, S0, Sa, Sb, Sc, Sd, and Irr onto the numerical values -5 to 10.

One can, therefore, train a supervised algorithm to assign T types to images for which measured parameters are available. Such parameters can be purely morphological, or include other information such as color. A series of papers by Lahav and collaborators\cite{storrielombardi:ann,lahav:annscience,naim:eyemorph,naim:annmorph,lahav:annmethods,collister:annz} do exactly this, by applying ANNs to predict the T type of galaxies at low redshift, and finding equal accuracy to human experts. ANNs have also been applied to higher redshift data to distinguish between normal and peculiar galaxies\cite{naim:peculiarmorph}, and the fundamentally topological and unsupervised SOM ANN has been used to classify galaxies from Hubble Space Telescope images\cite{naim:som}, where the initial distribution of classes is not known. Likewise, ANNs have been used to obtain morphological types from galaxy spectra.\cite{madgwick:morphspec}

Several authors study galaxy morphology at higher redshift by using the Hubble Deep Fields, where the galaxies are generally much more distant, fainter, less evolved, and morphologically peculiar. Three studies\cite{odewahn:hdfannmorph,windhorst:hizgals,cohen:hstmorph} use ANNs trained on surface brightness and light profiles to classify galaxies as E/S0, Sabc and Sd/Irr. Another application\cite{odewahn:fouriermorph} uses Fourier decomposition on galaxy images followed by ANNs to detect bars and assign T types.

Bazell \& Aha\cite{bazell:ensembles} uses ensembles of classifiers, including ANN and DT, to reduce the classification error, and Bazell\cite{bazell:features} studies the importance of various measured input attributes, finding that no single measured parameter fully reproduces the classifications. Ball \etal \cite{ball:ann} obtain similar results to Naim \etal \cite{naim:annmorph}, but updated for the SDSS. Ball \etal \cite{ball:bivlf} and Ball, Loveday \& Brunner\cite{ball:envt} utilize these classifications in studies of the bivariate luminosity function and the morphology-density relation in the SDSS, the first such studies to utilize both a digital sky survey of this size and detailed Hubble types.

\begin{figure}[!ht]
\quad \psfig{file=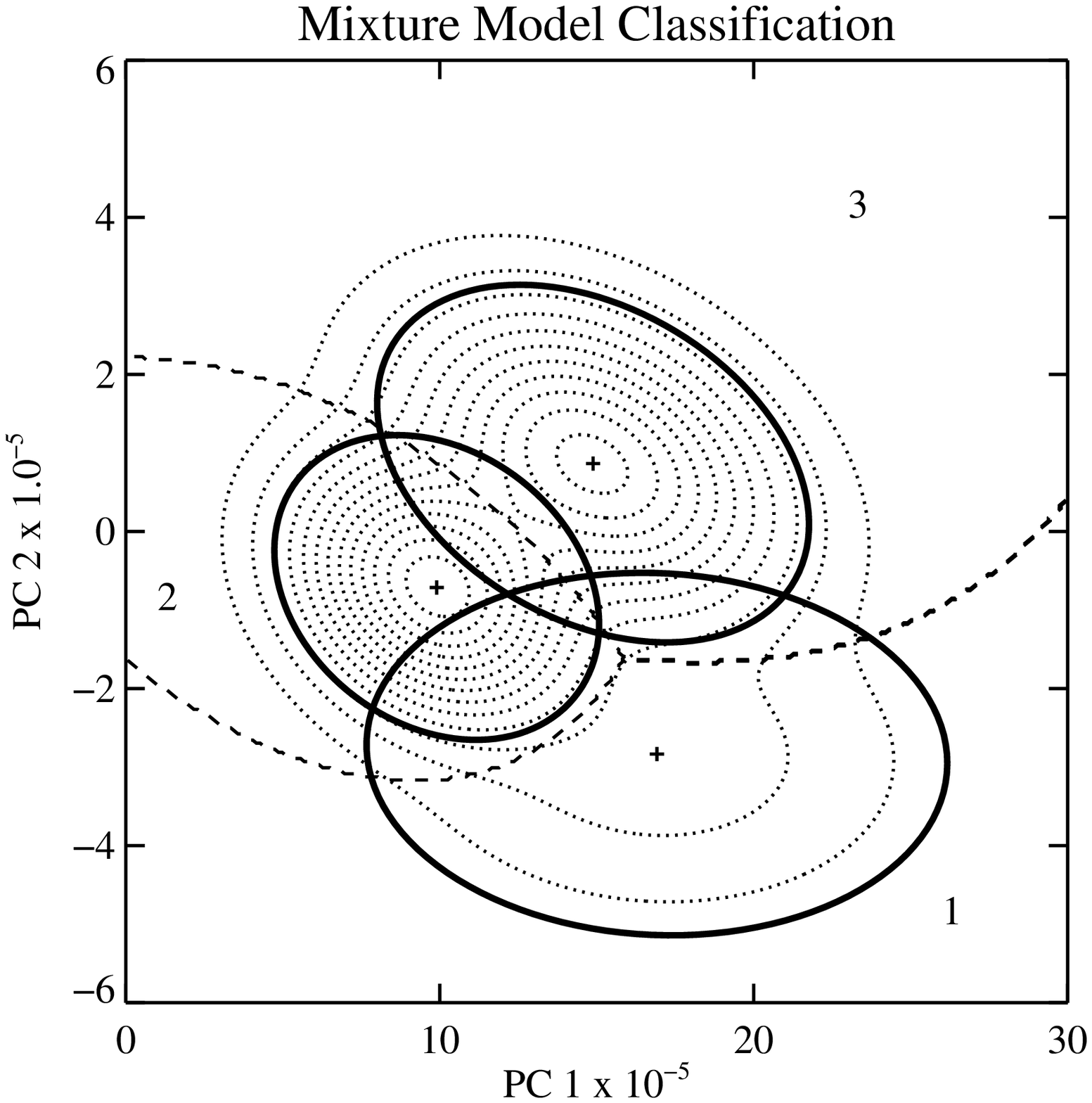,width=2.25in} \quad \psfig{file=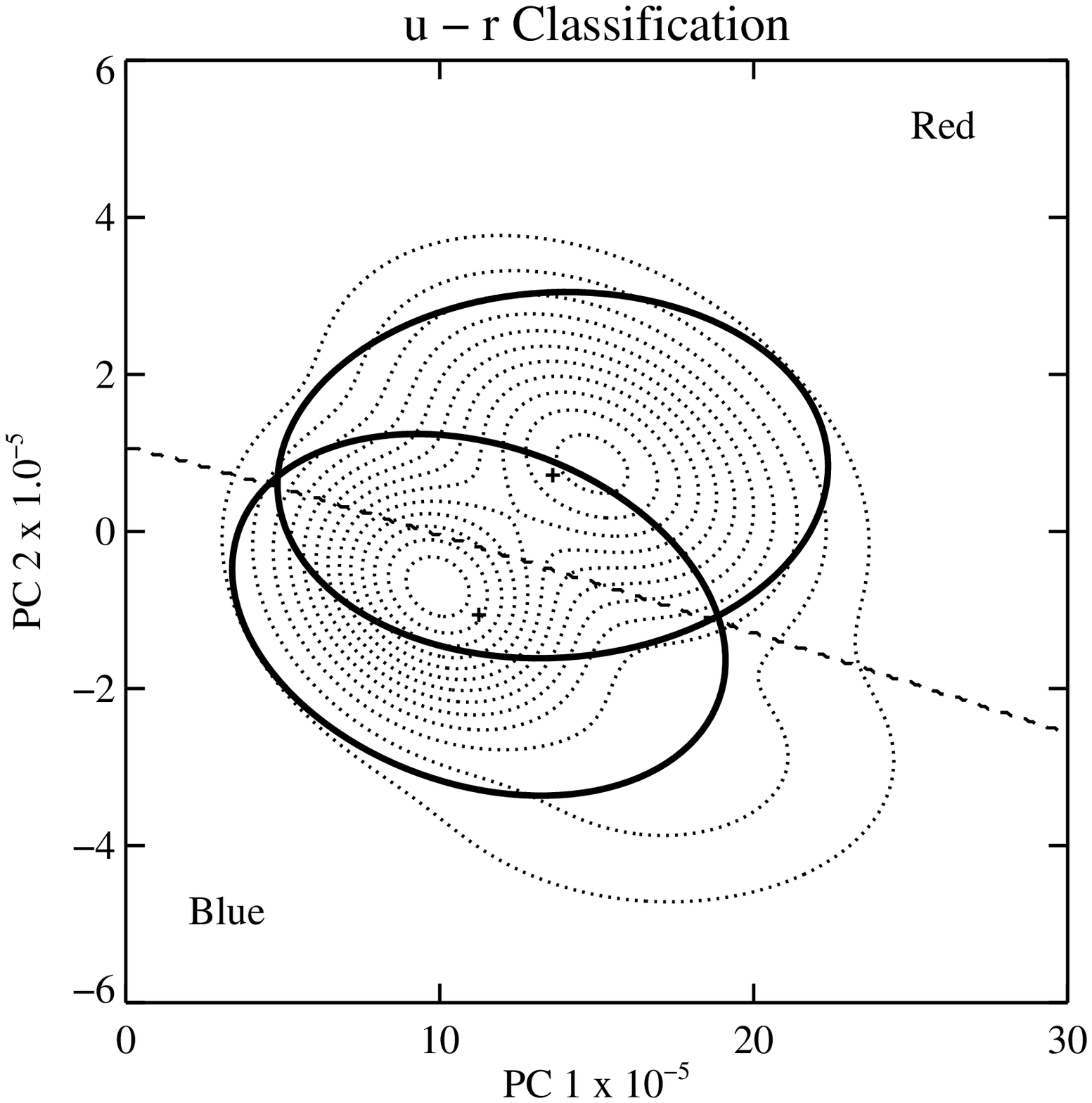,width=2.25in}
\vspace*{8pt}
\caption{Improvement in classification using a mixture model over that derived from the $u$ and $r$ passbands ($u-r$ color). In this case, the mixture model clearly delineates the third class, which is not seen using $u-r$. The axes are the first two principle components of the spectro-morphological parameter set (shapelet coefficients in five passbands) describing the galaxies. The light contours are the square root of the probability density from the mixture model fit, and the dark contours are the 95\% threshold for each class, in the right-hand panel fitted to the two classes by quadratic discriminant analysis. From Kelly \& McKay\protect \cite{kelly:shapelet2}. \label{Fig: Mixture}}
\end{figure}

Because of the complex nature of galaxy morphology and the plethora of available approaches, a large number of further studies exist: Kelly \& McKay\cite{kelly:shapelet2} (Fig. \ref{Fig: Mixture}) demonstrate improvement over a simple split in $u-r$ using mixture models, within a schema that incorporates morphology. Serra-Ricart \etal \cite{serraricart:annastroapps} use an encoder ANN to reduce the dimensionality of various datasets and perform several applications, including morphology. Adams \& Woolley\cite{adams:ann} use a committee of ANNs in a `waterfall' arrangement, in which the output from one ANN formed the input to another which produces more detailed classes, improving their results. Molinari \& Smareglia\cite{molinari:annesolf} use an SOM to identify E/S0 galaxies in clusters and measure their luminosity function. de Theije \& Katgert\cite{detheije:typekinematics} split E/S0 and spiral galaxies using spectral principal components and study their kinematics in clusters. Genetic algorithms have been employed\cite{cantupaz:evolving,kamath:bentdouble} for attribute selection and to evolve ANNs to classify `bent-double' galaxies in the FIRST\cite{becker:first} radio survey data. Radio morphology combines the compact nucleus of the radio galaxy and extremely long jets. Thus, the bent-double morphology indicates the presence of a galaxy cluster. de la Calleja \& Fuentes\cite{delacalleja:ann} combine ensembles of ANN and locally weighted regression. Beyond ANN, Spiekermann\cite{spiekermann:automorph} uses fuzzy algebra and heuristic methods, anticipating the importance of probabilistic studies (\S \ref{Subsec: PDF}) that are just now beginning to emerge. Owens, Griffiths \& Ratnatunga\cite{owens:dt} use oblique DTs, obtaining similar results to ANN. Zhang, Li \& Zhao\cite{zhang:morph} distinguish early and late types using k-means clustering. SVMs have recently been employed on the COSMOS survey by Huertas-Company \etal \cite{huertascompany:svmmorph,huertascompany:svmmorph2}, enabling early-late separation to $K_{AB} = 22$ mag twice as good as the CAS system. SVMs will also be used on data from the Gaia satellite\cite{tsalmantza:gaiaclassification}.

Recently, the popular {\it Galaxy Zoo} project\cite{lintott:galaxyzoo} has taken an alternative approach to morphological classification, employing {\it crowdsourcing}: an application was made available online in which members of the general public were able to view images from the SDSS and assign classifications according to an outlined scheme. The project was very successful, and in a period of six months over 100,000 people provided over 40 million classifications for a sample of 893,212 galaxies, mostly to a limiting depth of $r = 17.77$ mag. The classifications included categories not previously assigned in astronomical data mining studies, such as edge-on or the handedness of spiral arms, and the project has produced multiple scientific results. The approach represents a complementary one to automated algorithms, because, although humans can see things an algorithm will miss and will be subject to different systematic errors, the runtime is hugely longer: a trained ANN will produce the same 40 million classifications in a few minutes, rather than six months.

\subsubsection{Other Galaxy Classifications} \label{Subsubsec: Other Gal}

Many of the physical properties, and thus classification, of a galaxy are determined by its stellar population. The spectrum of a galaxy is therefore another method for classification\cite{humason:100redshifts,morgan:specclass}, and can sometimes produce a clearer link to the underlying physics than the morphology. Spectral classification is important because it is possible for a range of morphological types to have the same spectral type, and vice versa, because spectral types are driven by different underlying physical processes.

Numerous studies\cite{connolly:orthogonal,connolly:eclass,madgwick:parametrisation,yip:spectypes} have used PCA directly for spectral classification. PCA is also often used as a preprocessing step before the classification of spectral types using an ANN\cite{storrielombardi:annspec}. Folkes, Lahav \& Maddox\cite{folkes:annspec} predict morphological types for the 2dF Galaxy Redshift Survey (2dFGRS)\cite{colless:2dffinal} using spectra, and Ball \etal \cite{ball:ann} directly predict spectral types in the SDSS using an ANN. Slonim \etal \cite{slonim:ib} use the information bottleneck approach on the 2dFGRS spectra, which maximally preserves the spectral information for the desired number of classes. Lu \etal \cite{lu:ica} use ensemble learning for ICA on components of galaxy spectra. Abdalla \etal \cite{abdalla:emission} use ANN and locally weighted regression to directly predict emission line properties from photometry.

Bazell \& Miller\cite{bazell:classdisc} applied a semi-supervised method suitable for class discovery using ANNs to the ESO-LV\cite{lauberts:esouppsala} and SDSS Early Data Release (EDR) catalogs. They found that a reduction of up to 57\% in classification error was possible compared to purely supervised ANNs. The larger of the two catalogs, the SDSS EDR, represents a preliminary dataset about 6\% of the final data release of the SDSS, clearly indicating the as-yet untapped potential of this approach. The semi-supervised approach also resembles the hybrid empirical-template approach to photometric redshifts (\S \ref{Subsec: Photo-zs}), as both seek to utilize an existing training set where available even if it does not span the whole parameter space. However, the approach used by Bazell \& Miller is more general, because it allows new classes of objects to be added, whereas the hybrid approach can only iterate existing templates.

\subsubsection{Quasars/AGN} \label{Subsubsec: QSO Classification}

Most of the emitted electromagnetic radiation in the universe is either from stars, or the accretion disks surrounding supermassive black holes in active galactic nuclei (AGN). The latter phenomenon is particularly dramatic in the case of quasars, where the light from the central region can outshine the rest of the galaxy. Because supermassive black holes are thought to be fairly ubiquitous in large galaxies, and their fueling, and thus their intrinsic brightness, can be influenced by the environment surrounding the host galaxy, quasars and other AGN are important for understanding the formation and evolution of structure in the universe.

\begin{figure}[!ht]
\centerline{\psfig{file=figures/bamford.eps,width=3.25in,angle=270}}
\centerline{\psfig{file=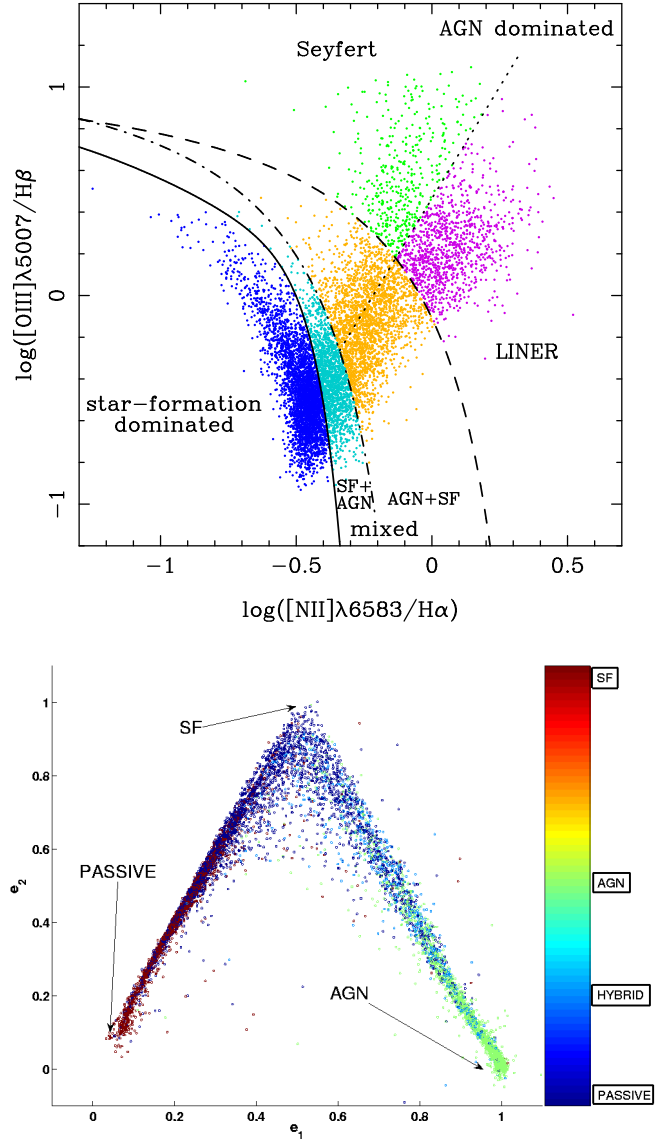,width=3.75in}}
\vspace*{8pt}
\caption{Upper panel: Baldwin-Philips-Terlevich diagram, which classifies active galactic nuclei (AGN) and star-forming galaxies but requires all four emission lines to be present in the spectrum. From Bamford \etal \protect \cite{bamford:nonparametric} (although it should be noted that the use of this diagram is not the basis of their study). The axes are the diagnostic emission line ratios from the spectra. Lower panel: AGN/star-forming/passive classification using an ANN, which has no such requirement. The axes are the two outputs from the ANN, $e_1$ and $e_2$ mapped onto $(e_1,e_2)=(e_1+e_2/2){\bf i} + e_2 {\bf j}$, where passive, AGN, star-forming, and hybrid are (0,0), (1,0), (0,1), and (0.5,0.5), respectively. From Abdalla \etal \protect \cite{abdalla:emission}. \label{Fig: BPT ANN}}
\end{figure}

The selection of quasars and other AGN from an astronomical survey is a well-known and important problem, and one well suited to a data mining approach. It is well-known that different wavebands (X-ray, optical, radio) will select different AGN, and that no one waveband can select them all. Traditionally, AGN are classified on the Baldwin-Phillips-Terlevich diagram\cite{baldwin:bpt}, in which sources are plotted on the two-dimensional space of the emission line ratios [O{\scriptsize{III}}] $\lambda$ 5007 / H$\beta$ and [N{\scriptsize{II}}] / H$\alpha$, that is separated by a single curved line into star-forming and AGN regions. Data mining not only improves on this by allowing a more refined or higher dimensional separation, but also by including passive objects in the same framework (Fig. \ref{Fig: BPT ANN}). This allows for the probability that an object contains an AGN to be calculated, and does not require all (or any) of the emission lines to be detected.

Several groups have used ANNs\cite{carballo:annqso,claeskens:gaiaqsos,carballo:annfirstqso} or DTs\cite{white:firstqso,suchkov:dt,ball:dtclassification,zhang:qso,zhang:decisiontable,zhao:dtactive,knigge:balqso} to select quasar candidates from surveys. White \etal \cite{white:firstqso} show that the DT method improves the reliability of the selection to 85\% compared to only 60\% for simpler criteria. Other algorithms employed include PCA\cite{yip:qsoclassification}, SVM and learning vector quantization\cite{zhang:methods}, kd-tree\cite{gao:svmkdqso}, clustering in the form of principal surfaces and negative entropy clustering\cite{abrusco:qsocandidates}, and kernel density estimation\cite{richards:dr6photoqso}. Many of these papers combine multiwavelength data, particularly X-ray, optical, and radio.

Similarly, one can select and classify candidates of all types of AGN\cite{zhao:agn}. If multiwavelength data are available, the characteristic data mining algorithm ability to form a model of the required complexity to extract the information could enable it to use the full information to extract more complete AGN samples. More generally, one can classify both normal and active galaxies in one system, differentiating between star formation and AGN. As one example, DTs have been used\cite{ball:dtclassification} to select quasar candidates in the SDSS, providing the probabilities P(star, galaxy, quasar). P(star formation, AGN) could be supplied in a similar framework. Bamford \etal \cite{bamford:nonparametric} combine mixture modeling and regression to perform non-parametric mixture regression, and is the first study to obtain such components and then study them versus environment. The components are passive, star-forming, and two types of AGN.

\subsubsection{Other Classifications} \label{Subsubsec: Other Classification}

Often, the first component of classification is the actual process of object detection, which often is done at some signal-to-noise threshold. Several statistical data mining algorithms have been employed, and software packages written, for this purpose, including the Faint Object Classification and Analysis System (FOCAS)\cite{jarvis:focas}, DAOPHOT\cite{stetson:daophot}, Source Extractor (SExtractor)\cite{bertin:sextractor}, maximum likelihood, wavelets, ICA\cite{maino:fastica}, mixture models\cite{guglielmetti:background}, and ANNs\cite{andreon:stargal}. Serra-Ricart \etal \cite{serraricart:faint} show that ANNs are able to classify faint objects as well as a Bayesian classifier but with considerable computational speedup.

Several studies are more general than star-galaxy separation or galaxy classification, and assign classifications of varying detail to a broad range of astrophysical objects. Goebel \etal \cite{goebel:bayesian} apply the AutoClass Bayesian classifier to the IRAS LRS atlas, finding new and scientifically interesting object classes. McGlynn \etal \cite{mcglynn:classx} use oblique DTs in a system called ClassX to classify X-ray objects into stars, white dwarfs, X-ray binaries, galaxies, AGN, and clusters of galaxies, concluding that the system has the potential to significantly increase the known populations of some rare object types. Suchkov, Hanisch \& Margon\cite{suchkov:dt} use the same system to classify objects in the SDSS. Bazell, Miller \& Subbarao\cite{bazell:subclasses} apply semi-supervised learning to SDSS spectra, including those classified as `unknown', finding two classes of objects consisting of over 50\% unknown.

Stellar classifications are necessarily either spectral or based on color, due to the pointlike nature of the source. This field has a long history and well established results such as the HR diagram and the OBAFGKM spectral sequence. The latter is extended to a two-dimensional system of spectral type and luminosity classes I--V to form the two-dimensional MK classification system of Morgan, Keenan \& Kellman\cite{morgan:mk}. Class I are supergiants, through to class V, dwarfs, or main-sequence stars. The spectral types correspond to the hottest and most massive stars, O, through to the coolest and least massive, M, and each class is subdivided into ten subclasses 0--9. Thus, the MK classification of the sun is G2V.

The use of automated algorithms to assign MK classes is analogous to that for assigning Hubble types to galaxies in several ways: before automated algorithms, stellar spectra were compared by eye to standard examples; the MK system is closely correlated to the underlying physics, but is ultimately based on observable quantities; the system works quite well but has been extended in numerous ways to incorporate objects that do not fit the main classes (e.g., L and T dwarfs, Wolf-Rayet stars, carbon stars, white dwarfs, and so on). Two differences from galaxy classification are the number of input parameters, in this case spectral indices, and the number of classes. In MK classification the numbers are generally higher, of order 50 or more input parameters, compared to of order 10 for galaxies.

Given a large body of work for galaxies that has involved the use of artificial neural networks, and the similarities just outlined, it is not surprising that similar approaches have been employed for stellar classification\cite{vonhippel:annstellar,weaver:annstellar,singh:annstellar,bailerjones:annstellar,gulati:annstellar,bazarghan:stellar}, with a typical accuracy of one spectral type and half a luminosity type. The relatively large number of object attributes and output classes compared to the number of objects in each class does not invalidate the approach, because the efforts described generally find that the number of principal components represented by the inputs is typically much lower. A well-known property of neural networks is that they are robust to a large number of redundant attributes (\S \ref{Subsubsec: Choice}).

Neural networks have been used for other stellar classifications schemes, e.g. Gupta \etal \cite{gupta:iras} define 17 classes for IRAS sources, including planetary nebulae and H{\scriptsize{II}} regions. Other methods have been employed; a recent example is Manteiga \etal \cite{manteiga:starmind}, who use a fuzzy logic knowledge-based system with a hierarchical tree of decision rules. Beyond the MK and other static classifications, variable stars have been extensively studied for many years, e.g., Wozniak \etal\cite{wozniak:mirasvm} use SVM to distinguish Mira variables. 

\begin{figure}[!ht]
\centerline{\psfig{file=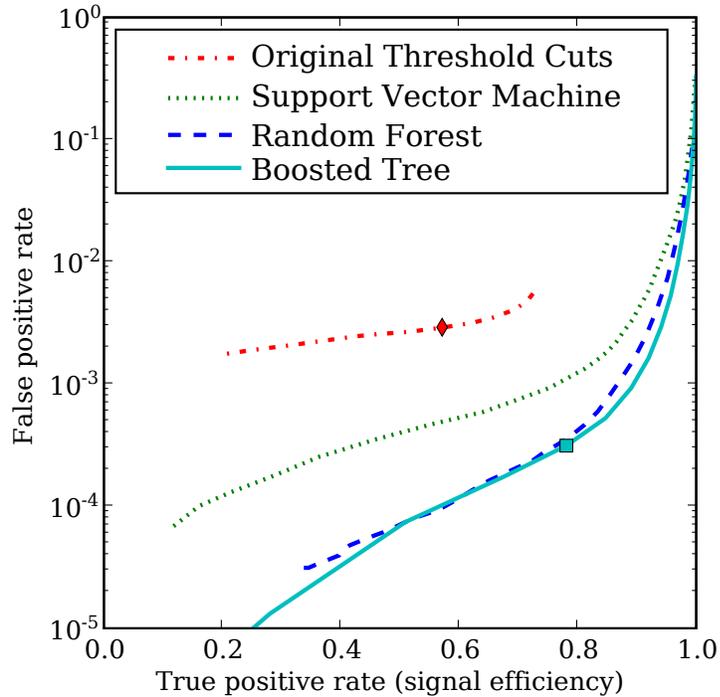,width=4in}}
\vspace*{8pt}
\caption{Improvement in the classification of supernovae using support vector machine and decision tree, compared to previously used threshold cuts. From Bailey \etal \protect \cite{bailey:sneclassification}. \label{Fig: SNe}}
\end{figure}

The detection and characterization of supernovae is important for both understanding the astrophysics of these events, and their use as standard candles in constraining aspects of cosmology such as the dark energy equation of state. Bailey \etal \cite{bailey:sneclassification} use boosted DTs, random forests, and SVMs to classify supernovae in difference images, finding a ten times reduction in the false-positive rate compared to standard techniques involving parameter thresholds (Fig. \ref{Fig: SNe}).

Given the general nature of the data mining approach, there are many further classification examples, including cosmic ray hits\cite{salzberg:dt,waniak:cosmicray}, planetary nebulae\cite{faundezabans:pne}, asteroids\cite{misra:asteroids}, and gamma ray sources\cite{chattopadhyay:gammaray,scaringi:isina}.

\subsection{Photometric redshifts} \label{Subsec: Photo-zs}

An area of astrophysics that has greatly increased in popularity in the last few years is the estimation of redshifts from photometric data (\phzs). This is because, although the distances are less accurate than those obtained with spectra, the sheer number of objects with photometric measurements can often make up for the reduction in individual accuracy by suppressing the statistical noise of an ensemble calculation.

Photo-zs were first demonstrated in the mid 20th century\cite{stebbins:photoz,baum:photoz}, and later in the 1980s\cite{koo:photoz,loh:photoz}. In the 1990s, the advent of the Hubble Space Telescope Deep fields resulted in numerous approaches\cite{gwyn:photoz,lanzetta:photoz,mobasher:photoz,sawicki:photoz,connolly:angcf,wang:photoz,benitez:photoz}, reviewed by Koo\cite{koo:photozhistory}. In the past decade, the advent of wide-field CCD surveys and multifiber spectroscopy have revolutionized the study of \phzs~to the point where they are indispensable for the upcoming next generation surveys, and a large number of studies have been made.

The two common approaches to \phzs~are the template method and the empirical training set method. The template approach has many complicating issues\cite{massarotti:template}, including calibration, zero-points, priors, multiwavelength performance (e.g., poor in the mid-infrared), and difficulty handling missing or incomplete training data. We focus in this review on the empirical approach, as it is an implementation of supervised learning. In the future, it is likely that a hybrid method incorporating both templates and the empirical approach will be used, and that the use of full probability density functions will become increasingly important. For many applications, knowing the error distribution in the redshifts is at least as important as the accuracy of the redshifts themselves, further motivating the calculation of PDFs.

\subsubsection{Galaxies} \label{Subsubsec: Galaxy Photo-zs}

At low redshifts, the calculation of photometric redshifts for normal galaxies is quite straightforward due to the break in the typical galaxy spectrum at 4000\AA. Thus, as a galaxy is redshifted with increasing distance, the color (measured as a difference in magnitudes) changes relatively smoothly. As a result, both template and empirical \phz~approaches obtain similar results, a root-mean-square deviation of $\sim 0.02$ in redshift, which is close to the best possible result given the intrinsic spread in the properties\cite{brunner:photoz}. This has been shown with ANNs\cite{firth:annphotoz,ball:ann,collister:annz,vanzella:hdfannphotoz,li:photoz,abrusco:sdssphotoz,collister:megazlrg,banerji:desvistaphotoz,oyaizu:photoz,niemack:stripe82galexphotoz,zhang:morph}, SVM\cite{wadadekar:svmphotoz,wang:novelphotoz}, DT\cite{carliles:photoz}, \knn\cite{ball:pdfphotoz}, empirical polynomial relations\cite{connolly:photoz,brunner:photoz,wang:photoz,sowardsemmerd:photoz,hsieh:photoz,lopes:photoz}, numerous template-based studies, and several other methods. At higher redshifts, obtaining accurate results becomes more difficult because the 4000\AA~break is shifted redward of the optical, galaxies are fainter and thus spectral data are sparser, and galaxies intrinsically evolve over time. The first explorations at higher redshift were the Hubble Deep Fields in the 1990s, described above (\S \ref{Subsec: Photo-zs}), and, more recently, new infrared data have become available, which allow the 4000\AA~break to be seen to higher redshift, which improves the results. Template-based algorithms work well, provided suitable templates into the infrared are available, and supervised algorithms simply incorporate the new data and work in the same manner as previously described.

While supervised learning has been successfully used, beyond the spectral regime the obvious limitation arises that in order to reach the limiting magnitude of the photometric portions of surveys, extrapolation would be required. In this regime, or where only small training sets are available, template-based results can be used, but without spectral information, the templates themselves are being extrapolated. However, the extrapolation of the templates is being done in a more physically motivated manner. It is likely that the more general hybrid approach of using empirical data to iteratively improve the templates,\cite{budavari:spectempl,csabai:photoz,csabai:edrphotoz,padmanabhan:photoz,brodwin:iracphotoz,budavari:unified} or the semi-supervised method described in \S \ref{Subsubsec: Semisupervised} will ultimately provide a more elegant solution. Another issue at higher redshift is that the available numbers of objects can become quite small (in the hundreds or fewer), thus reintroducing the curse of dimensionality by a simple lack of objects compared to measured wavebands. The methods of dimension reduction (\S \ref{Subsec: Attributes}) can help to mitigate this effect.

\subsubsection{Quasars/AGN} \label{Subsubsec: QSO Photo-zs}

Historically, the calculation of photometric redshifts for quasars and other AGN has been even more difficult than for galaxies, because the spectra are dominated by bright but narrow emission lines, which in broad photometric passbands can dominate the color. The color-redshift relation of quasars is thus subject to several effects, including degeneracy, one emission line appearing like another at a different redshift, an emission line disappearing between survey filters, and reddening. In addition, the filter sets of surveys are generally designed for normal galaxies and not quasars. The assignment of these quasar \phzs~is thus a complex problem that is amenable to data mining in a similar manner to the classification of AGN described in \S \ref{Subsubsec: QSO Classification}.

\begin{figure}[!ht]
\centerline{\psfig{file=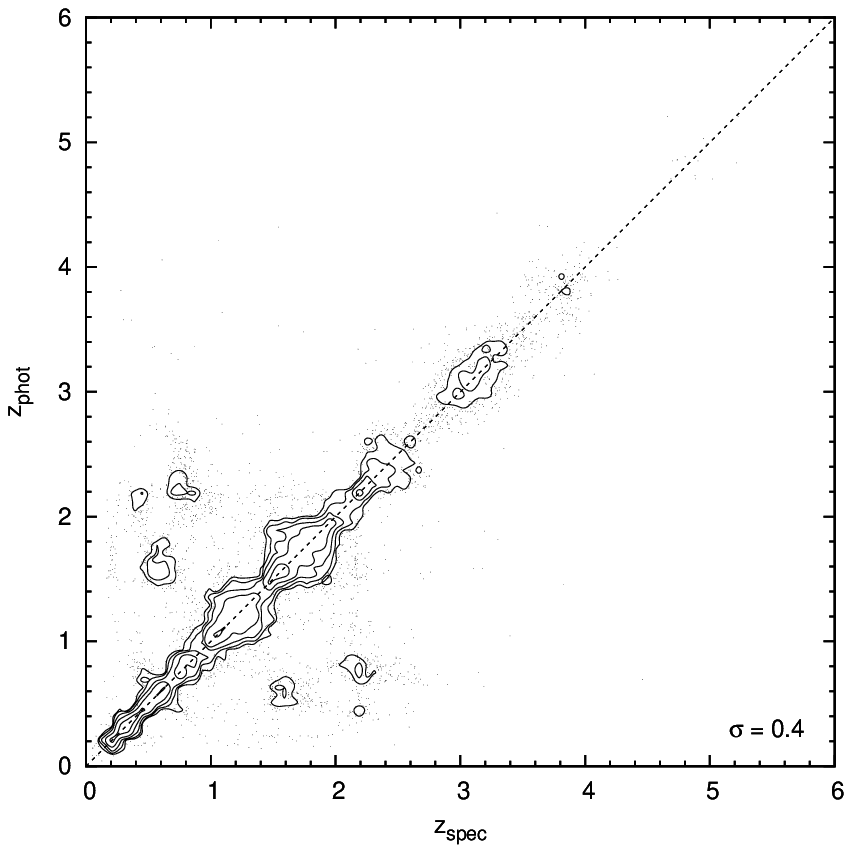,width=4.375in}}
\centerline{\psfig{file=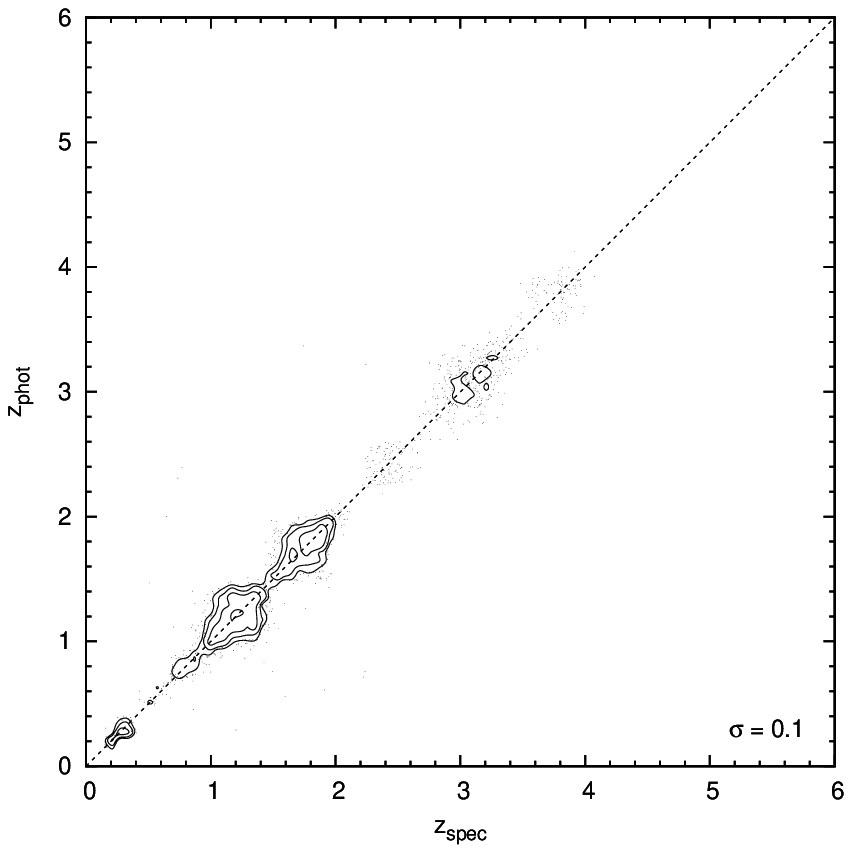,width=4.375in}}
\vspace*{8pt}
\caption{Photometric redshift, $z_p$, vs. spectroscopic redshift, $z_s$, for quasars in the Sloan Digital Sky Survey, showing, in the upper panel, catastrophic failures in which $z_p$ is very different from $z_s$. Each individual point represents one quasar, and the contours indicate areas of high areal point density. $\sigma$ is the root-mean-square dispersion between $z_p$ and $z_s$. The use of data mining techniques, including assigning full probability density functions in photometric redshift, enables the reduction or elimination of these catastrophics, as shown in the lower panel. Data based on that from Ball \etal \protect \cite{ball:pdfphotoz}. \label{Fig: Photo-z QSO}} 
\end{figure}

The calculation of quasar \phzs~has had some success using SDSS data\cite{budavari:qsophotoz,richards:qsophotoz,babbedge:impz,weinstein:qsophotoz,wu:qsophotoz,kitsonias:xrayagn}, but they suffer from {\it catastrophic failures}, in which, as shown in Fig. \ref{Fig: Photo-z QSO}, the photometric redshift for a subset of the objects is completely incorrect. However, data mining approaches have resulted in improvements to this situation. Ball \etal \cite{ball:ibphotoz} find that a single-neighbor \knn~gives a similar result to the templates, but multiple neighbors, or other supervised algorithms such as DT or ANN, pull in the regions of catastrophic failure and significantly decrease the spread in the results. Kumar\cite{kumar:ml} also shows this effect. Ball \etal \cite{ball:pdfphotoz} go further and are able to largely eliminate the catastrophics by selecting the subset of quasars with one peak in their redshift probability density function (\S \ref{Subsec: PDF}), a result confirmed by Wolf\cite{wolf:qsopdf}. Wolf \etal \cite{wolf:qsophotoz} also show significant improvement using the COMBO-17 survey, which has 17 filters compared to the five of the SDSS, but unfortunately the photometric sample is much smaller.

Beyond the spectral regime, template-based results are sufficient\cite{salvato:cosmosagnphotoz}, but again suffer from catastrophics. Given our physical understanding of the nature of quasars, it is in fact reasonable to extrapolate in magnitude when using colors as a training set, because while one is going to fainter magnitudes, one is not extrapolating in color. One could therefore quite reasonably assign empirical \phzs~for a full photometric sample of quasars.

\subsection{Other Astrophysical Applications} \label{Subsec: Other Uses}

Typically in data mining, information gathered from spectra has formed the training set to apply a predictive technique to objects with photometry. However, it is clear from this process that the spectrum itself contains a large amount of information, and data mining techniques may be used directly on the spectra to extract information that might otherwise remain hidden. Applications to galaxy spectral classification were described in \S \ref{Subsubsec: Other Gal}. In stellar work, besides the classification of stars into the MK system based on observable parameters, several studies have directly predicted physical parameters of stellar atmospheres using spectral indices. One example is Ramirez, Fuentes \& Gulati\cite{ramirez:ibga}, who utilize a genetic algorithm to select the appropriate input attributes, and predict the parameters using \knn. The attribute selection reduces run time and improves predictive accuracy. Solorio \etal \cite{solorio:ib} use \knn~to study stellar populations and improve the results by using active learning to populate sparse regions of parameter space, an alternative to dimension reduction. 

Although it has much potential for the future (\S \ref{Subsec: Time Domain}), the time domain is a field in which a lot of work has already been done. Examples include the classification of variable stars described in \S \ref{Subsubsec: Other Classification}, and, in order of distance, the interaction of the solar wind and the Earth's atmosphere, transient lunar phenomena, detection and classification of asteroids and other solar system objects by composition and orbit, solar system planetary atmospheres, stellar proper motions, extrasolar planets, novae, stellar orbits around the supermassive black hole at the Galactic center, microlensing from massive compact halo objects, supernovae, gamma ray bursts, and quasar variability. A good overview is provided by Becker\cite{becker:transientclassification}. The large potential of the time domain for novel discovery lies within the as yet unexplored parameter space defined by depth, sky coverage, and temporal resolution\cite{djorgovski:rare}. One constraining characteristic of the most variable sources beyond the solar system is that they are generally point sources. As a result, the timescales of interest are constrained by the light crossing time for the source.

The analysis of the cosmic microwave background (CMB) is amenable to several techniques, including Bayesian modeling, wavelets, and ICA. The latter, in particular via the FastICA algorithm\cite{maino:fastica}, has been used in removal of CMB foregrounds\cite{bottino:foreground}, and cluster detection via the Sunyaev-Zeldovich effect\cite{pires:sz}. Phillips \& Kogut\cite{phillips:nnparamest} use a committee of ANNs for cosmological parameter estimation in CMB datasets, by training them to identify parameter values in Monte Carlo simulations. This gives unbiased parameter estimation in considerably less processing time than maximum likelihood, but with comparable accuracy. 

One can use the fact that objects cross-matched between surveys will likely have correlated distributions in their measured attributes, for example, similar position on the sky, to improve cross-matching results using pattern classifiers. Rohde \etal \cite{rohde:matching} combine distribution estimates and probabilistic classifiers to produce such an improvement, and supply probabilistic outputs. 

Taylor \& Diaz\cite{taylor:evann} obtain empirical fits for Galactic metallicity using ANNs, whose architectures are evolved using genetic algorithms. This method is able to provide equations for metallicity from line ratios, mitigating the `black box' element common to ANNs, and, in addition, is potentially able to identify new metallicity diagnostics. 

Bogdanos \& Nesseris\cite{bogdanos:snega} analyze Type Ia supernovae using genetic algorithms to extract constraints on the dark energy equation of state. This method is non-parametric, which minimizes bias from the necessarily a priori assumptions of parametric models. 

Lunar and planetary science, space science, and solar physics also provide many examples of data mining uses. One example is Li \etal \cite{li:svmflare}, who demonstrate improvements in solar flare forecasting resulting from the use of a mixture of experts, in this case SVM and \knn. The analysis of the abundance of minerals or constituents in soil samples\cite{mustard:mixture} using mixture models is another example of direct data mining of spectra. 

Finally, data mining can be performed on astronomical simulations, as well as real datasets. Modern simulations can rival or even exceed real datasets in size and complexity, and as such the data mining approach can be appropriate. An example is the incorporation of theory\cite{lemson:votheory} into the Virtual Observatory (\S \ref{Subsec: VO}). Mining simulation data will present extra challenges compared to observations because in general there are fewer constraints on the type of data presented, e.g., observations are of the same universe, but simulations are not, simulations can probe many astrophysical processes that are not directly observable, such as stellar interiors, and they provide direct physical quantities as well as observational ones. Most of the largest simulations are cosmological, but they span many areas of astrophysics. A prominent cosmological simulation is the Millennium Run\cite{springel:millennium}, and over 200 papers have utilized its data\footnote{\url{http://www.mpa-garching.mpg.de/millennium}}.

\section{The Future} \label{Sec: Future}

We now turn to the future of data mining in astronomy. Several trends are apparent that indicate likely fruitful directions in the next few years. These trends can be used to make informed decisions about upcoming, very large surveys. This section assumes that the reader is somewhat familiar with the concepts in both \S\S \ref{Sec: Overview} and \ref{Sec: Uses}, namely, with both data mining and astronomy. We once again arrange the topics by data mining algorithm rather than by astronomical application, but we now interweave the algorithms with examples.

As in the past, it is likely that cross-fertilization with other fields will continue to be beneficial to astronomy, and of particular relevance here, the data mining efforts made by these fields. Examples include high energy physics, whose most obvious spinoff is the World Wide Web from CERN, but the subject has an extensive history of extremely large datasets from experiments such as particle colliders, and has provided well-known and commonly used data analysis software such as ROOT \cite{brun:root}, designed to cope with these data sizes and first developed in 1994. In the fields of biology and the geosciences, the concepts of {\it informatics}, the study of computer-based information systems, have been extensively utilized, creating the subfields of bio- and geoinformatics. The official recognition of an analogous subfield within astronomy, {\it astroinformatics}, has recently been recommended\cite{borne:datamining}.

\subsection{Probability Density Functions} \label{Subsec: PDF}

\begin{figure}[!ht]
\centerline{\psfig{file=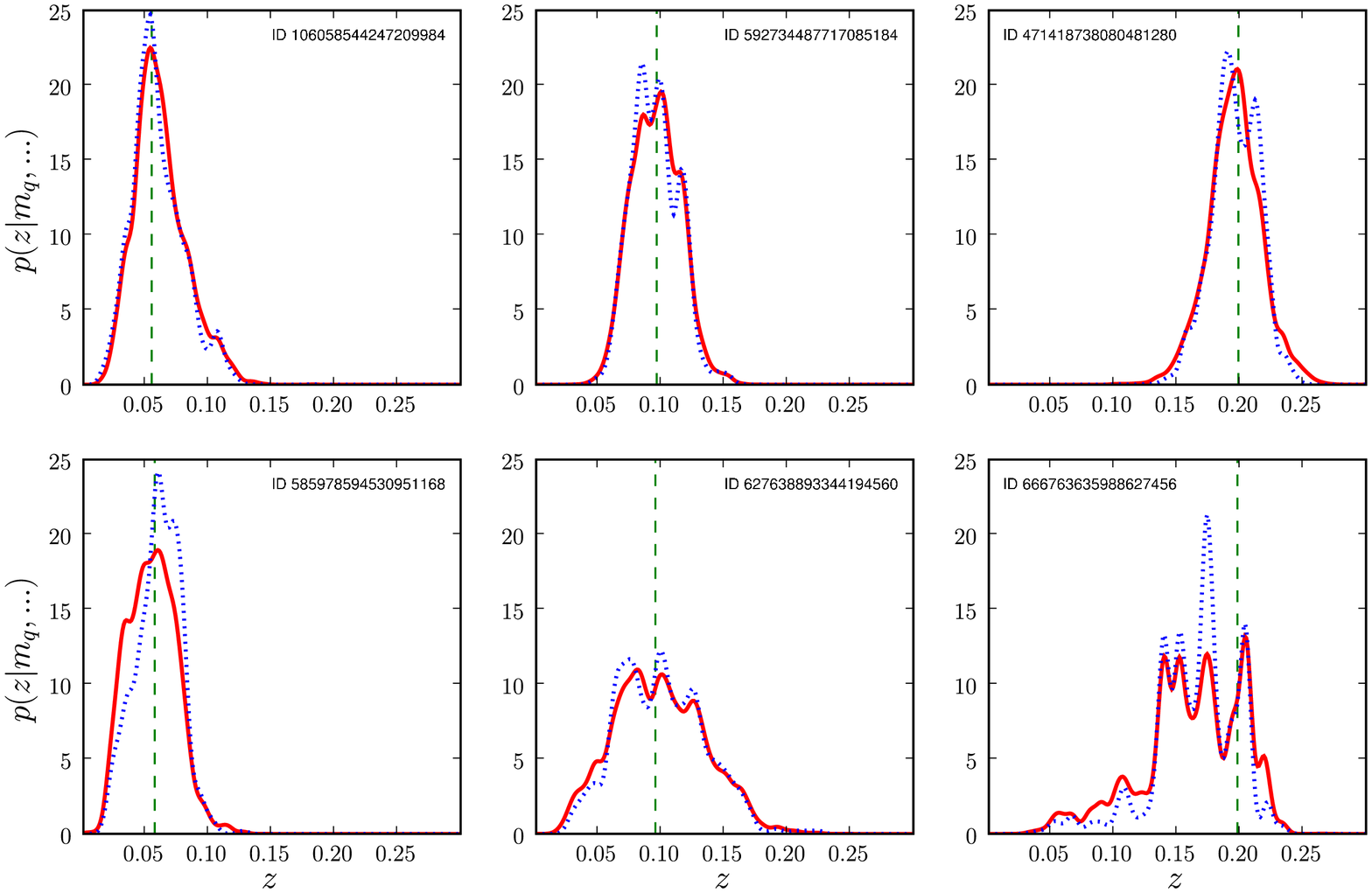,width=5.5in}}
\vspace*{8pt}
\caption{Example photometric redshift probability density functions (PDFs) for galaxies, showing the rich content of extra information compared to a single value, or value plus Gaussian error. The horizontal axes, $z$, are the photometric redshifts, and the vertical are the probability densities. The solid red and dotted blue lines are the PDF with and without the photometric uncertainties, respectively, and the vertical dashed green lines are, in these cases, the true distances. From Budav{\'a}ri\protect \cite{budavari:unified}. \label{Fig: PDFs}}
\end{figure}

A {\it probability density function} (PDF, Fig. \ref{Fig: PDFs}) is a function such that the probability that the value, $x$, is in the interval $a < x < b$, is the definite integral over the range: $$P(a < x < b) = \int_a^b f(x) dx.$$ Thus the total area under the function is one. PDFs are of great significance for data mining in astronomy because they retain information that is otherwise lost, and because they enable results with improved signal-to-noise from a given dataset. One can think of a PDF as a histogram in the limit of small bins but many objects. Approaches such as supervised learning are in general taking as input the information on objects and providing as output a prediction of properties. The most general way to do this is to work with the full PDFs at each stage. The formalism has recently been demonstrated in an astronomical context by Budav{\'a}ri\cite{budavari:unified}, and it is applicable to the prediction of any astronomical property. For inputs $a,b,c$,..., the output probabilities of a set of properties, $P(x,y,z,...)$ can be predicted. Fully probabilistic cross-matching of surveys has also been implemented by the same author\cite {budavari:crossid}. 

\begin{figure}[!ht]
\centerline{\psfig{file=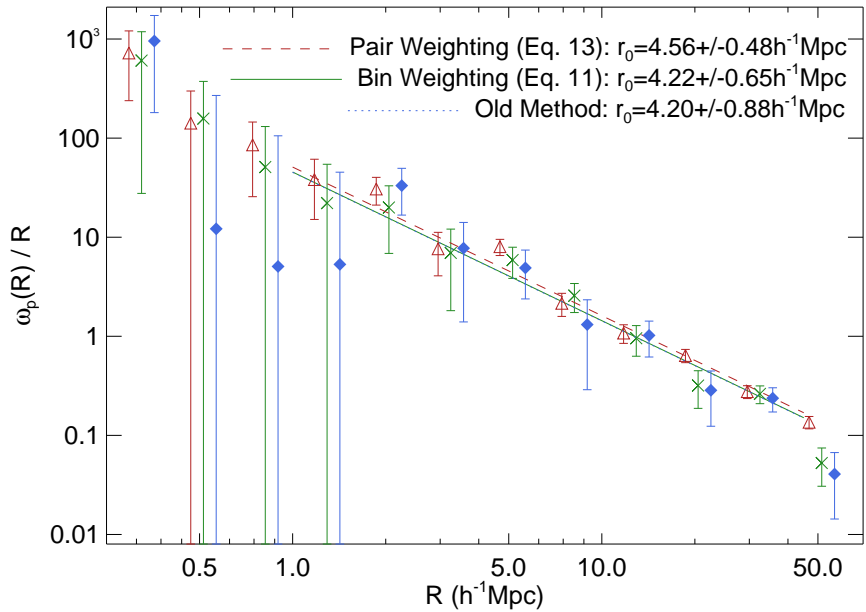,width=4.5in}}
\vspace*{8pt}
\caption{Improvement in the signal-to-noise ratio of the clustering signal of quasars enabled by PDFs. The improvements to the projected correlation function (vertical axis) enabled by utilizing PDFs are shown by the green crosses and red triangles, compared to the old method, based on single-valued photometric redshifts, shown by blue diamonds. The horizontal axis is the projected radial distance between objects. The diagonal lines are power-law fits, with scale length $r_0$, to the correlation function. The points are offset for clarity. From Myers, White \& Ball\protect \cite{myers:pdf}. \label{Fig: Clustering}}
\end{figure}

Results with PDFs in \phzs~are starting to appear, either with single values and a spread, or the full PDF. Cunha \etal \cite{cunha:nzpdfphotoz} show that full PDFs help reduce bias. Margoniner \& Wittman\cite{margoniner:photoz} show that they enable subsamples with improved signal-to-noise, and Wittman\cite{wittman:pzerror} also demonstrates reduction in error. Ball \etal \cite{ball:pdfphotoz} show that generating full \phz~PDFs for quasars allows subsection of a sample virtually free of catastrophic failures, the first time this has been demonstrated, and an important result for their use as tracers of the large scale structure in the universe. Wolf\cite{wolf:qsopdf} confirms a similar result. Myers, White \& Ball\cite{myers:pdf} show that using the full PDF for clustering measurements will improve the signal-to-noise by four to five times for a given dataset without any alteration of the data (Fig. \ref{Fig: Clustering}). This method is applicable to the clustering of any astronomical object. Full PDFs have also been shown to improve performance in the photometric detection of galaxy clusters\cite{vanbreukelen:cluster}, again due to the increased signal-to-noise ratio. Several further efforts use a single \phz~and a spread, but not the full PDF. However, the method of Myers, White \& Ball shows that it is the full PDF that will give the most benefit. PDFs will also be important for weak lensing\cite{margoniner:photoz}.

As well as \phzs, predicting properties naturally incorporates probabilistic classification. Progress has been made, e.g., the SDSS has been classified according to P(galaxy, star, neither)\cite{ball:dtclassification}. Similar classifications that could be made are P(star formation, AGN) and P(quasar, not quasar). Bailer-Jones \etal \cite{bailerjones:gaiaquasars} implement probabilistic classification that emphasizes finding very rare objects, in this case quasars among the stars that will be seen by Gaia.

Ball \etal \cite{ball:pdfphotoz} generate a PDF by perturbing inputs for a single-neighbor \knn. The idea of perturbing data has been studied in the field of Privacy Preserving Data Mining\cite{vaidya:ppdm,aggarwal:ppdm}, but here the aim is to generate a PDF using the errors on the input attributes in a way that is computationally scalable to upcoming datasets. The approach appears to work well despite the fact that at present, survey photometric errors are generally poorly characterized\cite{scranton:sdsscovariance}. Proper characterization of errors will be of great importance to future surveys as the probabilistic approach becomes more important. Scalability may be best implemented either by using kd-tree like data structures, or by direct implementation on novel supercomputing hardware such as FPGA, GPU, or Cell processors (\S \ref{Subsec: Hardware}), which can provide enormous performance benefits for applications that require a large number of distance calculations.

\subsection{Real-Time Processing and the Time Domain} \label{Subsec: Time Domain}

The time domain is already a significant area of study and will become increasingly important over the next decade with the advent of large scale synoptic surveys such as the Large Synoptic Survey Telescope (LSST)\cite{ivezic:lsst}. A large number of temporal resolved observations over large areas of the sky remains an unexplored area, and the historical precedent suggests that many interesting phenomena remain to be discovered\cite{djorgovski:rare}.

However, as one might expect, this field presents a number of challenges not encountered in the data mining of static objects. These include (i) how to handle multiple observations of objects that can vary in irregular and unpredictable ways, both intrinsic and due to the observational equipment, (ii) objects in difference images (the static background is subtracted, leaving the variation), (iii) the necessarily extremely rapid response to certain events such as gamma ray bursts where physical information can be lost mere seconds after an event becomes detectable, (iv) robust classification of large streams of data in real time, (v) lack of previous information on several phenomena, and (vi) the volume and storage of time domain information in databases. Other challenges are seen in static data, but will assume increased importance as real-time accuracy is needed. For example, the removal of artifacts\cite{donalek:synoptic} that might otherwise be flagged as unusual objects and incur expensive follow-up telescope time. Variability will be both photometric, a change in brightness, and astrometric, because objects can move. While some astronomical phenomena, such as certain types of variable stars, vary in a regular way, others vary in a nonlinear, irregular, stochastic, or chaotic manner, and the variability itself can change with time (heteroskedasticity)\cite{studenmund:econometrics}. Time series analysis is a well developed area of statistics, and many of these techniques will be useful.

The combination of available information, but incomplete coverage of the possible phenomena suggests that a probabilistic (\S \ref{Subsec: PDF}) approach\cite{mahabal:probabilistic}, either involving priors, or semi-supervised (\S \ref{Subsubsec: Semisupervised}) will in general be the most appropriate. This is because the algorithms can use the existing information, but objectively interpret new phenomena. Supervised learning will perform better for problems where more information and larger datasets are available, and unsupervised or Bayesian priors will perform better when there are fewer observations. Many events will still require followup observations, but since there will be far more events than can ever be followed up in detail, data mining algorithms will help ensure that the observations made are optimal in terms of the targeted scientific results.

As a confederation of data archives and interoperable standards of many of the world's existing telescopes, the Virtual Observatory (VO, \S \ref{Subsec: VO}) will be crucial in meeting the challenge of the time domain, and significant infrastructure for the VO already exists. The VOEventNet\cite{drake:voeventnet} is a system for the rapid handling of real time events, and provides an online federated data stream of events from several telescopes. It can be followed by both human observers and robotic telescopes.

Numerous next-generation wide-field surveys in the planning or construction stages will be synoptic. The largest such survey in the optical is the LSST, which will observe the entire sky, visible from its location, every three nights. These observations will provide a data stream exceeding one petabyte per year, and, as a result, they anticipate many of the challenges described here\cite{ivezic:lsstclassification}. Like LSST\cite{borne:lsstmining}, the Gaia satellite\cite{perryman:gaia} has working groups dedicated to data mining. The Classification Working Group has employed several data mining techniques, and developed new approaches\cite{bailerjones:domain,bailerjones:gaiaquasars} to be used when the survey comes online. Other ongoing or upcoming synoptic surveys include Palomar-Quest\cite{djorgovski:pq}, the Catalina Real-Time Transient Survey\cite{drake:catalina}, Pan-STARRS\cite{hodapp:panstarrs}, and those at other wavelengths such as instruments leading up to and including the Square Kilometer Array\cite{johnston:askap}. 

The time domain will not only provide challenges to existing methods of data mining, but will open up new avenues for the extraction of information, such as using the variability of objects for classification\cite{eyer:variability} or photometric redshift\cite{kaczmarczik:astrometricphotoz}. Because they are due to a relatively compact source in the center of galaxies, active galactic nuclei vary on much shorter timescales than normal galaxies. This variability has been proposed as a mechanism to select quasar and other AGN candidates. Other events are suspected theoretically but have not been observed\cite{mahabal:transients}. But given the dataset sizes, automated detection of such events at some level is clearly required. The computational demands of real time processing of the enormous data streams from these surveys is significant, and will likely be met by the use of newly emerging specialized computing hardware (\S \ref{Subsec: Hardware}).

\subsection{Petascale Computing} \label{Subsec: Petascale}

The current state of the art in supercomputing consists of terabyte-sized files and teraflop computing speeds, which is conveniently encapsulated in the term {\it terascale computing}. Following Moore's law\cite{moore:law}, in which computer performance has increased exponentially for the last several decades, the coming decade will feature the similarly-derived {\it petascale computing}\cite{bader:petascale}. Much of the performance increase in the past decade has been driven by increases in processor (CPU) clock frequency, but this rate has now slowed due to physical limitations on the sizes of components, and more importantly power consumption and energy (heat) dissipation. It has therefore become more economical to manufacture chips with multiple processor cores.

The typical supercomputer today is a cluster, which consists of a large number of conventional CPUs connected by a specialized interconnect system, a distributed or shared memory, a shared filesystem, and hosting the Linux operating system. Many systems are heterogeneous because this is scalable and cost-effective, but coordinating and making effective such a system can be challenging. In particular, it will be vital that the system is properly balanced between processing power and disk input/output (I/O) to supply the data. Combined with the increasing number of processor cores, this means that {\it parallel and distributed computing} is rapidly increasing in importance.

A useful set of `rules of thumb' for parallel and other aspects of computing were formulated by Amdahl in the 1960s\cite{amdahl:law}, and they remain true today. One of these is that roughly 50,000 CPU cycles are required per byte of data. Most scientific datasets require far fewer cycles than this, and it is thus likely that future performance will be I/O limited, unless sufficient disks are provided in parallel. Bell, Gray \& Szalay\cite{bell:petascale} estimate that a petascale system will require 100,000 one TB disks. The exact details of how to distribute the data for best performance are likely to be system-dependent\cite{dolence:lci}. The available CPU speed should scale to the data size, although it will not scale to most na{\"i}vely implemented data mining algorithms (\S \ref{Subsec: Parallel}).

An example of an upcoming petascale system whose uses will include astronomical data mining is the {\it Blue Waters}\footnote{\url{http://www.ncsa.uiuc.edu/BlueWaters}} system at the National Center for Supercomputing Applications (NCSA), which is due to come online in 2011. Specifications include 200,000 compute cores with 4 GHz 8 core processors, 1 PB of main memory, 10 PB of user disk storage, 500 PB of archival storage, and 400 GB ${\mathrm s}^{-1}$ bandwidth connectivity to provide sustained petascale compute power. It will implement the IBM PERCS (Productive, Easy-to-use, Reliable Computer System)\cite{ebcioglu:percs}, which will integrate their CPU, operating system, parallel programming, and file systems. This provides a method of addressing the issues of running real-world applications at the petascale by balancing the CPU, I/O, networking, and so on. Similarly, a considerable investment of effort is being carried out in the years leading up to deployment in 2011 on the development of applications for the system, in consultation with the scientists who will run them. Several astronomical applications are included, mostly simulations, but also data mining in the form of the analysis of LSST datasets.

Not all petascale computing will be done on systems as large as Blue Waters. In the US, the National Science Foundation Office of Cyberinfrastructure has been advised\cite{bell:petascale} to implement a power-law type system, with a small number of very large systems, of order ten times more regional centers, and ten times more local facilities (Tiers 1--3). Such local facilities, for example Beowulf clusters, are already common in university departments, and consist of typically a few dozen commodity machines. A recent trend matching the increasing requirements for data-intensive as opposed to CPU-intensive computing is the GrayWulf cluster\cite{szalay:graywulf}, which implements the idea of data `storage bricks': cheap, modular, and portable versions of a balanced system which when added together provide petascale computation.

\subsection{Parallel and Distributed Data Mining} \label{Subsec: Parallel}

As indicated in \S \ref{Subsec: Petascale} above, because of the slowing increase in raw speed of individual CPUs, processors are becoming increasingly parallelized, both in terms of the number of processor cores on a single chip, and increasing amounts of these chips being deployed in parallel on supercomputing clusters. Providing appropriately scaled systems (CPU, I/O, etc.) is one challenge, but most data mining algorithms not only will be required to run on petascale data, but their na{\"i}ve implementations scale as $N^2$, or worse. It has been suggested\cite{szalay:petabyte} that any algorithm that scales beyond $N {\mathrm{log}} N$ will rapidly be rendered infeasible.

McConnell and Skillicorn\cite{mcconnell:ddm} have promoted parallel and distributed data mining\cite{freitas:parallel,kargupta:dpkd,zaki:parallel,bhaduri:ddmbib}, which is well-known in the data mining field, but virtually unused in astronomy. In this approach, the algorithms explicitly take advantage of available parallelism. The simplest example is task-farming, or the embarrassingly parallel approach, in which a task is divided into many mutually-independent subtasks, each of which is allocated to a single processor. This can be done on an array of ordinary desktop machines as well as a supercomputer. A more complex challenge is when many parts of the data must be accessed, or when an algorithm relies on the outputs from calculations distributed across multiple compute nodes. For a large dataset the hardware required likely includes shared memory (\S \ref{Subsec: Petascale}), thus shared memory parallelization\cite{jin:shared} can be important. Many algorithms exist for the implementation of data mining on parallel computer systems beyond simple task farming, but these are not widely used within science, as compared to the commercial sector. The application programming interfaces MPI and OpenMP have been widely used on distributed and shared memory systems, respectively, for simulation and some data analysis, but they do not offer the semantic capabilities\cite{gray:semantic} needed for data mining, i.e., the metadata describing the meaning of the data being processed and the results produced are not easily incorporated.

Parallel data mining is challenging, as not only must the algorithm be implemented on the hardware, but many algorithms simply cannot be ported as-is to such a system. Instead, parallelization requires that the algorithm itself, as encapsulated in the code, must often be fundamentally altered at the pseudocode level. This can be a time-consuming and counterintuitive process, especially to scientists who are generally not trained or experienced in parallel programming. Progress is slowly being made in astronomy, including a parallel implementation of kd-trees\cite{gardner:paralleltree}, cosmological simulations requiring datasets larger than the node memory size\cite{norman:enzo}, and parallelization of algorithms\cite{brunner:fpganpcf}.

An alternative approach is grid computing, in which the exact resource used is unimportant to the user, although not all data mining algorithms lend themselves to this paradigm. A variant of grid computing is crowdsourcing, in which members of the public volunteer their spare CPU cycles to process data for a project. The most well-known project of this type is SETI@Home, and more recently, the Galaxy Zoo project, which employed large numbers of people to successfully classify galaxies in SDSS images. Such crowdsourcing is likely to become even more important in the future, particularly in combination with greatly improved outreach via astronomical applications on social networking sites such as Facebook\cite{gomez:socialnetworking}.

Scalability is also helped on conventional CPUs by the employment of tree structures, such as the kd-tree, which partition the data. This enables a search to access any data value without searching the whole dataset. Kd-trees have been used for many astronomical applications, including speeding up N-point correlation functions\cite{moore:npcf}; cross-matching, classification, and photometric redshifts\cite{gao:kdtree}. They can be extended to more sophisticated structures, for example, the multi-tree\cite{gray:multitree}. However, implementation of such tree structures on parallel hardware or computational accelerators (\S \ref{Subsec: Hardware}) remains difficult\cite{gardner:paralleltree}.

\subsection{The Virtual Observatory} \label{Subsec: VO}

The Virtual Observatory (VO) is an analogous concept to a physical observatory, but instead of telescopes, various centers house data archives. The VO consists of numerous national-level organizations, and the International Virtual Observatory Alliance. Within the national organizations there are various data centers that house large datasets, computing facilities to process and analyze them, and people with considerable expertise in the datasets stored at that particular center.

Common data standards and web services are necessary for the VO to work. Such standards have emerged, including web services using XML and SOAP, a data format, VOTable\cite{ochsenbein:votable}, a query language based on SQL, the Astronomical Data Query Language\cite{shirasaki:adql}, image access protocols for images (SIAP\cite{shirasaki:adql}), and spectra (SSAP)\footnote{\url{http://www.ivoa.net/Documents}}, VOEventNet\cite{drake:voeventnet} for the time domain, plus various standards of interoperability and ways of describing resources such as the Unified Content Descriptor\cite{derriere:ucd}. Large numbers of high level tools for working with data are also available\footnote{\url{http://www.us-vo.org}}.

An example of the emerging data standards for archiving is the Common Archive Observation Model\cite{dowler:caom2008} (CAOM) of the Canadian Astronomical Data Center (CADC). Given that it is likely that the future VO will continue to consist of a number of data centers like the CADC, this model represents a useful and realistic way in which data can be made meaningfully accessible, but not so rigidly presented as to prevent the desired analysis of future researchers with as yet unforeseen science goals. This model consists of the components Artifact, Plane, SimpleObservation, and CompositeObservation, which describe logical parts of the data from individual files to logical sets of observations such as spectra, and forms the basis of all archiving activity at the CADC.

The increasing immobility of large datasets as described in \S \ref{Subsec: Petascale} will render it uneconomical in terms of time and money to download large datasets to local machines. Rather than bringing the data to the analysis, it will become more sensible to take the analysis to the data\cite{gray:decade}. To be able to perform complicated data mining analyses, it is necessary that the data be organized well enough to make this tractable, and that the center archiving the data must have sufficient computing power and web services to perform the analyses. The organizational requirement means that the data must be stored as a database with the sophistication found in the commercial sector, where mining of terascale databases is routine. Commercial software and computer science expertise will help, but the task is non-trivial because astronomical data analysis can require particular data types and structures not usually found in commercial software, such as time series observations. An example of such a database already in place is the SDSS, and its underlying schema\cite{gray:sdss} has been used and copied by other surveys such as GALEX.

Nevertheless, it is likely that considerable analyses will continue to be carried out on smaller subsets of the data, and this data may well continue to be downloaded and analyzed locally, as it has been to date. If one anticipates working exclusively with one survey, it may still be more efficient to implement a GrayWulf-like cluster locally and download the complete dataset.

Another difficult problem faced by the VO is that a significant future scientific benefit from large datasets will be in the cross-matching of multiple datasets, in particular, multiwavelength data. But if such data are distributed among different data centers and are difficult to move, such work may be intractable. What can be done, however, is to make available as part of the VO web services, tools for cross-matching datasets at a given center. A common data format and description, combined with the fact that much of the science is done from small subsets of large datasets, means that this is certainly tractable. As a result, it is not surprising that there is significant demand for such tools\cite{vignali:vo}.

An important consideration for the VO is that many astronomers, indeed many scientists in general, will want to run their own software on the data, and not simply a higher level tool that involves trusting someone else's code. This will be true even if the source code is available. Or, a scientist might wish to complete an analysis that is not available in a higher level tool. It is thus important that the data are available at a low level of processing so that one can set one's own requirements as needed. NASA has a categorization of data where 0 is raw, 1 is calibrated, and 2 is a derived product, such as a catalog. An ideal data archive would have available well documented and accessible level 2 catalogs, similarly documented and accessible level 1 data, and perhaps not online but stored level 0 data, to enable, for example, a re-reduction.

Data have been released using the VO publishing interfaces\cite{gonzalezsolares:iphas}, data mining algorithms such as ANNs have been implemented\cite{brescia:voneural}, and applications for analyses with web interfaces are online\cite{kitching:cloudcosmology}. Multiwavelength analyses are becoming more feasible and useful\cite{vignali:vo}, and it is therefore now possible, but still time-consuming, to perform scientific analyses using VO tools\cite{chilingarian:vo}. We expect this will be an area where considerable work will still need to be done, however, in order to fully enable the full exploitation of the archives of astronomy data in the future.

\subsection{Visualization} \label{Subsec: Future Visualization}

Visualization of data is an important part of the scientific process, and the combination of terascale computing and data mining poses obvious challenges. Common plotting codes presently in use in astronomy include SuperMongo\footnote{\url{http://www.astro.princeton.edu/~rhl/sm}}, PGPlot\footnote{\url{http://www.astro.caltech.edu/~tjp/pgplot}}, Gnuplot\footnote{\url{http://www.gnuplot.info}}, and IDL\footnote{\url{http://idlastro.gsfc.nasa.gov}} \cite{landsman:idlastro}, but these are stand-alone codes that do not easily cope with data that cannot be completely loaded into the available memory space. Newer tools, such as TOPCAT\cite{taylor:topcat}, VisIVO\cite{comparato:viz}, and VOMegaPlot\cite{urunkar:vomegaplot} support the Virtual Observatory standards such as VOTable and PLASTIC\cite{taylor:plastic} for interoperability between programs. The full library on which the TOPCAT program is based, STILTS\cite{taylor:stilts}, is able to plot arbitrarily-sized datasets.

As with hardware, software, and data analysis, collaboration with computer scientists and other disciplines has resulted in progress in various areas of scientific visualization. At Harvard, the AstroMed project at the Initiative for Innovative Computing (IIC) has collaborated with medical imaging teams\cite{borkin:astromed}. The rendering of complex multi-dimensional volumetric and surficial data is a common desire of both fields, and the medical imaging software was considerably more advanced than was typical in astronomy in terms of graphical capability. As with the creation and curation of databases for large datasets, collaboration with the IT sector has enabled significant progress and the use of tools beyond the scope of those that could be created by astronomers alone, such as Google Sky\cite{scranton:googlesky}. It is likely that such collaboration will continue to increase in importance.

The program S2Plot\cite{barnes:s2plot}, developed at Swinburne, is motivated by the idea of making three-dimensional plots as easy to transfer from one medium to another (interchange) as two-dimensional plots. The existing familiar interface of a plotting code, in this case PGPlot, has been extended\cite{fluke:interchanging} to enable rendering of multi-dimensional data on several media, including desktop machines, PDF files, Powerpoint-style slides, or web pages. Systems in which the user is able to interact directly with the data are also likely to play a significant role. Partiview\cite{levy:partiview}, developed at NCSA, enables the visualization of particulate data and some isosurfaces either on a desktop or in an immersive CAVE system, and several astronomical datasets have been visualized. Szalay, Springel \& Lemson\cite{szalay:gpuvis} describe using graphical processing units (\S \ref{Subsec: Hardware}) to aid visualization, in which the data are preprocessed to hierarchical levels of detail, and only rendered to the resolution required to appear to the eye as if the whole dataset is being rendered. Paraview\footnote{\url{http://www.paraview.org}} is a program designed for parallel processing on large datasets using distributed memory systems, or on smaller data on a desktop.

Finally, in recent years, numerous online virtual worlds have become popular, the most well-known of which is Second Life. Hut\cite{hut:virtual} and Djorgovski\footnote{\url{http://blogs.discovermagazine.com/cosmicvariance/2008/11/03/guest-post-george-djorgovski-a-new-world-overture}} describe their interaction within these worlds, both with other astronomers in the form of avatars in meetings, and with datasets. While it may initially seem to be just a gimmicky way to have a meeting, the interaction with other avatars is described as `fundamentally visceral', much more so than one would expect. This suggests that, along with social networks for outreach, such interaction among astronomers may become more common, as one will be able to attend a meeting without having to travel physically.

\subsection{Novel Supercomputing Hardware} \label{Subsec: Hardware}

For the final part of \S \ref{Sec: Future}, we turn to novel supercomputing hardware. This is a rapidly developing area, but it has enormous potential to speed up existing analyses, and render previously impossible questions tractable. Specialized hardware has been used in astronomy for many years, but until recently only in limited contexts and applications, such as the GRAPE\cite{ebisuzaki:grape} systems designed specifically for $n$-body calculations, or direct processing of data in instrument-specific hardware. Here, we describe three hardware formats that have emerged in recent years as viable solutions to a more general range of astronomical problems: graphical processing units (GPUs), field-programmable gate arrays (FPGAs), and the Cell processor.

As described in \S \ref{Subsec: Petascale}, the increasing speed of CPU clock cycles has now been largely replaced by increasing parallelism as the main method for continuing improvements in computing power. The methods described there implement {\it coarse-grained} parallelism, which is at the level of separate pieces of hardware or application processes. The hardware described here implements {\it fine-grained} parallelism, in which, at the instruction level, a calculation that would require multiple operations on a CPU is implemented in one operation. The hardware forms an intermediary between the previously-used application-specific integrated circuits (ASIC), and the general purpose CPU.

Future petascale machines (\S \ref{Subsec: Petascale}) are likely to include some or all of these three, either as highly integrated components in a cluster-type system, or as part of the heterogeneous hardware making up a distributed grid-like system that has overall petascale performance.

Spurred by the computer gaming industry, the GPUs on graphics cards within desktop-scale computers have increased in performance much more rapidly than conventional processors (CPUs). They are specially designed to be very fast at carrying out huge numbers of operations that are used in the rendering of graphics, by using vector datatypes and streaming the data. Vector processors have been used before in supercomputing, but GPUs have become of great interest to the scientific community due to their commodity-level pricing, which results from their widespread commercial use, and the increasing ease of use for more general operations than certain graphical processes.

At first, GPUs dealt only with fixed-point numbers, but now single-precision floating point and even double-precision are becoming more common. Thus the chips are no longer simply specialized graphics engines, but are becoming much more general-purpose (GPGPUs). Double-precision is required or highly desirable for many scientific applications. The ease of use of GPUs has been increased thanks to NVidia's Compute Unified Device Architecture development environment (CUDA)\footnote{\url{http://www.nvidia.com/cuda}} for its cards, and will be further aided by the Open Computing Language (OpenCL)\footnote{\url{http://www.khronos.org/opencl}} for heterogeneous environments. These enable the GPU functions to be called in a similar way to a C library, and are becoming a de facto standard. CUDA has also been ported to other higher level languages, including PyCUDA in Python.

GPUs are beginning to be used in astronomy, and several applications have appeared. GPUs can reproduce the functionality of the GRAPE hardware for n-body simulations\cite{gaburov:sapporo}, and CUDA implementations have been shown to outperform GRAPE in some circumstances\cite{belleman:cuda}. GPUs are beginning to be used for real-time processing of data from next generation instruments\cite{ord:gpu} as part of the Data Intensive Science Consortium at the Harvard IIC. Significant speedup has been demonstrated of a $k$ nearest neighbor search on a GPU compared to a kd-tree implemented in C on a CPU\cite{garcia:knngpu}.

FPGAs\cite{brown:fpga,buell:hprc} are another form of hardware that has become viable for somewhat general-purpose scientific computing. While FPGAs have been widely used as specialized hardware for many years, including in telescopes for data processing or adaptive optics, it is only in the past few years that their speed, cost, capacity, and ease of use have made them viable for more general use by non-specialists. As with GPUs, the ability to work with full double precision floating point numbers is also increasing, and their use is via libraries and development environments that enable the FPGA portion of the code to appear as just another function call in C or a C-like language. These tools implement the hardware description language to program the FPGA, which need not be known by the user.

An FPGA consists of a grid of logic gates which must be programmed via software to implement a specific set of functions before running code (hence field-programmable). If the calculation to be performed can be fully represented in this way on the available gates, this enables a throughput speed of one whole calculation of a function per clock cycle, which given a modern FPGA's clock speed of 100 MHz or more, is 100 million per second. In practice, however, the actual speed is often limited by the I/O.

One recent example is the direct mapping of an ANN onto an FPGA\cite{won:annfpga}, which can then in principle classify one object per clock cycle, or 100 million objects per second at 100 MHz. FPGAs will continue to be widely used as specialized components for astronomical systems, for example in providing real-time processing of the next generation synoptic surveys. Brunner, Kindratenko \& Myers\cite{brunner:fpganpcf} demonstrated a significant speedup of the N-point correlation function using FPGAs. Freeman, Weeks \& Austin\cite{freeman:similarity} directly implement distance calculations, such as required by the \knn~data mining algorithm, on an FPGA.

Finally, the IBM Cell processor\cite{scarpino:cell} is a chip containing a conventional CPU and and array of eight more powerful coprocessors for hardware acceleration in a similar manner to the GPU and FPGA. Like the NVidia GPU, it has been widely used in mass-production machines such as the Playstation 3, and is or will be incorporated into several `hybrid' petascale machines, including IBM's Roadrunner, and possibly Blue Waters. Unfortunately, also like the GPU, it is not yet as easy to use as desired for large scale scientific use, but progress in the area is continuing. 

Further novel supercomputing hardware such as ClearSpeed may become viable for science and widely used. It is an area of exciting developments and considerable potential. As with many new developments, however, one must be somewhat careful, in this case because the continued development of the hardware is driven by large commercial companies (NVidia, IBM, etc.), and not the scientific community. Nevertheless, the potential scientific gains are so large that it is certainly worth keeping an eye on.

\section{Conclusions} \label{Sec: Conclusions}

In this review, we have introduced data mining in astronomy, given an overview of its implementation in the form of knowledge discovery in databases, reviewed its application to various science problems, and discussed its future. Throughout, we have tried to emphasize data mining as a tool to enable improved science, not as an end in itself, and to highlight areas where improvements have been made over previous analyses, where they might yet be made, and limitations of this approach.

An astronomer is not a cutting-edge expert in data mining algorithms any more than they are in statistics, databases, hardware, software, etc., but they will need to know enough to usefully apply such approaches to the science problem they wish to address. It is likely that such progress will be made via collaboration with people who are experts in these areas, particularly within large projects, that will employ specialists and have working groups dedicated to data mining. Fully implemented, commercial-level databases will be required since the data will be too big to organize, download, or analyze in any other way.

The available infrastructure should, therefore, be designed so that this data mining approach to research is maximally enabled. The raw or minimally-processed data should be made available in a manner so one can apply user-specific codes either locally or using computational resources local to the data if data size necessitates it. It is unlikely that most researchers will either require or trust the exact resources made available by higher level tools. Instead, they will be useful for exploratory work, but ultimately one must be able to run personal or trusted code on the data, from the level of re-reduction upwards.

A problem arises when one wishes to utilize multiple or distributed datasets, for example in cross-matching data for multi-wavelength studies. Therefore, datasets that can be easily made interoperable via a standard storage schema should be made available. In this manner, a user can bring computing power and algorithms to tackle their particular science question. This problem is particularly acute when large datasets are held at widely separated sites, because transfer of such data across the network is currently impractical. A great deal of science is done on small subsets of the full data, so data will still be frequently downloaded and analyzed locally, but the paradigm of downloading entire datasets is not sustainable.

\section*{Acknowledgments}

We thank the referee for a useful and comprehensive report.

The authors acknowledge support from NASA through grants NN6066H156 and NNG06GF89G, from Microsoft Research, and from the University of Illinois.

The authors made extensive use of the storage and computing facilities at the National Center for Supercomputing Applications and thank the technical staff for their assistance in enabling this work.

This research has made use of the SAO/NASA Astrophysics Data System.


\begin{thebibliography}{100}

\bibitem{bell:petascale}
G.~{Bell}, J.~{Gray} and A.~{Szalay}, {\em IEEE Computer} {\bf 39}, 110
  (2006).

\bibitem{bell:deluge}
G.~{Bell}, T.~{Hey} and A.~{Szalay}, {\em Science} {\bf 323}, 1297  (2009).

\bibitem{hey:4thparadigm} T.~{Hey}, S.~{Tansley} and K.~{Talle} (eds.), {\em {The Fourth Paradigm: Data-Intensive Scientific
Discovery}} (Microsoft Research, Redmond, WA, 2009).

\bibitem{hand:illusion}
D.~J. {Hand}, {\em Statistical Science} {\bf 21}, p.~1  (2006).

\bibitem{witten:datamining2nd}
I.~H. {Witten} and E.~{Frank}, {\em {Data Mining: Practical Machine Learning
  Tools and Techniques}},
  Morgan Kaufmann Series in Data Management Systems, 2nd edn. (Morgan Kaufmann,
  San Francisco, 2005).

\bibitem{bishop:pattern}
C.~M. {Bishop}, {\em {Pattern Recognition and Machine Learning}} (Springer, New
  York, 2007).

\bibitem{hastie:learning2nd}
T.~{Hastie}, R.~{Tibshirani} and J.~{Friedman}, {\em {The Elements of
  Statistical Learning: Data Mining, Inference, and Prediction}}, Springer Series in Statistics, 2nd edn. (Springer, New York,
  2009).

\bibitem{borne:datamining}
K.~{Borne}, {\em {Scientific Data Mining in Astronomy}}, Data Mining and
  Knowledge Discovery Series Data Mining and Knowledge Discovery Series,
  (Taylor \& Francis: CRC Press, Boca Raton, FL, 2009),
  pp. 91--114.

\bibitem{wells:fits}
D.~C. {Wells}, E.~W. {Greisen} and R.~H. {Harten}, {\em \aaps} {\bf 44}, p. 363
   (1981).

\bibitem{ochsenbein:votable}
F.~{Ochsenbein} {\em et~al.}, {VOTable: Tabular Data for the Virtual
  Observatory}, in {\em Toward an International Virtual Observatory\/},  eds.
  P.~J. {Quinn} and K.~M. {G{\'o}rski} (2004).

\bibitem{pyle:dataprep}
D.~{Pyle}, {\em {Data Preparation for Data Mining}}, Morgan Kaufmann Series in Data Management Systems
  (Morgan Kaufmann, San Francisco, 1999).

\bibitem{hogg:histogram}
D.~W. {Hogg}, preprint, [arXiv/0807.4820]   (2008).

\bibitem{karhunen:kl}
K.~{Karhunen}, {\em Annales Academiae Scientiarum Fennicae Series A. I.
  Mathematica-Physica} {\bf 37}, 3  (1947).

\bibitem{loeve:kl}
M.~M. {Lo{\`e}ve}, {\em {Fonctions Al{\'e}atoires de Second Ordre}}, in {\em
  Processus Stochastiques et Mouvement Brownien\/},  ed. P.~{Levy} (Hermann,
  Paris, 1948), Paris.

\bibitem{jolliffe:pca}
I.~T. {Jolliffe}, {\em {Principal Component Analysis}}, Springer Series in Statistics, 2nd edn. (Springer, New York,
  2002).

\bibitem{arya:epsilon}
S.~{Arya}, D.~M. {Mount}, N.~S. {Netanyahu}, R.~{Silverman} and A.~Y. {Wu},
  {\em Journal of the Association for Computing Machinery} {\bf 45}, 891
  (1998).

\bibitem{tishby:ib}
N.~{Tishby}, F.~C. {Pereira} and W.~{Bialek}, {The Information Bottleneck
  Method}, in {\em The 37th annual Allerton Conference on Communication,
  Control, and Computing\/}, 1999.

\bibitem{fisher:matrix}
R.~A. {Fisher}, {\em Annals of Eugenics} {\bf 7}, 179  (1936).

\bibitem{hyvarinen:ica}
A.~{Hyv\"arinen}, J.~{Karhunen} and E.~{Oja}, {\em {Independent Component
  Analysis}} (John Wiley \& Sons, New York, 2001).

\bibitem{lilly:zcosmos}
S.~J. {Lilly} {\em et~al.}, {\em \apjs} {\bf 172}, 70  (2007).

\bibitem{york:sdss}
D.~G. {York} {\em et~al.}, {\em \aj} {\bf 120}, 1579  (2000).

\bibitem{jeffrey:annealing}
W.~{Jeffrey} and R.~{Rosner}, {\em \apj} {\bf 310}, 473  (1986).

\bibitem{bishop:ann}
C.~M. {Bishop}, {\em {Neural Networks for Pattern Recognition}} (Oxford
  University Press, Oxford, 1995).

\bibitem{ripley:ann}
B.~D. {Ripley}, {\em {Pattern Recognition and Neural Networks}} (Cambridge
  University Press, Cambridge, UK, 2008).

\bibitem{duda:pattern}
R.~O. {Duda}, P.~E. {Hart} and D.~G. {Stork}, {\em {Pattern Classification}},
  2nd edn. (Cambridge University Press, Cambridge, UK, 2000).

\bibitem{mcculloch:ann}
W.~S. {McCulloch} and W.~H. {Pitts}, {\em Bulletin of Mathematical Biophysics}
  {\bf 5}, 115  (1943).

\bibitem{hopfield:ann}
J.~J. {Hopfield} and D.~W. {Tank}, {\em Science} {\bf 233}, 625  (1986).

\bibitem{werbos:backprop}
P.~J. {Werbos}, {Beyond regression: new tools for prediction and analysis in
  the behavioural sciences}, PhD thesis, Harvard, (Cambridge, MA, 1974).

\bibitem{parker:backprop}
D.~B. {Parker}, {\em {Learning Logic}}, Tech. Rep. TR-47, Center for
  Computational Research in Economics and Management Science, MIT (Cambridge,
  MA, 1985).

\bibitem{rumelhart:backprop}
D.~E. {Rumelhart}, G.~E. {Hinton} and R.~J. {Williams}, {\em Nature} {\bf 323},
  533  (1986).

\bibitem{levenberg:levmarq}
K.~{Levenberg}, {\em Quarterly of Applied Mathematics} {\bf 2}, p. 164  (1944).

\bibitem{marquardt:levmarq}
D.~W. {Marquardt}, {\em Journal of the Society of Industrial and Applied
  Mathematics} {\bf 2}, p. 431  (1963).

\bibitem{firth:annphotoz}
A.~E. {Firth}, O.~{Lahav} and R.~S. {Somerville}, {\em \mnras} {\bf 339}, 1195
  (2003).

\bibitem{morgan:dt}
J.~N. Morgan and J.~A. Sonquist, {\em Journal of the American Statistical
  Association} {\bf 58}, 415  (1963).

\bibitem{breiman:dt}
L.~{Breiman}, J.~{Friedman}, R.~{Olshen} and C.~{Stone}, {\em {Classification
  and Regression Trees}} (Wadsworth, 1984).

\bibitem{quinlan:dt}
J.~R. {Quinlan}, {\em Machine Learning} {\bf 1}, p.~81  (1986).

\bibitem{quinlan:dtbook}
J.~R. {Quinlan}, {\em {C4.5: Programs for Machine Learning}} (Morgan Kaufmann,
  San Francisco, 1993).

\bibitem{rokach:dt}
L.~{Rokach} and O.~{Maimon}, {\em {Data Mining with Decision Trees: Theory and
  Applications}} (World Scientific, New York, 2008).

\bibitem{salzberg:dt}
S.~{Salzberg}, R.~{Chandar}, H.~{Ford}, S.~K. {Murthy} and R.~{White}, {\em
  \pasp} {\bf 107}, 279  (1995).

\bibitem{cortes:svm}
C.~{Cortes} and V.~{Vapnik}, {\em Machine Learning} {\bf 20}, 273  (1995).

\bibitem{burges:svm}
C.~J.~C. {Burges}, {\em Knowledge Discovery and Data Mining} {\bf 2}, 121
  (1998).

\bibitem{vapnik:svm}
V.~{Vapnik}, {\em {The Nature of Statistical Learning Theory}}, 2nd edn.
  (Springer, New York, 1999).

\bibitem{cristianini:svm}
N.~{Cristianini} and J.~{Shawe-Taylor}, {\em {An Introduction to Support Vector
  Machines and Other Kernel-based Learning Methods}} (Cambridge University
  Press, 2000).

\bibitem{kecman:svm}
V.~{Kecman}, {\em {Learning and Soft Computing: Support Vector Machines, Neural
  Networks, and Fuzzy Logic Models}} (MIT Press, Cambridge, MA, 2001).

\bibitem{schlkopf:svm}
B.~{Schlkopf} and A.~J. {Smola}, {\em {Learning with Kernels: Support Vector
  Machines, Regularization, Optimization, and Beyond}} (MIT Press, Cambridge,
  MA, 2001).

\bibitem{abe:svm}
S.~{Abe}, {\em {Support Vector Machines for Pattern Classification}} (Springer,
  New York, 2005).

\bibitem{wang:svm}
L.~{Wang}, {\em {Support Vector Machines: Theory and Applications}} (Springer,
  New York, 2005).

\bibitem{steinwart:svm}
I.~{Steinwart} and A.~{Christmann}, {\em {Support Vector Machines}} (Springer,
  New York, 2008).

\bibitem{aizerman:kerneltrick}
M.~A. {Aizerman}, E.~M. {Braverman} and L.~I. {Rozonoer}, {\em Automation and
  Remote Control} {\bf 25}, 1175  (1964).

\bibitem{huertascompany:svmmorph}
M.~{Huertas-Company}, D.~{Rouan}, L.~{Tasca}, G.~{Soucail} and O.~{Le
  F{\`e}vre}, {\em \aap} {\bf 478}, 971  (2008).

\bibitem{fix:knn}
E.~{Fix} and J.~{Hodges Jr.}, {\em {Discriminatory analysis: non-parametric
  discrimination: Consistency properties.}}, Tech. Rep. Report No. 4, USAF
  School of Aviation Medicine (Randolph Field, TX, 1951).

\bibitem{cover:nn}
T.~M. {Cover} and P.~E. {Hart}, {\em IEEE Transactions on Information Theory}
  {\bf 13}, 21  (1967).

\bibitem{aha:ib}
D.~W. {Aha}, D.~{Kibler} and M.~K. {Albert}, {\em Machine Learning} {\bf 6}, 37
   (1991).

\bibitem{dasarathy:knn}
B.~{Dasarathy}, {\em {Nearest Neighbor Pattern Classification Techniques}}
  (IEEE Computer Society Press, New York, 1991).

\bibitem{shakhnarovich:knn}
G.~{Shakhnarovich}, T.~{Darrell} and P.~{Indyk} (eds.), {\em {Nearest-Neighbor
  Methods in Learning and Vision: Theory and Practice}} (MIT Press, Cambridge,
  MA, 2006).

\bibitem{parzen:kde}
E.~{Parzen}, {\em Annals of Mathematical Statistics} {\bf 33}, 1065  (1962).

\bibitem{duda:kde}
R.~O. {Duda} and P.~E. {Hart}, {\em {Pattern Classification and Scene
  Analysis}} (John Wiley, New York, 1973).

\bibitem{silverman:kde}
B.~W. {Silverman}, {\em {Density Estimation for Statistics and Data
  Analysis}}, Monographs on
  Statistics and Applied Probability (CRC Press, Boca Raton, FL, 1986).

\bibitem{scott:kde}
D.~W. {Scott}, {\em {Multivariate Density Estimation: Theory, Practice, and
  Visualization}}, Wiley Series in
  Probability and Statistics (Wiley-Interscience, New York, 1992).

\bibitem{taylor:kde}
C.~{Taylor}, {\em Vistas in Astronomy} {\bf 41}, 411  (1997).

\bibitem{wasserman:kde}
L.~{Wasserman}, {\em {All of Statistics: a Concise Course in Statistics}}
  (Springer, New York, 2005).

\bibitem{klemela:kde}
J.~{Klemel\"a}, {\em {Smoothing of Multivariate Data: Density Estimation and
  Visualization}}, Wiley Series in
  Probability and Statistics (John Wiley \& Sons, New York, 2009).

\bibitem{steinhaus:kmeans}
H.~{Steinhaus}, {\em Bull. Acad. Polon. Sci.} {\bf 4}, 801  (1956).

\bibitem{macqueen:kmeans}
J.~{MacQueen}, {Some Methods for Classification and Analysis of Multivariate
  Observations}, in {\em Proceedings of the Fifth Berkeley Symposium on
  Mathematical Statistics and Probability\/},  eds. L.~M. {LeCam} and J.~Neyman
  (University of California Press, Berkeley, 1967).

\bibitem{titterington:mixture}
D.~M. {Titterington}, A.~F.~M. {Smith} and U.~E. {Makov}, {\em {Statistical
  Analysis of Finite Mixture Distributions}} (John Wiley, New York, 1985).

\bibitem{mclachlan:mixture}
G.~J. {McLachlan} and D.~{Peel}, {\em {Finite Mixture Models}}, Wiley Series in Probability and Statistics
  (Wiley-Interscience, New York, 2000).

\bibitem{connolly:fast}
A.~J. {Connolly}, C.~{Genovese}, A.~W. {Moore}, R.~C. {Nichol}, J.~{Schneider}
  and L.~{Wasserman}, preprint, [arXiv:astro-ph/0008187]   (2000).

\bibitem{dolence:lci}
J.~{Dolence} and R.~J. {Brunner}, {Fast Two-Point Correlations of Extremely
  Large Data Sets}, {\em The 9th LCI International Conference on High-Performance
  Clustered Computing}, Urbana-Champaign, IL,  (2008).

\bibitem{dempster:em}
A.~{Dempster}, N.~{Laird} and D.~{Rubin}, {\em Journal of the Royal Statistical
  Society B} {\bf 39}, 1  (1977).

\bibitem{watanabe:em}
M.~{Watanabe} and K.~{Yamaguchi} (eds.), {\em {The EM Algorithm and Related
  Statistical Models}},
  Statistics: a Series of Textbooks and Monographs (CRC Press, Boca Raton, FL,
  2003).

\bibitem{mclachlan:em}
G.~J. {McLachlan} and T.~{Krishnan}, {\em {The EM Algorithm and
  Extensions}}, Wiley Series in
  Probability and Statistics (John Wiley \& Sons, New York, 2008).

\bibitem{kohonen:somprev}
T.~{Kohonen}, {\em Biological Cybernetics} {\bf 43}, 59  (1982).

\bibitem{kohonen:som}
T.~{Kohonen}, {\em {Self-Organizing Maps, 3rd extended edition}}, Springer
  Series in Information Sciences, Vol.~30, 3rd edn. (Springer, Berlin, 2001).

\bibitem{naim:som}
A.~{Naim}, K.~U. {Ratnatunga} and R.~E. {Griffiths}, {\em \apjs} {\bf 111}, p.
  357  (1997).

\bibitem{kohonen:lvq}
T.~{Kohonen}, {\em {Self-Organization and Associative Memory}}, 3rd edn.
  (Springer-Verlag, Berlin, 1989).

\bibitem{comon:ica}
P.~{Comon}, {\em Signal Processing} {\bf 36}, 287  (1994).

\bibitem{lee:ica}
T.~{Lee}, {\em {Independent Component Analysis - Theory and Applications}}
  (Kluwer Academic Publishers, New York, 1998).

\bibitem{roberts:ica}
S.~{Roberts} and R.~{Everson} (eds.), {\em {Independent Component Analysis:
  Principles and Practice}} (Cambridge University Press, Cambridge, UK, 2001).

\bibitem{stone:ica}
J.~V. {Stone}, {\em {Independent Component Analysis: A Tutorial Introduction}}
  (MIT Press, Cambridge, MA, 2004).

\bibitem{chapelle:semisupervised}
O.~{Chapelle}, B.~{Sch\"olkopf} and A.~{Zien} (eds.), {\em {Semi-Supervised
  Learning}} (MIT Press, Cambridge, MA, 2006).

\bibitem{zhu:semisupervised}
X.~{Zhu}, A.~{Goldberg}, R.~{Brachman} and T.~{Dietterich}, {\em {Introduction
  to Semi-supervised Learning}}, Synthesis Lectures on Artificial Intelligence and
  Machine Learning (Morgan \& Claypool, San Rafael, CA, 2009).

\bibitem{bazell:classdisc}
D.~{Bazell} and D.~J. {Miller}, {\em \apj} {\bf 618}, 723  (2005).

\bibitem{holland:genetic}
J.~H. {Holland}, {\em {Adaptation in Natural and Artificial Systems: An
  Introductory Analysis with Applications to Biology, Control and Artificial
  Intelligence}} (The University of Michigan Press, Ann Arbor, MI, 1975).

\bibitem{goldberg:genetic}
D.~E. {Goldberg}, {\em {Genetic Algorithms in Search, Optimization, and Machine
  Learning}} (Addison-Wesley, Reading, MA, 1989).

\bibitem{coley:ga}
D.~A. {Coley}, {\em {An Introduction to Genetic Algorithms for Scientists and
  Engineers}} (World Scientific, New York, 1997).

\bibitem{mitchell:ga}
M.~{Mitchell}, {\em {An Introduction to Genetic Algorithms}} (MIT Press,
  Cambridge, MA, 1998).

\bibitem{haupt:genetic2nd}
R.~L. {Haupt} and S.~E. {Haupt}, {\em {Practical Genetic Algorithms}}, 2nd edn.
  (Wiley Inter-Science, New York, 2004).

\bibitem{sivanandam:ga}
S.~N. {Sivanandam} and S.~N. {Deepa}, {\em {Introduction to Genetic
  Algorithms}} (Springer, New York, 2007).

\bibitem{goldberg:design}
{{Goldberg}, D.~E.}, {\em Design of innovation: Lessons from and for competent
  genetic algorithms} (Kluwer Academic Publishers, Boston, MA, 2002).

\bibitem{adamo:associationrules}
J.~M. {Adamo}, {\em {Data Mining for Association Rules and Sequential Patterns:
  Sequential and Parallel Algorithms}} (Springer, New York, 2000).

\bibitem{zhang:associationrules}
C.~{Zhang} and S.~{Zhang}, {\em {Association Rule Mining: Models and
  Algorithms}}, Lecture Notes in Computer
  Science (Springer, New York, 2002).

\bibitem{benjamini:fdr}
Y.~{Benjamini} and Y.~{Hochberg}, {\em Journal of the Royal Statistical Society
  B} {\bf 57}, 289  (1995).

\bibitem{welge:d2k}
M.~{Welge}, W.~H. {Hsu}, L.~S. {Auvil}, T.~M. {Redman} and D.~{Tcheng},
  {High-Performance Knowledge Discovery and Data Mining Systems Using
  Workstation Clusters}, in {\em {12th National Conference on High Performance
  Networking and Computing (SC99)}\/}, 1999.

\bibitem{salzberg:classifiercritique}
S.~L. {Salzberg}, {\em Data Mining and Knowledge Discovery} {\bf 1}, 1  (1995).

\bibitem{kirkpatrick:annealing}
S.~{Kirkpatrick}, C.~D. {Gelatt} and M.~P. {Vecchi}, {\em Science} {\bf 220},
  671  (1983).

\bibitem{cerny:annealing}
V.~{{\v C}ern{\'y}}, {\em Journal of Optimization Theory and Applications} {\bf
  45}, 41  (1985).

\bibitem{vanlaarhoven:annealing}
P.~J. {van Laarhiven} and E.~H. {Aarts}, {\em {Simulated Annealing: Theory and
  Applications}} (Springer, New York, 1987).

\bibitem{aarts:annealing}
E.~{Aarts} and J.~{Korst}, {\em {Simulated Annealing and Boltzmann Machines: A
  Stochastic Approach to Combinatorial Optimization and Neural Computing}}
  (Wiley, New York, 1989).

\bibitem{breiman:bagging}
L.~{Breiman}, {\em Machine Learning} {\bf 26}, 123  (1996).

\bibitem{breiman:randomforest}
L.~{Breiman}, {\em Machine Learning} {\bf 45}, 5  (2001).

\bibitem{bentley:kdtree}
J.~L. {Bentley}, {\em Communications of the ACM} {\bf 18}, 509  (1975).

\bibitem{gardner:paralleltree}
J.~P. {Gardner}, A.~{Connolly} and C.~{McBride}, {Enabling rapid development of
  parallel tree search applications}, in {\em CLADE '07: Proceedings of the 5th
  IEEE workshop on Challenges of large applications in distributed
  environments\/},  (ACM, New York, 2007).

\bibitem{miller:annapps}
A.~S. {Miller}, {\em Vistas in Astronomy} {\bf 36}, 141  (1993).

\bibitem{lahav:annmethods}
O.~{Lahav}, A.~{Naim}, L.~{Sodr{\'e}} and M.~C. {Storrie-Lombardi}, {\em
  \mnras} {\bf 283}, p. 207  (1996).

\bibitem{bailerjones:ann}
C.~A.~L. {Bailer-Jones}, R.~{Gupta} and H.~P. {Singh}, {An Introduction to
  Artificial Neural Networks}, in {\em Automated Data Analysis in Astronomy\/},
   eds. R.~{Gupta}, H.~P. {Singh} and C.~A.~L. {Bailer-Jones} (2002).

\bibitem{li:annapps}
L.-L. {Li}, Y.-X. {Zhang}, Y.-H. {Zhao} and D.-W. {Yang}, {\em Progress in
  Astronomy} {\bf 24}, 285  (2006).

\bibitem{tagliaferri:nnast} R.~{Tagliaferri} {\em et~al.}, Neural Networks,
{\bf 16}, 297 (2003).

\bibitem{white:dts}
R.~L. {White}, {Astronomical Applications of Oblique Decision Trees}, American
  Institute of Physics Conference Series Vol.~1082 (2008).

\bibitem{charbonneau:ga}
P.~{Charbonneau}, {\em \apjs} {\bf 101}, p. 309  (1995).

\bibitem{bailerjones:stellar}
C.~A.~L. {Bailer-Jones}, {Automated Stellar Classification for Large Surveys: A
  Review of Methods and Results}, in {\em Automated Data Analysis in
  Astronomy\/},  eds. R.~{Gupta}, H.~P. {Singh} and C.~A.~L.
  {Bailer-Jones} (2002).

\bibitem{weir:skicat}
N.~{Weir}, U.~M. {Fayyad}, S.~G. {Djorgovski} and J.~{Roden}, {\em \pasp} {\bf
  107}, p. 1243  (1995).

\bibitem{burl:jartool}
M.~C. {Burl}, L.~{Asker}, P.~{Smyth}, U.~{Fayyad}, P.~{Perona}, J.~{Aubele} and
  L.~{Crumpler}, {\em Machine Learning} {\bf 30}, 165  (1998).

\bibitem{burl:diamondeye}
M.~C. {Burl}, C.~{Fowlkes}, J.~{Roden}, A.~{Stechert} and S.~{Mukhtar},
  {Diamond Eye: a distributed architecture for image data mining}, Society of
  Photo-Optical Instrumentation Engineers (SPIE) Conference Series
  Vol.~3695 (1999).

\bibitem{kamath:sapphire}
C.~{Kamath}, {\em Journal of Physics Conference Series} {\bf 125}, 012094
  (2008).

\bibitem{kamath:scientific}
C.~{Kamath}, {\em {Scientific data mining: a practical perspective}} (Society
  for Industrial and Applied Mathematics, Philadelphia, PA, 2009).

\bibitem{maddox:apmstargal}
S.~J. {Maddox}, G.~{Efstathiou}, W.~J. {Sutherland} and J.~{Loveday}, {\em
  \mnras} {\bf 243}, 692  (1990).

\bibitem{djorgovski:dposs}
S.~G. {Djorgovski}, R.~R. {Gal}, S.~C. {Odewahn}, R.~R. {de Carvalho},
  R.~{Brunner}, G.~{Longo} and R.~{Scaramella}, {The Palomar Digital Sky Survey
  (DPOSS)}, in {\em Wide Field Surveys in Cosmology\/},  eds. S.~{Colombi},
  Y.~{Mellier} and B.~{Raban} (1998).

\bibitem{odewahn:autostargal}
S.~C. {Odewahn}, E.~B. {Stockwell}, R.~L. {Pennington}, R.~M. {Humphreys} and
  W.~A. {Zumach}, {\em \aj} {\bf 103}, 318  (1992).

\bibitem{odewahn:annstargal}
S.~C. {Odewahn} and M.~L. {Nielsen}, {\em Vistas in Astronomy} {\bf 38}, 281
  (1994).

\bibitem{bazell:preprocessing}
D.~{Bazell} and Y.~{Peng}, {\em \apjs} {\bf 116}, p.~47  (1998).

\bibitem{andreon:stargal}
S.~{Andreon}, G.~{Gargiulo}, G.~{Longo}, R.~{Tagliaferri} and N.~{Capuano},
  {\em \mnras} {\bf 319}, 700  (2000).

\bibitem{philip:dbnn}
N.~S. {Philip}, Y.~{Wadadekar}, A.~{Kembhavi} and K.~B. {Joseph}, {\em \aap}
  {\bf 385}, 1119  (2002).

\bibitem{odewahn:dpossstargal}
S.~C. {Odewahn}, R.~R. {de Carvalho}, R.~R. {Gal}, S.~G. {Djorgovski},
  R.~{Brunner}, A.~{Mahabal}, P.~A.~A. {Lopes}, J.~L.~K. {Moreira} and
  B.~{Stalder}, {\em \aj} {\bf 128}, 3092  (2004).

\bibitem{collister:megazlrg}
A.~{Collister} {\em et~al.}, {\em \mnras} {\bf 375}, 68  (2007).

\bibitem{weir:stargal}
N.~{Weir}, U.~M. {Fayyad} and S.~{Djorgovski}, {\em \aj} {\bf 109}, p. 2401
  (1995).

\bibitem{ball:dtclassification}
N.~M. {Ball}, R.~J. {Brunner}, A.~D. {Myers} and D.~{Tcheng}, {\em \apj} {\bf
  650}, p. 497  (2006).

\bibitem{qin:stargal}
D.-M. {Qin}, P.~{Guo}, Z.-Y. {Hu} and Y.-H. {Zhao}, {\em Chinese Journal of
  Astronomy and Astrophysics} {\bf 3}, 277  (2003).

\bibitem{miller:som}
A.~S. {Miller} and M.~J. {Coe}, {\em \mnras} {\bf 279}, 293  (1996).

\bibitem{hubble:extragalnebulae}
E.~P. {Hubble}, {\em \apj} {\bf 64}, 321  (1926).

\bibitem{hubble:realm}
E.~P. {Hubble}, {\em {Realm of the Nebulae}} (Yale University Press, Newhaven,
  CT, 1936).

\bibitem{sandage:hubbatlas}
A.~{Sandage}, {\em {The Hubble atlas of galaxies}} (Carnegie Institution,
  Washington, DC, 1961).

\bibitem{sandage:carnegieatlas}
A.~{Sandage} and J.~{Bedke}, {\em {The Carnegie atlas of galaxies}} (Carnegie
  Institution of Washington with The Flintridge Foundation, Washington, DC,
  1994).

\bibitem{vandenbergh:morph}
S.~{van den Bergh}, {\em {Galaxy morphology and classification}} (Cambridge
  University Press, Cambridge, UK, 1998).

\bibitem{sandage:classfnhistory}
A.~{Sandage}, {\em \araa} {\bf 43}, 581  (2005).

\bibitem{roberts:hubble}
M.~S. {Roberts} and M.~P. {Haynes}, {\em \araa} {\bf 32}, 115  (1994).

\bibitem{firmani:hubble}
C.~{Firmani} and V.~{Avila-Reese}, {Physical processes behind the morphological
  Hubble sequence}, Revista Mexicana de Astronomia y Astrofisica Conference
  Series Vol.~17 (2003).

\bibitem{morgan:ci}
W.~W. {Morgan}, {\em \pasp} {\bf 70}, p. 364  (1958).

\bibitem{morgan:ci2}
W.~W. {Morgan}, {\em \pasp} {\bf 71}, p. 394  (1959).

\bibitem{devaucouleurs:ci}
G.~{de Vaucouleurs}, {Qualitative and Quantitative Classifications of
  Galaxies.}, in {\em Evolution of Galaxies and Stellar Populations\/},  eds.
  B.~M. {Tinsley} and R.~B. {Larson} (1977).

\bibitem{devaucouleurs:devprofile}
G.~{de Vaucouleurs}, {\em Annales d'Astrophysique} {\bf 11}, p. 247  (1948).

\bibitem{patterson:expprofile}
F.~S. {Patterson}, {\em Harvard College Observatory Bulletin} {\bf 914}, 9
  (1940).

\bibitem{freeman:expprofile}
K.~C. {Freeman}, {\em \apj} {\bf 160}, p. 811  (1970).

\bibitem{sersic:australes}
J.~L. {S\'ersic}, {\em {Atlas de galaxias australes}} (Observatorio
  Astronomico, Cordoba, Argentina, 1968).

\bibitem{graham:sersic}
A.~W. {Graham} and S.~P. {Driver}, {\em \pasa} {\bf 22}, 118  (2005).

\bibitem{vandenbergh:lumclass}
S.~{van den Bergh}, {\em \apj} {\bf 131}, p. 215  (1960).

\bibitem{vandenbergh:lumclassb}
S.~{van den Bergh}, {\em \apj} {\bf 131}, p. 558  (1960).

\bibitem{vandenbergh:ddo}
S.~{van den Bergh}, {\em \apj} {\bf 206}, 883  (1976).

\bibitem{conselice:cas}
C.~J. {Conselice}, {\em \apjs} {\bf 147}, 1  (2003).

\bibitem{devaucouleurs:tsystem}
G.~{de Vaucouleurs}, {\em Memoirs of the Commonwealth Observatory, Mount
  Stromlo} {\bf 3}  (1956).

\bibitem{devaucouleurs:ttype}
G.~{de Vaucouleurs}, {\em Handbuch der Physik} {\bf 53}, p. 275  (1959).

\bibitem{barden:ferengi}
M.~{Barden}, K.~{Jahnke} and B.~{H{\"a}u{\ss}ler}, {\em \apjs} {\bf 175}, 105
  (2008).

\bibitem{storrielombardi:ann}
M.~C. {Storrie-Lombardi}, O.~{Lahav}, L.~{Sodr\'e} and L.~J.
  {Storrie-Lombardi}, {\em \mnras} {\bf 259}, p.~8P  (1992).

\bibitem{lahav:annscience}
O.~{Lahav} {\em et~al.}, {\em Science} {\bf 267}, 859  (1995).

\bibitem{naim:eyemorph}
A.~{Naim} {\em et~al.}, {\em \mnras} {\bf 274}, 1107  (1995).

\bibitem{naim:annmorph}
A.~{Naim}, O.~{Lahav}, L.~{Sodr\'e} and M.~C. {Storrie-Lombardi}, {\em \mnras}
  {\bf 275}, 567  (1995).

\bibitem{collister:annz}
A.~A. {Collister} and O.~{Lahav}, {\em \pasp} {\bf 116}, 345  (2004).

\bibitem{naim:peculiarmorph}
A.~{Naim}, K.~U. {Ratnatunga} and R.~E. {Griffiths}, {\em \apj} {\bf 476}, p.
  510  (1997).

\bibitem{madgwick:morphspec}
D.~S. {Madgwick}, {\em \mnras} {\bf 338}, 197  (2003).

\bibitem{odewahn:hdfannmorph}
S.~C. {Odewahn}, R.~A. {Windhorst}, S.~P. {Driver} and W.~C. {Keel}, {\em
  \apjl} {\bf 472}, p. L13  (1996).

\bibitem{windhorst:hizgals}
R.~{Windhorst}, S.~{Odewahn}, C.~{Burg}, S.~{Cohen} and I.~{Waddington}, {\em
  \apss} {\bf 269}, 243  (1999).

\bibitem{cohen:hstmorph}
S.~H. {Cohen}, R.~A. {Windhorst}, S.~C. {Odewahn}, C.~A. {Chiarenza} and S.~P.
  {Driver}, {\em \aj} {\bf 125}, 1762  (2003).

\bibitem{odewahn:fouriermorph}
S.~C. {Odewahn}, S.~H. {Cohen}, R.~A. {Windhorst} and N.~S. {Philip}, {\em
  \apj} {\bf 568}, 539  (2002).

\bibitem{bazell:ensembles}
D.~{Bazell} and D.~W. {Aha}, {\em \apj} {\bf 548}, 219  (2001).

\bibitem{bazell:features}
D.~{Bazell}, {\em \mnras} {\bf 316}, 519  (2000).

\bibitem{ball:ann}
N.~M. {Ball}, J.~{Loveday}, M.~{Fukugita}, O.~{Nakamura}, S.~{Okamura},
  J.~{Brinkmann} and R.~J. {Brunner}, {\em \mnras} {\bf 348}, 1038  (2004).

\bibitem{ball:bivlf}
N.~M. {Ball}, J.~{Loveday}, R.~J. {Brunner}, I.~K. {Baldry} and J.~{Brinkmann},
  {\em \mnras} {\bf 373}, 845  (2006).

\bibitem{ball:envt}
N.~M. {Ball}, J.~{Loveday} and R.~J. {Brunner}, {\em \mnras} {\bf 383}, 907
  (2008).

\bibitem{kelly:shapelet2}
B.~C. {Kelly} and T.~A. {McKay}, {\em \aj} {\bf 129}, 1287  (2005).

\bibitem{serraricart:annastroapps}
M.~{Serra-Ricart}, X.~{Calbet}, L.~{Garrido} and V.~{Gaitan}, {\em \aj} {\bf
  106}, 1685  (1993).

\bibitem{adams:ann}
A.~{Adams} and A.~{Woolley}, {\em Vistas in Astronomy} {\bf 38}, 273  (1994).

\bibitem{molinari:annesolf}
E.~{Molinari} and R.~{Smareglia}, {\em \aap} {\bf 330}, 447  (1998).

\bibitem{detheije:typekinematics}
P.~A.~M. {de Theije} and P.~{Katgert}, {\em \aap} {\bf 341}, 371  (1999).

\bibitem{cantupaz:evolving}
E.~{Cant{\'u}-Paz} and C.~{Kamath}, {Evolving Neural Networks For The
  Classification Of Galaxies}, in {\em GECCO '02: Proceedings of the Genetic
  and Evolutionary Computation Conference\/},  (Morgan Kaufmann Publishers
  Inc., San Francisco, 2002).

\bibitem{kamath:bentdouble}
C.~{Kamath}, E.~{Cant{\'u}-Paz}, I.~K. {Fodor} and N.~I. {Tang}, {\em Computing
  in Science and Engineering} {\bf 4}, 52  (2002).

\bibitem{becker:first}
R.~H. {Becker}, R.~L. {White} and D.~J. {Helfand}, {\em \apj} {\bf 450}, p. 559
   (1995).

\bibitem{delacalleja:ann}
J.~{de la Calleja} and O.~{Fuentes}, {\em \mnras} {\bf 349}, 87  (2004).

\bibitem{spiekermann:automorph}
G.~{Spiekermann}, {\em \aj} {\bf 103}, 2102  (1992).

\bibitem{owens:dt}
E.~A. {Owens}, R.~E. {Griffiths} and K.~U. {Ratnatunga}, {\em \mnras} {\bf
  281}, 153  (1996).

\bibitem{zhang:morph}
Y.~{Zhang}, L.~{Li} and Y.~{Zhao}, {\em \mnras} {\bf 392}, 233  (2009).

\bibitem{huertascompany:svmmorph2}
M.~{Huertas-Company} {\em et~al.}, {\em \aap} {\bf 497}, 743  (2009).

\bibitem{tsalmantza:gaiaclassification}
P.~{Tsalmantza} {\em et~al.}, {\em \aap} {\bf 470}, 761  (2007).

\bibitem{lintott:galaxyzoo}
C.~J. {Lintott} {\em et~al.}, {\em \mnras} {\bf 389}, 1179  (2008).

\bibitem{humason:100redshifts}
M.~L. {Humason}, {\em \apj} {\bf 83}, p.~10  (1936).

\bibitem{morgan:specclass}
W.~W. {Morgan} and N.~U. {Mayall}, {\em \pasp} {\bf 69}, p. 291  (1957).

\bibitem{connolly:orthogonal}
A.~J. {Connolly}, A.~S. {Szalay}, M.~A. {Bershady}, A.~L. {Kinney} and
  D.~{Calzetti}, {\em \aj} {\bf 110}, p. 1071  (1995).

\bibitem{connolly:eclass}
A.~J. {Connolly} and A.~S. {Szalay}, {\em \aj} {\bf 117}, 2052  (1999).

\bibitem{madgwick:parametrisation}
D.~{Madgwick}, O.~{Lahav}, K.~{Taylor} and {The 2dFGRS Team}, {Parameterisation
  of Galaxy Spectra in the 2dF Galaxy Redshift Survey}, in {\em Mining the
  Sky\/},  eds. A.~J. {Banday}, S.~{Zaroubi} and M.~{Bartelmann} (2001).

\bibitem{yip:spectypes}
C.~W. {Yip} {\em et~al.}, {\em \aj} {\bf 128}, 585  (2004).

\bibitem{storrielombardi:annspec}
M.~C. {Storrie-Lombardi}, M.~J. {Irwin}, T.~{von Hippel} and L.~J.
  {Storrie-Lombardi}, {\em Vistas in Astronomy} {\bf 38}, 331  (1994).

\bibitem{folkes:annspec}
S.~R. {Folkes}, O.~{Lahav} and S.~J. {Maddox}, {\em \mnras} {\bf 283}, 651
  (1996).

\bibitem{colless:2dffinal}
M.~{Colless} {\em et~al.}, preprint, [arXiv:astro-ph/0306581]   (2003).

\bibitem{slonim:ib}
N.~{Slonim}, R.~{Somerville}, N.~{Tishby} and O.~{Lahav}, {\em \mnras} {\bf
  323}, 270  (2001).

\bibitem{lu:ica}
H.~{Lu}, H.~{Zhou}, J.~{Wang}, T.~{Wang}, X.~{Dong}, Z.~{Zhuang} and C.~{Li},
  {\em \aj} {\bf 131}, 790  (2006).

\bibitem{abdalla:emission}
F.~B. {Abdalla}, A.~{Mateus}, W.~A. {Santos}, L.~{Sodr{\`e}}, Jr.,
  I.~{Ferreras} and O.~{Lahav}, {\em \mnras} {\bf 387}, 945  (2008).

\bibitem{lauberts:esouppsala}
A.~{Lauberts} and E.~A. {Valentijn}, {\em {The surface photometry catalogue of
  the ESO-Uppsala galaxies}} (European Southern Observatory, Garching, Germany,
  1989).

\bibitem{baldwin:bpt}
J.~A. {Baldwin}, M.~M. {Phillips} and R.~{Terlevich}, {\em \pasp} {\bf 93}, 5
  (1981).

\bibitem{carballo:annqso}
R.~{Carballo}, A.~S. {Cofi{\~n}o} and J.~I. {Gonz{\'a}lez-Serrano}, {\em
  \mnras} {\bf 353}, 211  (2004).

\bibitem{claeskens:gaiaqsos}
J.-F. {Claeskens}, A.~{Smette}, L.~{Vandenbulcke} and J.~{Surdej}, {\em \mnras}
  {\bf 367}, 879  (2006).

\bibitem{carballo:annfirstqso}
R.~{Carballo}, J.~I. {Gonz{\'a}lez-Serrano}, C.~R. {Benn} and
  F.~{Jim{\'e}nez-Luj{\'a}n}, {\em \mnras} {\bf 391}, 369  (2008).

\bibitem{white:firstqso}
R.~L. {White} {\em et~al.}, {\em \apjs} {\bf 126}, 133  (2000).

\bibitem{suchkov:dt}
A.~A. {Suchkov}, R.~J. {Hanisch} and B.~{Margon}, {\em \aj} {\bf 130}, 2439
  (2005).

\bibitem{zhang:qso}
Y.-X. {Zhang} and Y.-H. {Zhao}, {\em Chinese Journal of Astronomy and
  Astrophysics} {\bf 7}, 289  (2007).

\bibitem{zhang:decisiontable}
Y.~{Zhang}, Y.~{Zhao} and D.~{Gao}, {\em Advances in Space Research} {\bf 41},
  1949  (2008).

\bibitem{zhao:dtactive}
Y.~{Zhao} and Y.~{Zhang}, {\em Advances in Space Research} {\bf 41}, 1955
  (2008).

\bibitem{knigge:balqso}
C.~{Knigge}, S.~{Scaringi}, M.~R. {Goad} and C.~E. {Cottis}, {\em \mnras} {\bf
  386}, 1426  (2008).

\bibitem{yip:qsoclassification}
C.~W. {Yip} {\em et~al.}, {\em \aj} {\bf 128}, 2603  (2004).

\bibitem{zhang:methods}
Y.~{Zhang} and Y.~{Zhao}, {\em \pasp} {\bf 115}, 1006  (2003).

\bibitem{gao:svmkdqso}
D.~{Gao}, Y.-X. {Zhang} and Y.-H. {Zhao}, {\em \mnras} {\bf 386}, 1417  (2008).

\bibitem{abrusco:qsocandidates}
R.~{D'Abrusco}, G.~{Longo} and N.~A. {Walton}, {\em \mnras} {\bf 396}, 223
  (2009).

\bibitem{richards:dr6photoqso}
G.~T. {Richards} {\em et~al.}, {\em \apjs} {\bf 180}, 67  (2009).

\bibitem{zhao:agn}
M.-F. {Zhao}, C.~{Wu}, A.~{Luo}, F.-C. {Wu} and Y.-H. {Zhao}, {\em Chinese
  Astronomy and Astrophysics} {\bf 31}, 352  (2007).

\bibitem{bamford:nonparametric}
S.~P. {Bamford}, A.~L. {Rojas}, R.~C. {Nichol}, C.~J. {Miller}, L.~{Wasserman},
  C.~R. {Genovese} and P.~E. {Freeman}, {\em \mnras} {\bf 391}, 607  (2008).

\bibitem{jarvis:focas}
J.~F. {Jarvis} and J.~A. {Tyson}, {\em \aj} {\bf 86}, 476  (1981).

\bibitem{stetson:daophot} P.~B.~{Stetson}, {\em \pasp} {\bf 99}, 191 (1987).

\bibitem{bertin:sextractor}
E.~{Bertin} and S.~{Arnouts}, {\em \aaps} {\bf 117}, 393  (1996).

\bibitem{maino:fastica}
D.~{Maino} {\em et~al.}, {\em \mnras} {\bf 334}, 53  (2002).

\bibitem{guglielmetti:background}
F.~{Guglielmetti}, R.~{Fischer} and V.~{Dose}, {\em \mnras} {\bf 396}, 165
  (2009).

\bibitem{serraricart:faint}
M.~{Serra-Ricart}, V.~{Gaitan}, L.~{Garrido} and I.~{Perez-Fournon}, {\em
  \aaps} {\bf 115}, p. 195  (1996).

\bibitem{goebel:bayesian}
J.~{Goebel}, J.~{Stutz}, K.~{Volk}, H.~{Walker}, F.~{Gerbault}, M.~{Self},
  W.~{Taylor} and P.~{Cheeseman}, {\em \aap} {\bf 222}, L5  (1989).

\bibitem{mcglynn:classx}
T.~A. {McGlynn} {\em et~al.}, {\em \apj} {\bf 616}, 1284  (2004).

\bibitem{bazell:subclasses}
D.~{Bazell}, D.~J. {Miller} and M.~{SubbaRao}, {\em \apj} {\bf 649}, 678
  (2006).

\bibitem{morgan:mk}
W.~W. {Morgan}, P.~C. {Keenan} and E.~{Kellman}, {\em {An atlas of stellar
  spectra, with an outline of spectral classification}} (The University of
  Chicago press, 1943).

\bibitem{vonhippel:annstellar}
T.~{von Hippel}, L.~J. {Storrie-Lombardi}, M.~C. {Storrie-Lombardi} and M.~J.
  {Irwin}, {\em \mnras} {\bf 269}, p.~97  (1994).

\bibitem{weaver:annstellar}
W.~B. {Weaver} and A.~V. {Torres-Dodgen}, {\em \apj} {\bf 487}, p. 847  (1997).

\bibitem{singh:annstellar}
H.~P. {Singh}, R.~K. {Gulati} and R.~{Gupta}, {\em \mnras} {\bf 295}, 312
  (1998).

\bibitem{bailerjones:annstellar}
C.~A.~L. {Bailer-Jones}, M.~{Irwin} and T.~{von Hippel}, {\em \mnras} {\bf
  298}, 361  (1998).

\bibitem{gulati:annstellar}
R.~K. {Gulati} and L.~{Altamirano}, {\em \apss} {\bf 273}, 73  (2000).

\bibitem{bazarghan:stellar}
M.~{Bazarghan} and R.~{Gupta}, {\em \apss} {\bf 315}, 201  (2008).

\bibitem{gupta:iras}
R.~{Gupta}, H.~P. {Singh}, K.~{Volk} and S.~{Kwok}, {\em \apjs} {\bf 152}, 201
  (2004).

\bibitem{manteiga:starmind}
M.~{Manteiga}, I.~{Carricajo}, A.~{Rodr{\'{\i}}guez}, C.~{Dafonte} and
  B.~{Arcay}, {\em \aj} {\bf 137}, 3245  (2009).

\bibitem{wozniak:mirasvm}
P.~R. {Wo{\'z}niak}, S.~J. {Williams}, W.~T. {Vestrand} and V.~{Gupta}, {\em
  \aj} {\bf 128}, 2965  (2004).

\bibitem{bailey:sneclassification}
S.~{Bailey}, C.~{Aragon}, R.~{Romano}, R.~C. {Thomas}, B.~A. {Weaver} and
  D.~{Wong}, {\em \apj} {\bf 665}, 1246  (2007).

\bibitem{waniak:cosmicray}
W.~{Waniak}, {\em Experimental Astronomy} {\bf 21}, 151  (2006).

\bibitem{faundezabans:pne}
M.~{Faundez-Abans}, M.~I. {Ormeno} and M.~{de Oliveira-Abans}, {\em \aaps} {\bf
  116}, 395  (1996).

\bibitem{misra:asteroids}
A.~{Misra} and S.~J. {Bus}, {Artificial Neural Network Classification of
  Asteroids in the Sloan Digital Sky Survey}, in {\em AAS/Division for
  Planetary Sciences Meeting Abstracts\/}, 2008.

\bibitem{chattopadhyay:gammaray}
T.~{Chattopadhyay}, R.~{Misra}, A.~K. {Chattopadhyay} and M.~{Naskar}, {\em
  \apj} {\bf 667}, 1017  (2007).

\bibitem{scaringi:isina}
S.~{Scaringi}, A.~J. {Bird}, D.~J. {Clark}, A.~J. {Dean}, A.~B. {Hill}, V.~A.
  {McBride} and S.~E. {Shaw}, {\em \mnras} {\bf 390}, 1339  (2008).

\bibitem{stebbins:photoz}
J.~{Stebbins} and A.~E. {Whitford}, {\em \apj} {\bf 108}, p. 413  (1948).

\bibitem{baum:photoz}
W.~A. {Baum}, {Photoelectric Magnitudes and Red-Shifts}, in {\em {IAU Symp. 15:
  Problems of Extra-Galactic Research}\/}, 1962.

\bibitem{koo:photoz}
D.~C. {Koo}, {\em \aj} {\bf 90}, 418  (1985).

\bibitem{loh:photoz}
E.~D. {Loh} and E.~J. {Spillar}, {\em \apj} {\bf 303}, 154  (1986).

\bibitem{gwyn:photoz}
S.~D.~J. {Gwyn} and F.~D.~A. {Hartwick}, {\em \apjl} {\bf 468}, p. L77  (1996).

\bibitem{lanzetta:photoz}
K.~M. {Lanzetta}, A.~{Yahil} and A.~{Fernandez-Soto}, {\em \nat} {\bf 381}, 759
   (1996).

\bibitem{mobasher:photoz}
B.~{Mobasher}, M.~{Rowan-Robinson}, A.~{Georgakakis} and N.~{Eaton}, {\em
  \mnras} {\bf 282}, L7  (1996).

\bibitem{sawicki:photoz}
M.~J. {Sawicki}, H.~{Lin} and H.~K.~C. {Yee}, {\em \aj} {\bf 113}, 1  (1997).

\bibitem{connolly:angcf}
A.~J. {Connolly}, A.~S. {Szalay} and R.~J. {Brunner}, {\em \apjl} {\bf 499}, p.
  L125  (1998).

\bibitem{wang:photoz}
Y.~{Wang}, N.~{Bahcall} and E.~L. {Turner}, {\em \aj} {\bf 116}, 2081  (1998).

\bibitem{benitez:photoz}
N.~{Ben{\'{\i}}tez}, {\em \apj} {\bf 536}, 571  (2000).

\bibitem{koo:photozhistory}
D.~C. {Koo}, {Overview - Photometric Redshifts: A Perspective from an
  Old-Timer[!] on their Past, Present, and Potential}, in {\em Photometric
  Redshifts and the Detection of High Redshift Galaxies\/},  eds. R.~{Weymann},
  L.~{Storrie-Lombardi}, M.~{Sawicki} and R.~{Brunner}, Astronomical Society of
  the Pacific Conference Series, Vol.~191 (1999).

\bibitem{massarotti:template}
M.~{Massarotti}, A.~{Iovino} and A.~{Buzzoni}, {\em \aap} {\bf 368}, 74
  (2001).

\bibitem{brunner:photoz}
R.~J. {Brunner}, A.~J. {Connolly}, A.~S. {Szalay} and M.~A. {Bershady}, {\em
  \apjl} {\bf 482}, p. L21  (1997).

\bibitem{vanzella:hdfannphotoz}
E.~{Vanzella} {\em et~al.}, {\em \aap} {\bf 423}, 761  (2004).

\bibitem{li:photoz}
L.-L. {Li}, Y.-X. {Zhang}, Y.-H. {Zhao} and D.-W. {Yang}, {\em Chinese Journal
  of Astronomy and Astrophysics} {\bf 7}, 448  (2007).

\bibitem{abrusco:sdssphotoz}
R.~{D'Abrusco}, A.~{Staiano}, G.~{Longo}, M.~{Brescia}, M.~{Paolillo}, E.~{De
  Filippis} and R.~{Tagliaferri}, {\em \apj} {\bf 663}, 752  (2007).

\bibitem{banerji:desvistaphotoz}
M.~{Banerji}, F.~B. {Abdalla}, O.~{Lahav} and H.~{Lin}, {\em \mnras} {\bf 386},
  1219  (2008).

\bibitem{oyaizu:photoz}
H.~{Oyaizu}, M.~{Lima}, C.~E. {Cunha}, H.~{Lin}, J.~{Frieman} and E.~S.
  {Sheldon}, {\em \apj} {\bf 674}, 768  (2008).

\bibitem{niemack:stripe82galexphotoz}
M.~D. {Niemack}, R.~{Jimenez}, L.~{Verde}, F.~{Menanteau}, B.~{Panter} and
  D.~{Spergel}, {\em \apj} {\bf 690}, 89  (2009).

\bibitem{wadadekar:svmphotoz}
Y.~{Wadadekar}, {\em \pasp} {\bf 117}, 79  (2005).

\bibitem{wang:novelphotoz}
D.~{Wang}, Y.-X. {Zhang}, C.~{Liu} and Y.-H. {Zhao}, {\em Chinese Journal of
  Astronomy and Astrophysics} {\bf 8}, 119  (2008).

\bibitem{carliles:photoz}
S.~{Carliles}, T.~{Budav{\'a}ri}, S.~{Heinis}, C.~{Priebe} and A.~{Szalay},
  {Photometric Redshift Estimation on SDSS Data Using Random Forests}, in {\em
  Astronomical Data Analysis Software and Systems XVII\/},  eds. R.~W.
  {Argyle}, P.~S. {Bunclark} and J.~R. {Lewis}, Astronomical Society of the
  Pacific Conference Series, Vol.~394 (2008).

\bibitem{ball:pdfphotoz}
N.~M. {Ball}, R.~J. {Brunner}, A.~D. {Myers}, N.~E. {Strand}, S.~L. {Alberts}
  and D.~{Tcheng}, {\em \apj} {\bf 683}, 12  (2008).

\bibitem{connolly:photoz}
A.~J. {Connolly}, I.~{Csabai}, A.~S. {Szalay}, D.~C. {Koo}, R.~G. {Kron} and
  J.~A. {Munn}, {\em \aj} {\bf 110}, p. 2655  (1995).

\bibitem{sowardsemmerd:photoz}
D.~{Sowards-Emmerd}, J.~A. {Smith}, T.~A. {McKay}, E.~{Sheldon}, D.~L. {Tucker}
  and F.~J. {Castander}, {\em \aj} {\bf 119}, 2598  (2000).

\bibitem{hsieh:photoz}
B.~C. {Hsieh}, H.~K.~C. {Yee}, H.~{Lin} and M.~D. {Gladders}, {\em \apjs} {\bf
  158}, 161  (2005).

\bibitem{lopes:photoz}
P.~A.~A. {Lopes}, {\em \mnras} {\bf 380}, 1608  (2007).

\bibitem{budavari:spectempl}
T.~{Budav{\'a}ri}, A.~S. {Szalay}, A.~J. {Connolly}, I.~{Csabai} and
  M.~{Dickinson}, {\em \aj} {\bf 120}, 1588  (2000).

\bibitem{csabai:photoz}
I.~{Csabai}, A.~J. {Connolly}, A.~S. {Szalay} and T.~{Budav{\'a}ri}, {\em \aj}
  {\bf 119}, 69  (2000).

\bibitem{csabai:edrphotoz}
I.~{Csabai} {\em et~al.}, {\em \aj} {\bf 125}, 580  (2003).

\bibitem{padmanabhan:photoz}
N.~{Padmanabhan} {\em et~al.}, {\em \mnras} {\bf 359}, 237  (2005).

\bibitem{brodwin:iracphotoz}
M.~{Brodwin} {\em et~al.}, {\em \apj} {\bf 651}, 791  (2006).

\bibitem{budavari:unified}
T.~{Budav{\'a}ri}, {\em \apj} {\bf 695}, 747  (2009).

\bibitem{budavari:qsophotoz}
T.~{Budav{\'a}ri} {\em et~al.}, {\em \aj} {\bf 122}, 1163  (2001).

\bibitem{richards:qsophotoz}
G.~T. {Richards} {\em et~al.}, {\em \aj} {\bf 122}, 1151  (2001).

\bibitem{babbedge:impz}
T.~S.~R. {Babbedge} {\em et~al.}, {\em \mnras} {\bf 353}, 654  (2004).

\bibitem{weinstein:qsophotoz}
M.~A. {Weinstein} {\em et~al.}, {\em \apjs} {\bf 155}, 243  (2004).

\bibitem{wu:qsophotoz}
X.-B. {Wu}, W.~{Zhang} and X.~{Zhou}, {\em Chinese Journal of Astronomy and
  Astrophysics} {\bf 4}, 17  (2004).

\bibitem{kitsonias:xrayagn}
S.~{Kitsionas}, E.~{Hatziminaoglou}, A.~{Georgakakis} and I.~{Georgantopoulos},
  {\em \aap} {\bf 434}, 475  (2005).

\bibitem{ball:ibphotoz}
N.~M. {Ball}, R.~J. {Brunner}, A.~D. {Myers}, N.~E. {Strand}, S.~{Alberts},
  D.~{Tcheng} and X.~{Llor{\`a}}, {\em \apj} {\bf 663}, p. 774  (2007).

\bibitem{kumar:ml}
N.~D. {Kumar}, {Machine learning techniques for astrophysical modelling and
  photometric redshift estimation of quasars in optical sky surveys}, Master's
  thesis, Oxford University  (2008).

\bibitem{wolf:qsopdf}
C.~{Wolf}, {\em \mnras} {\bf 397}, 520  (2009).

\bibitem{wolf:qsophotoz}
C.~{Wolf}, L.~{Wisotzki}, A.~{Borch}, S.~{Dye}, M.~{Kleinheinrich} and
  K.~{Meisenheimer}, {\em \aap} {\bf 408}, 499  (2003).

\bibitem{salvato:cosmosagnphotoz}
M.~{Salvato} {\em et~al.}, {\em \apj} {\bf 690}, 1250  (2009).

\bibitem{ramirez:ibga}
J.~F. {Ram{\'{\i}}rez}, O.~{Fuentes} and R.~K. {Gulati}, {\em Experimental
  Astronomy} {\bf 12}, 163  (2001).

\bibitem{solorio:ib}
T.~{Solorio}, O.~{Fuentes}, R.~{Terlevich} and E.~{Terlevich}, {\em \mnras}
  {\bf 363}, 543  (2005).

\bibitem{becker:transientclassification}
A.~C. {Becker}, {\em Astronomische Nachrichten} {\bf 329}, p. 280  (2008).

\bibitem{djorgovski:rare}
S.~G. {Djorgovski}, A.~A. {Mahabal}, R.~J. {Brunner}, R.~R. {Gal}, S.~{Castro},
  R.~R. {de Carvalho} and S.~C. {Odewahn}, {Searches for Rare and New Types of
  Objects}, in {\em Virtual Observatories of the Future\/},  eds. R.~J.
  {Brunner}, S.~G. {Djorgovski} and A.~S. {Szalay}, Astronomical Society of the
  Pacific Conference Series, Vol.~225 (2001).

\bibitem{bottino:foreground}
M.~{Bottino}, A.~J. {Banday} and D.~{Maino}, {\em \mnras} {\bf 389}, 1190
  (2008).

\bibitem{pires:sz}
S.~{Pires}, J.~B. {Juin}, D.~{Yvon}, Y.~{Moudden}, S.~{Anthoine} and
  E.~{Pierpaoli}, {\em \aap} {\bf 455}, 741  (2006).

\bibitem{phillips:nnparamest}
N.~G. {Phillips} and A.~Kogut, preprint, [arXiv:astro-ph/0108234]   (2001).

\bibitem{rohde:matching}
D.~J. {Rohde}, M.~R. {Gallagher}, M.~J. {Drinkwater} and K.~A. {Pimbblet}, {\em
  \mnras} {\bf 369}, 2  (2006).

\bibitem{taylor:evann}
M.~{Taylor} and A.~I. {Diaz}, {On the Deduction of Galactic Abundances with
  Evolutionary Neural Networks}, in {\em From Stars to Galaxies: Building the
  Pieces to Build Up the Universe\/},  eds. A.~{Vallenari}, R.~{Tantalo},
  L.~{Portinari} and A.~{Moretti}, Astronomical Society of the Pacific
  Conference Series, Vol.~374 (2007).

\bibitem{bogdanos:snega}
C.~{Bogdanos} and S.~{Nesseris}, {\em Journal of Cosmology and Astro-Particle
  Physics} {\bf 5}, p.~6  (2009).

\bibitem{li:svmflare}
R.~{Li}, Y.~{Cui}, H.~{He} and H.~{Wang}, {\em Advances in Space Research} {\bf
  42}, 1469  (2008).

\bibitem{mustard:mixture}
J.~F. {Mustard}, L.~{Li} and G.~{He}, {\em \jgr} {\bf 103}, 19419  (1998).

\bibitem{lemson:votheory}
G.~{Lemson} and J.~{Zuther}, {\em Memorie della Societa Astronomica Italiana}
  {\bf 80}, 342  (2009).

\bibitem{springel:millennium}
V.~{Springel} {\em et~al.}, {\em \nat} {\bf 435}, 629  (2005).

\bibitem{brun:root}
R.~{Brun} and F.~{Rademakers}, {\em Nuclear Instruments and Methods in Physics
  Research A} {\bf 389}, 81  (1997).

\bibitem{budavari:crossid}
T.~{Budav{\'a}ri} and A.~S. {Szalay}, {\em \apj} {\bf 679}, 301  (2008).

\bibitem{cunha:nzpdfphotoz}
C.~E. {Cunha}, M.~{Lima}, H.~{Oyaizu}, J.~{Frieman} and H.~{Lin}, {\em \mnras}
  {\bf 396}, 2379  (2009).

\bibitem{margoniner:photoz}
V.~E. {Margoniner} and D.~M. {Wittman}, {\em \apj} {\bf 679}, 31  (2008).

\bibitem{wittman:pzerror}
D.~{Wittman}, {\em \apjl} {\bf 700}, L174  (2009).

\bibitem{myers:pdf}
A.~D. {Myers}, M.~{White} and N.~M. {Ball}, {\em \mnras} {\bf 399}, 2279
  (2009).

\bibitem{vanbreukelen:cluster}
C.~{van Breukelen} and L.~{Clewley}, {\em \mnras} {\bf 395}, 1845  (2009).

\bibitem{bailerjones:gaiaquasars}
C.~A.~L. {Bailer-Jones}, K.~W. {Smith}, C.~{Tiede}, R.~{Sordo} and
  A.~{Vallenari}, {\em \mnras} {\bf 391}, 1838  (2008).

\bibitem{vaidya:ppdm}
J.~{Vaidya}, C.~{Clifton} and M.~{Zhu}., {\em {Privacy Preserving Data Mining}}
  (Springer, New York, 2005).

\bibitem{aggarwal:ppdm}
C.~C. {Aggarwal} and P.~S. {Yu} (eds.), {\em {Privacy-Preserving Data Mining:
  Models and Algorithms}} (Springer, New York, 2008).

\bibitem{scranton:sdsscovariance}
R.~{Scranton}, A.~J. {Connolly}, A.~S. {Szalay}, R.~H. {Lupton}, D.~{Johnston},
  T.~{Budav{\'a}ri}, J.~{Brinkmann} and M.~{Fukugita}, preprint,
  [arXiv:astro-ph/0508564]   (2005).

\bibitem{ivezic:lsst}
{\v Z}.~{Ivezi{\'c}}, J.~A. {Tyson}, R.~{Allsman}, J.~{Andrew}, R.~{Angel} and
  {for the LSST Collaboration}, preprint, [arXiv/0805.2366]   (2008).

\bibitem{donalek:synoptic}
C.~{Donalek}, A.~{Mahabal}, S.~G. {Djorgovski}, S.~{Marney}, A.~{Drake},
  E.~{Glikman}, M.~J. {Graham} and R.~{Williams}, {New Approaches to Object
  Classification in Synoptic Sky Surveys}, American Institute of Physics
  Conference Series Vol.~1082 (2008).

\bibitem{studenmund:econometrics}
A.~H. {Studenmund}, {\em {Using Econometrics}}, 2nd edn. (Addison-Wesley, New
  York, 2005).

\bibitem{mahabal:probabilistic}
A.~{Mahabal}, S.~G. {Djorgovski}, M.~{Turmon}, J.~{Jewell}, R.~R. {Williams},
  A.~J. {Drake}, M.~G. {Graham}, C.~{Donalek}, E.~{Glikman} and {Palomar-QUEST
  team}, {\em Astronomische Nachrichten} {\bf 329}, 288  (2008).

\bibitem{drake:voeventnet}
A.~J. {Drake}, R.~{Williams}, M.~J. {Graham}, A.~{Mahabal}, S.~G. {Djorgovski},
  R.~R. {White}, W.~T. {Vestrand} and J.~{Bloom}, {VOEventNet: An Open Source
  of Transient Alerts for Astronomers.}, Bulletin of the American Astronomical
  Society Vol.~38 (2007).

\bibitem{ivezic:lsstclassification}
{\v Z}.~{Ivezi{\'c}} {\em et~al.}, {Parametrization and Classification of 20
  Billion LSST Objects: Lessons from SDSS}, American Institute of Physics
  Conference Series Vol.~1082 (2008).

\bibitem{borne:lsstmining}
K.~{Borne}, J.~{Becla}, I.~{Davidson}, A.~{Szalay} and J.~A. {Tyson}, {The LSST
  Data Mining Research Agenda}, American Institute of Physics Conference Series
  Vol.~1082 (2008).

\bibitem{perryman:gaia}
M.~A.~C. {Perryman} {\em et~al.}, {\em \aap} {\bf 369}, 339  (2001).

\bibitem{bailerjones:domain}
C.~A.~L. {Bailer-Jones}, {A Method for Exploiting Domain Information in
  Astrophysical Parameter Estimation}, in {\em Astronomical Data Analysis
  Software and Systems XVII\/},  eds. R.~W. {Argyle}, P.~S. {Bunclark} and
  J.~R. {Lewis}, Astronomical Society of the Pacific Conference Series,
  Vol.~394 (2008).

\bibitem{djorgovski:pq}
S.~G. {Djorgovski} {\em et~al.}, {\em Astronomische Nachrichten} {\bf 329}, p.
  263  (2008).

\bibitem{drake:catalina}
A.~J. {Drake} {\em et~al.}, {\em \apj} {\bf 696}, 870  (2009).

\bibitem{hodapp:panstarrs}
K.~W. {Hodapp} {\em et~al.}, {\em Astronomische Nachrichten} {\bf 325}, 636
  (2004).

\bibitem{johnston:askap}
S.~{Johnston} {\em et~al.}, {\em Publications of the Astronomical Society of
  Australia} {\bf 24}, 174  (2007).

\bibitem{eyer:variability}
L.~{Eyer} {\em et~al.}, {Variability type classification of multi-epoch
  surveys}, American Institute of Physics Conference Series Vol.~1082 (2008).

\bibitem{kaczmarczik:astrometricphotoz}
M.~C. {Kaczmarczik}, G.~T. {Richards}, S.~S. {Mehta} and D.~J. {Schlegel}, {\em
  \aj} {\bf 138}, 19  (2009).

\bibitem{mahabal:transients}
A.~{Mahabal} {\em et~al.}, {Towards Real-time Classification of Astronomical
  Transients}, American Institute of Physics Conference Series Vol.~1082 (2008).

\bibitem{moore:law}
G.~E. {Moore}, {\em Electronics} {\bf 38}, 114  (1965).

\bibitem{bader:petascale}
D.~A. {Bader} (ed.), {\em {Petascale Computing: Algorithms and
  Applications}}, Computational Science Series (CRC
  Press, Boca Raton, FL, 2007).

\bibitem{amdahl:law}
G.~{Amdahl}, {Validity of the Single Processor Approach to Achieving
  Large-Scale Computing Capabilities}, in {\em Spring Joint Computer
  Conference\/},  (AFIPS Press, Atlantic City, N.J., 1967).

\bibitem{ebcioglu:percs}
K.~{Ebcioglu}, V.~{Saraswat} and V.~{Sarkar}, {The IBM PERCS Project and New
  Opportunities for Compiler-Driven Performance via a New Programming Model},
  {\em Compiler-Driven Performance Workshop (CASCON 2004)},  (2004).

\bibitem{szalay:graywulf}
A.~S. {Szalay} {\em et~al.}, {GrayWulf: Scalable Clustered Architecture for
  Data Intensive Computing}, {\em Hawaii International Conference on System
  Sciences} (IEEE Computer Society, Los Alamitos, CA, 2009).

\bibitem{szalay:petabyte}
A.~S. {Szalay}, J.~{Gray} and {Vandenberg, J.}, {Petabyte Scale Data Mining:
  Dream or Reality?}, {\em SPIE Astronomy Telescopes and Instruments}, Waikoloa,
  Hawaii,  (2002).

\bibitem{mcconnell:ddm}
S.~M. {McConnell} and D.~B. {Skillicorn}, {Distributed Data Mining for
  Astrophysical Datasets}, in {\em Astronomical Data Analysis Software and
  Systems XIV\/},  eds. P.~{Shopbell}, M.~{Britton} and R.~{Ebert},
  Astronomical Society of the Pacific Conference Series, Vol.~347 (2005).

\bibitem{freitas:parallel}
A.~A. {Freitas} and S.~H. {Lavington}, {\em Mining Very Large Databases with
  Parallel Processing} (Kluwer Academic Publishers, 1998).

\bibitem{kargupta:dpkd}
H.~{Kargupta} and P.~{Chan}, {\em {Advances in Distributed and Parallel
  Knowledge Discovery}} (AAAI/MIT Press, Cambridge, MA, 2000).

\bibitem{zaki:parallel}
M.~J. {Zaki} and C.~{Ho} (eds.), {\em Large-scale Parallel Data Mining},
  Lecture Notes in Artificial Intelligence, State-of-the-Art-Survey, Vol.~1759
  (Springer, New York, 2002).

\bibitem{bhaduri:ddmbib}
K.~{Bhaduri}, K.~{Liu}, H.~{Kargupta} and J.~{Ryan}, {Distributed Data Mining
  Bibliography} Online bibliography,  (2006).

\bibitem{jin:shared}
R.~{Jin}, G.~{Yang} and G.~{Agrawal}, {\em IEEE Transactions On Knowledge and
  Data Engineering} {\bf 17}, 71  (2005).

\bibitem{gray:semantic}
N.~{Gray}, {The Fact and Future of Semantic Astronomy}, in {\em Astronomical
  Data Analysis Software and Systems XVII\/},  eds. R.~W. {Argyle}, P.~S.
  {Bunclark} and J.~R. {Lewis}, Astronomical Society of the Pacific Conference
  Series, Vol.~394 (2008).

\bibitem{norman:enzo}
M.~L. {Norman}, G.~L. {Bryan}, R.~{Harkness}, J.~{Bordner}, D.~{Reynolds},
  B.~{O'Shea} and R.~{Wagner}, preprint, [arXiv/0705.1556]   (2007).

\bibitem{brunner:fpganpcf}
R.~J. {Brunner}, V.~{Kindratenko} and A.~D. {Myers}, {\em {Developing and
  Deploying Advanced Algorithms to Novel Supercomputing Hardware}}, tech. rep.,
  NASA  (2007).

\bibitem{gomez:socialnetworking}
E.~L. {Gomez}, H.~L. {Gomez} and J.~{Yardley}, preprint, [arXiv/0903.0266]
   (2009).

\bibitem{moore:npcf}
A.~W. {Moore} {\em et~al.}, {Fast Algorithms and Efficient Statistics: N-Point
  Correlation Functions}, in {\em Mining the Sky\/},  eds. A.~J. {Banday},
  S.~{Zaroubi} and M.~{Bartelmann} (2001).

\bibitem{gao:kdtree}
D.~{Gao}, Y.~{Zhang} and Y.~{Zhao}, {The Application of kd-tree in Astronomy},
  in {\em Astronomical Data Analysis Software and Systems XVII\/},  eds. R.~W.
  {Argyle}, P.~S. {Bunclark} and J.~R. {Lewis}, Astronomical Society of the
  Pacific Conference Series, Vol.~394 (2008).

\bibitem{gray:multitree}
A.~G. {Gray}, A.~W. {Moore}, R.~C. {Nichol}, A.~J. {Connolly}, C.~{Genovese}
  and L.~{Wasserman}, {Multi-Tree Methods for Statistics on Very Large Datasets
  in Astronomy}, in {\em Astronomical Data Analysis Software and Systems
  (ADASS) XIII\/},  eds. F.~{Ochsenbein}, M.~G. {Allen} and D.~{Egret},
  Astronomical Society of the Pacific Conference Series, Vol.~314 (2004).

\bibitem{shirasaki:adql}
Y.~{Shirasaki}, M.~{Ohishi}, Y.~{Mizumoto}, M.~{Tanaka}, S.~{Honda}, M.~{Oe},
  N.~{Yasuda} and Y.~{Masunaga}, {Structured Query Language for Virtual
  Observatory}, in {\em Astronomical Data Analysis Software and Systems XIV\/},
   eds. P.~{Shopbell}, M.~{Britton} and R.~{Ebert}, Astronomical Society of the
  Pacific Conference Series, Vol.~347 (2005).

\bibitem{derriere:ucd}
S.~{Derriere} {\em et~al.}, {UCD in the IVOA context}, in {\em Astronomical
  Data Analysis Software and Systems (ADASS) XIII\/},  eds. F.~{Ochsenbein},
  M.~G. {Allen} and D.~{Egret}, Astronomical Society of the Pacific Conference
  Series, Vol.~314 (2004).

\bibitem{dowler:caom2008}
P.~{Dowler}, S.~{Gaudet}, D.~{Durand}, R.~{Redman}, N.~{Hill} and S.~{Goliath},
  {Common Archive Observation Model}, in {\em Astronomical Data Analysis
  Software and Systems XVII\/},  eds. R.~W. {Argyle}, P.~S. {Bunclark} and
  J.~R. {Lewis}, Astronomical Society of the Pacific Conference Series,
  Vol.~394 (2008).

\bibitem{gray:decade}
J.~{Gray}, D.~T. {Liu}, M.~{Nieto-Santisteban}, A.~S. {Szalay}, D.~{DeWitt} and
  G.~{Heber}, {\em {Scientific Data Management in the Coming Decade}},
  Technical Report MSR-TR-2005-10, Microsoft Research  (2005).

\bibitem{gray:sdss}
J.~{Gray}, A.~S. {Szalay}, A.~R. {Thakar}, P.~Z. {Kunszt}, C.~{Stoughton},
  D.~{Slutz} and J.~{vandenBerg}, preprint, [arXiv:cs/0202014]   (2002).

\bibitem{vignali:vo}
C.~{Vignali}, F.~{Fiore}, A.~{Comastri}, M.~{Brusa}, R.~{Gilli},
  N.~{Cappelluti}, F.~{Civano} and G.~{Zamorani}, {Multi-wavelength data
  handling in current and future surveys: the possible role of Virtual
  Observatory}, in {\em Multi-wavelength Astronomy and Virtual Observatory\/},
  ed. {D.~Baines \& P.~Osuna} (2009).

\bibitem{gonzalezsolares:iphas}
E.~A. {Gonz{\'a}lez-Solares} {\em et~al.}, {\em \mnras} {\bf 388}, 89  (2008).

\bibitem{brescia:voneural}
M.~{Brescia} {\em et~al.}, {\em Memorie della Societa Astronomica Italiana}
  {\bf 80}, p. 565  (2009).

\bibitem{kitching:cloudcosmology}
T.~{Kitching}, A.~{Amara}, A.~{Rassat} and A.~{Refregier}, preprint,
  [arXiv/0901.3143]   (2009).

\bibitem{chilingarian:vo}
I.~V. {Chilingarian}, {Virtual Observatory for Astronomers: Where Are We Now?},
  in {\em Multi-wavelength Astronomy and Virtual Observatory\/},  ed.
  {D.~Baines \& P.~Osuna} (2009).

\bibitem{landsman:idlastro}
W.~B. {Landsman}, {The IDL Astronomy User's Library}, in {\em Astronomical Data
  Analysis Software and Systems II\/},  eds. R.~J. {Hanisch}, R.~J.~V.
  {Brissenden} and J.~{Barnes}, Astronomical Society of the Pacific Conference
  Series, Vol.~52 (1993).

\bibitem{taylor:topcat}
M.~B. {Taylor}, {TOPCAT \& STIL: Starlink Table/VOTable Processing Software},
  in {\em Astronomical Data Analysis Software and Systems XIV\/},  eds.
  P.~{Shopbell}, M.~{Britton} and R.~{Ebert}, Astronomical Society of the
  Pacific Conference Series, Vol.~347 (2005).

\bibitem{comparato:viz}
M.~{Comparato}, U.~{Becciani}, A.~{Costa}, B.~{Larsson}, B.~{Garilli},
  C.~{Gheller} and J.~{Taylor}, {\em \pasp} {\bf 119}, 898  (2007).

\bibitem{urunkar:vomegaplot}
N.~{Urunkar}, A.~K. {Kembhavi}, A.~{Navelkar}, J.~{Pandya}, V.~{Moosani},
  P.~{Nair} and M.~{Shaikh}, {\em Highlights of Astronomy} {\bf 14}, 620
  (2007).

\bibitem{taylor:plastic}
J.~D. {Taylor}, T.~{Boch}, M.~{Comparato}, M.~{Taylor}, N.~{Winstanley} and
  R.~G. {Mann}, {Binding Applications Together with PLASTIC}, in {\em
  Astronomical Data Analysis Software and Systems XVI\/},  eds. R.~A. {Shaw},
  F.~{Hill} and D.~J. {Bell}, Astronomical Society of the Pacific Conference
  Series, Vol.~376 (2007).

\bibitem{taylor:stilts}
M.~B. {Taylor}, {STILTS - A Package for Command-Line Processing of Tabular
  Data}, in {\em Astronomical Data Analysis Software and Systems XV\/},  eds.
  C.~{Gabriel}, C.~{Arviset}, D.~{Ponz} and S.~{Enrique}, Astronomical Society
  of the Pacific Conference Series, Vol.~351 (2006).

\bibitem{borkin:astromed}
M.~{Borkin}, A.~{Goodman}, M.~{Halle} and D.~{Alan}, {Application of Medical
  Imaging Software to 3D Visualization of Astronomical Data}, in {\em
  Astronomical Data Analysis Software and Systems XVI\/},  eds. R.~A. {Shaw},
  F.~{Hill} and D.~J. {Bell}, Astronomical Society of the Pacific Conference
  Series, Vol.~376 (2007).

\bibitem{scranton:googlesky}
R.~{Scranton} {\em et~al.}, preprint, [arXiv/0709.0752]   (2007).

\bibitem{barnes:s2plot}
D.~G. {Barnes}, C.~J. {Fluke}, P.~D. {Bourke} and O.~T. {Parry}, {\em
  Publications of the Astronomical Society of Australia} {\bf 23}, 82  (2006).

\bibitem{fluke:interchanging}
C.~J. {Fluke}, D.~G. {Barnes} and N.~T. {Jones}, {\em Publications of the
  Astronomical Society of Australia} {\bf 26}, 37  (2009).

\bibitem{levy:partiview}
S.~{Levy}, {Interactive 3-D visualization of particle systems with Partiview},
  in {\em Astrophysical Supercomputing using Particle Simulations\/},  eds.
  J.~{Makino} and P.~{Hut}, IAU Symposium, Vol.~208 (2003).

\bibitem{szalay:gpuvis}
T.~{Szalay}, V.~{Springel} and G.~{Lemson}, preprint, [arXiv/0811.2055]
  (2008).

\bibitem{hut:virtual}
P.~{Hut}, {Virtual Laboratories and Virtual Worlds}, IAU Symposium
  Vol.~246 (2008).

\bibitem{ebisuzaki:grape}
T.~{Ebisuzaki}, J.~{Makino}, T.~{Fukushige}, M.~{Taiji}, D.~{Sugimoto},
  T.~{Ito} and S.~K. {Okumura}, {\em \pasj} {\bf 45}, 269  (1993).

\bibitem{gaburov:sapporo}
E.~{Gaburov}, S.~{Harfst} and S.~{Portegies Zwart}, {\em New Astronomy} {\bf
  14}, 630  (2009).

\bibitem{belleman:cuda}
R.~G. {Belleman}, J.~{B{\'e}dorf} and S.~F. {Portegies Zwart}, {\em New
  Astronomy} {\bf 13}, 103  (2008).

\bibitem{ord:gpu}
S.~{Ord}, L.~{Greenhill}, R.~{Wayth}, D.~{Mitchell}, K.~{Dale}, H.~{Pfister}
  and R.~G. {Edgar}, preprint, [arXiv/0902.0915]   (2009).

\bibitem{garcia:knngpu}
V.~{Garcia}, E.~{Debreuve} and M.~{Barlaud}, preprint, [arXiv/0804.1448]
  (2008).

\bibitem{brown:fpga}
S.~D. {Brown}, R.~J. {Francis}, J.~{Rose} and Z.~G. {Vranesic}, {\em
  {Field-Programmable Gate Arrays}}, The Springer International Series in
  Engineering and Computer Science (Springer, New York, 1992).

\bibitem{buell:hprc}
D.~{Buell}, T.~{El-Ghazawi}, K.~{Gaj} and V.~{Kindratenko}, {\em Computer} {\bf
  40}, 23  (2007).

\bibitem{won:annfpga}
E.~{Won}, {\em Nuclear Instruments and Methods in Physics Research A} {\bf
  581}, 816  (2007).

\bibitem{freeman:similarity}
M.~{Freeman}, M.~{Weeks} and J.~{Austin}, {Hardware implementation of
  similarity functions}, in {\em IADIS AC\/}, 2005.

\bibitem{scarpino:cell}
M.~{Scarpino}, {\em {Programming the Cell Processor: For Games, Graphics, and
  Computation}} (Prentice Hall PTR, New York, 2008).

\end{thebibliography}



\end{document}